\documentclass[prx,reprint,floatfix,amsmath,amssymb]{revtex4-2}

\usepackage[utf8]{inputenc}
\usepackage{enumitem,dsfont,bm,bbold}
\usepackage{hyperref}
\usepackage{array,multirow,braket}
\usepackage{graphicx,xcolor}

\graphicspath{{./}{./figures/}}
 
\newcommand{\vk}{\vec k}
\newcommand{\ve}{\vec e}
\newcommand{\vb}{\vec b}

\renewcommand{\vr}{\vec r}
\newcommand{\vtau}{\vec \tau}

\newcommand{\vR}{\vec R}
\newcommand{\vG}{\vec G}

\newcommand{\vh}{\vec h}

\newcommand{\vsigma}{\mbox{\boldmath $\sigma$}}

\renewcommand{\vec}[1]{\mathbf{#1}}
\renewcommand{\vtau}{\boldsymbol{\tau}}

\renewcommand\a{ \mathrm{a} }
\renewcommand\b{ \mathrm{b} }
\renewcommand\c{ \mathrm{c} }
\renewcommand\d{ \mathrm{d} }
\newcommand\g{ \mathrm{g} }

\newcommand\G\Gamma

\newcommand\M{ {\mathrm{M}} }

\newcommand\X{ {\mathrm{X}} }
\newcommand\Y{ {\mathrm{Y}} }

\newcommand\sG{{\overline\G}}
\newcommand\sM{{\overline\M}}

\newcommand\C{ {\mathrm{C}} }
\newcommand\D{ {\mathrm{D}} }
\newcommand\I{ {\mathrm{I}} }
\newcommand\T{ {\mathrm{T}} }

\newcommand\cket[1]{\ket{#1}_\text{c}}
\newcommand\Rshift{ {\delta\vR} }

\newcommand{\dg}{\dagger}
\newcommand{\pdg}{{\phantom{\dagger}}}
\newcommand\e{ \mathrm{e} }
\renewcommand\i{ \mathrm{i} }
\newcommand\wt[1]{\widetilde{#1}}

\newcommand{\abs}[1]{\left| #1 \right|}

\newcommand\diag{\text{diag}}

\newcommand{\sgn}{{\rm sgn}\,}
\newcommand\id{\openone}

\newcommand\ie{\textit{i.e.}}
\newcommand\eg{\textit{e.g.}}
\newcommand\viz{\textit{viz}}

\newcommand{\U}{\mathrm{U}}

\renewcommand{\O}{\mathrm{O}}
\newcommand{\SO}{\mathrm{SO}}
\newcommand{\Sp}{\mathrm{Sp}}

\newcommand{\ZZ}{\mathbb{Z}}
\newcommand{\NN}{\mathbb{N}}
\newcommand{\RR}{\mathbb{R}}
\newcommand{\CC}{\mathbb{C}}
\newcommand\FF{\mathbb{F}}
\newcommand\HH{\mathbb{H}}

\newcommand\cwc{ \mathbb{D} }
\newcommand\Xsp{\mathcal{X}}
\newcommand\Gr{\mathrm{Gr}}

\newcommand\wilson{ {W} }

\newcommand{\Chern}{\mathrm{C}}
\newcommand\Euler{\mathrm{Eu}}
\newcommand\SW{\mathrm{SW}}
\newcommand\Hopf{\mathrm{H}}

\newcommand\trans{\Lambda}
\newcommand\symm{C}
\newcommand\sqrtC{B}
\newcommand{\Q}{{\cal Q}}
\newcommand\uop{\mathcal{U}}

\newcommand\nuc{n}

\newcommand\ksurf{\vk_\perp}

\newcommand\rotn{C}
\newcommand\mirror{M}
\renewcommand\o{ \mathrm{o} }

\newcommand\highlight[1]{{\color{blue}#1}}
\newcommand\para[1]{\noindent  \emph{#1 ---} }

\begin{document}

\title{Homotopic classification of band structures: Stable, fragile, delicate, and stable representation-protected topology}
\author{Piet W. Brouwer}
\author{Vatsal Dwivedi}
\affiliation{Dahlem Center for Complex Quantum Systems and Physics Department, Freie Universit\"at Berlin, Arnimallee 14, 14195 Berlin, Germany}
\date{\today}

\begin{abstract}
	The topological classification of gapped band structures depends on the particular definition of topological equivalence. For translation-invariant systems, stable equivalence is defined by a lack of restrictions on the numbers of occupied and unoccupied bands, while imposing restrictions on one or both leads to ``fragile'' and ``delicate'' topology, respectively. In this article, we describe a homotopic classification of band structures --- which captures the topology beyond the stable equivalence --- in the presence of additional lattice symmetries. 	As examples, we present complete homotopic classifications for spinless band structures with twofold rotation, fourfold rotation and fourfold dihedral symmetries, both in presence and absence of time-reversal symmetry. Whereas the rules of delicate and fragile topology do not admit a bulk-boundary correspondence, we identify a version of stable topology, which restricts the representations of bands, but not their numbers, which does allow for anomalous states at  symmetry-preserving boundaries, which are associated with nontrivial bulk topology.
\end{abstract}

\maketitle

\section{Introduction}
The concept of topological equivalence, combined with symmetry constraints, has been an important tool to classify gapped band structures of non-interacting fermions \cite{hasan2010,qi2011,bernevig2013,ando2015}. Topological gapped band structures include the quantized Hall effect, the quantum spin-Hall effect, topological insulators, and, if crystalline symmetries are imposed, topological crystalline insulators. A nontrivial topology of the band structure implies the existence of anomalous, gapless states on the sample boundary and vice versa, provided the boundary respects the symmetry constraints that are imposed on the bulk. 

Various notions of topological equivalence of band structures have been employed in the literature, which differ in the possible transformations that are allowed to deform two topologically equivalent Hamiltonians into each other. The standard definition in condensed matter physics has been the ``stable equivalence'' \cite{kitaev2009}, which allows addition of an arbitrary number of ``trivial'' bands which respect the symmetry requirements. This definition ensures that the topology realized in mathematical models, often consisting of only a few bands, is realized in physical systems consisting of atoms with infinitely many orbitals. Indeed, a plethora of stable topological phases \cite{hasan2010,qi2011,bernevig2013,ando2015}, that were originally discovered theoretically, have been realized experimentally, such as the quantum Hall effect, the quantum spin Hall insulator and the three dimensional topological insulators.  

A conceptually simpler, although physically less useful, definition of topological equivalence does away with the freedom to add of extra bands, resulting in ``delicate'' topology \cite{kennedy2015homotopy,nelson2021multicellularity}. The paradigmatic example of such a band structure is the so-called Hopf insulator \cite{moore2008,deng2013hopf,kennedy2015homotopy,kennedy2016hopf}, realized by two-band models in three dimensions without any additional symmetries. A third intermediate possibility, which involves the freedom to add unoccupied bands while keeping the occupied subspace unchanged, results in ``fragile'' topology \cite{bradlyn2019,po2018,hwang2019,liu2019shift}. Nontrivial fragile topology has been invoked to explain, {\em e.g.}, the obstruction to the construction of real-space models for the lowest two bands of magic-angle twisted bilayer graphene \cite{zou2018,song2019magic,po2019,ahn2019nielsen,zelatel2019}. In general, topology that is trivial according to the rules of stable topology, but nontrivial following the delicate or fragile schemes is referred to as ``unstable'

The classification of stable topological phases with only local symmetries leads to the well-known ``periodic table of topological insulators and superconductors'' \cite{kitaev2009,schnyder2008,schnyder2009}. Stable stable topological insulators with additional crystalline symmetries have been comprehensively classified for many crystalline symmetries \cite{turner2012,chiu2013,shiozaki2014,trifunovic2019,cornfeld2019}. However, delicate and fragile band structures have only been identified systematically in the absence of crystalline symmetries \cite{kennedy2016} or, with crystalline symetries, using partial classifications or symmetry-specific arguments \cite{bradlyn2019,po2018,hwang2019,liu2019shift,bouhon2020,song2020,song2020b,kobayashi2021}. 

In this article, we show how delicate band structures can be classified systematically using the theory of homotopy on CW complexes \cite{hatcher2002}. Similar techniques have previously been used by Moore and Balents to rederive the quantum spin Hall effect and to obtain the classification of topological insulators in three dimensions \cite{moore2007}. Our classification procedure is defined for a fixed number of occupied and empty bands, thereby yielding the classification of delicate topological phases. Classifications for the stable and fragile equivalence schemes naturally follow by taking the limit of a large number of conduction and/or valence bands. We illustrate the classification procedure by (re)deriving delicate classifications of spinless band structures in two and three dimensions without crystalline symmetry, and with $\C_2 \T$, $\C_2$, $\C_4$, and $\D_4$ symmetries, where $\C_n$ and $\D_n$ refer to $n$-fold rotation and dihedral symmetries and $\T$ is time-reversal. Band structures with $\C_2 \T$, $\C_2$, and $\C_4$ symmetry have been used as paradigmatic examples of unstable topology protected by crystalline symmetries \cite{kennedy2015homotopy,fang2015nonsymmorphic,bouhon2020,kobayashi2021,alexandradinata2014,alexandradinata2016,alexandradinata2020,song2020,song2020b}, which are systematized by our classification. 

Stable topology is an essential prerequisite for the existence of anomalous gapless boundary states. This follows from elementary considerations: As the rules of fragile and delicate topology restrict the number of occupied and/or unoccupied bands, the size of the unit cell must be fixed, which requires the existence of a discrete translation symmetry in all spatial directions. Since a boundary necessarily breaks the translation symmetry, it follows that anomalous boundary states cannot exist without stable protection \footnote{
	Anomalous boundary states also exist for the so-called {\em weak} \cite{fu2007ti} topology, whereby the topological invariant for a $d$-dimensional system is protected by discrete translation symmetry in less than $d$ dimensions. The anomalous boundary modes occur only at boundaries that respect the discrete translation symmetry. 
}. 
A more subtle route to protected gapless boundary states beyond conventional stable topology was pointed out by Song {\em et al.} \cite{song2020b}, Alexandradinata  {\em et al.} \cite{alexandradinata2020}, and Kobayashi and Furusaki \cite{kobayashi2021} for a $\C_4$-symmetric lattice model originally proposed by Fu \cite{fu2011}. In this case, anomalous surface states are protected by topology that remains stable under addition of bands that are not only constrained by a symmetry, but also by its particular representation (\ie, the types of orbitals). We term this {\em representation-protected} stable topology. Following our systematic classification, we identify such phases and the associated anomalous boundary states not only for $\C_4$-symmetric band structures, but also for $\C_2$ and $\D_4$ symmetries. For the latter, the representation-protected phase is intimately linked to the existence of a two-dimensional irreducible representation of the discrete nonabelian group $\D_4$.

Despite the absence of anomalous boundary states, unstable topology has physical relevance, first and foremost, because it may signal an obstruction to the construction of a real-space basis with localized orbitals within the available number of bands, as in the case for magic-angle twisted bilayer graphene \cite{zou2018,song2019magic,po2019,ahn2019nielsen,zelatel2019}. Unstable topology has been shown to affect bulk physics, such as the appearance of Bloch oscillations with an extended period \cite{hoeller2018} and connected, unbounded Hofstadter butterflies in the presence of a magnetic field \cite{lian2020,liu2019landau,lu2021}. Various \emph{boundary signatures} beyond anomalous gapless boundary states have also been proposed \footnote{Some of these signatures can be interpreted as manifestations of various other forms of stable topology, \viz, that of topologically localized insulators \cite{lapierre2022} for Hopf insulators and that of obstructed atomic limits for the twisted boundary signature and the fractional corner charges.}, 
such as a nonzero Chern number for the surface bands in the case of Hopf insulators\cite{alexandradinata2021,lapierre2021}, boundary states for a pre-defined crystal termination \cite{nelson2021multicellularity}, gap closing under twisted boundary condition \cite{song2020,peri2020}, and fractional corner charges \cite{benalcazar2019,ahn2019nielsen}. In this article, we do not venture into such manifestations of unstable topology and focus exclusively on the classification problem and the anomalous boundary states associated with various forms of stable topology. 

The remainder of this article is organized as follows: In Sec.~\ref{sec:gen}, we describe the general classification procedure and the required mathematical background. In Sec.\ \ref{sec:C0} we apply the classification procedure to band structures without crystalline symmetries. Section \ref{sec:outline} contains general observations concerning the classification of band structures with crystalline symmetries, which is followed by specific examples of increasing complexity: Band structures with $\C_2 \T$ symmetry in Sec.\ \ref{sec:C2T}, with $\C_2$ symmetry in Sec.\ \ref{sec:C2}, with $\C_4$ symmetry in Sec.\ \ref{sec:C4}, and with $\D_4$ symmetry in Sec.\ \ref{sec:D4}. We conclude with a discussion in Sec.\ \ref{sec:concl}. Various technical details of the classification procedure have been relegated to the appendices.

\section{Classification strategy}
\label{sec:gen}
In this section, we describe the homoptopic classification for band structures with arbitrary lattice symmetries, which is based on well-established methods of algebraic topology for CW complexes. Our classification procedure is applicable to all ten Altland-Zirnbuaer classes \cite{altland1997}; however, to avoid overburdenening the present exposition, we restrict the explicit discussion to Bloch Hamiltonians without antisymmetry requirements, corresponding to the Altland-Zirnbauer classes A, AI, and AII. These classes include all topological crystalline insulators, but not topological superconductors or lattice models with a sublattice symmetry. In the examples of Secs.\ \ref{sec:C0}-\ref{sec:D4} we further restrict to symmetry classes A and AI, describing insulators with spin-rotation symmetry with and without time-reversal symmetry.

\subsection{Stable, fragile, delicate, and representation protected topologies} 
\label{sec:gen_top}
We begin with some general remarks on various notions of topological equivalence for translation invariant systems, which are completely described by a finite-dimensional Bloch Hamiltonian $H(\vk)$. Two Bloch Hamiltonians $H_{1,2}$ are considered \emph{homotopically} equivalent,  
\begin{equation}
	H_1 \sim H_2,
	\label{eq:delicate}
\end{equation}
if $H_1$ can be continuously deformed to $H_2$ without closing the gap or violating the symmetry constraints imposed on $H_{1,2}$. A looser notion is that of \emph{stable} equivalence, whereby two Bloch Hamiltonians $H_{1,2}$ are considered topologically equivalent if there exists a gapped Bloch Hamiltonian $H$, subject to the same symmetry constraints as $H_{1,2}$, so that
\begin{equation}
	H_1 \oplus H \sim H_2 \oplus H
	\label{eq:stable}
\end{equation}
in the homotopic sense \footnote{
	This definition requires that $H_{1,2}$ have equal number of conduction and valence bands. However, if the symmetries allow for the definition of a canonical ``trivial'' band, then this definition can be further relaxed to allow comparison of systems with different number of bands by adding the trivial bands.
}.
Finally, \emph{fragile} equivalence is defined by Eq.~(\ref{eq:stable}) with the further constraint on $H$ that it consist only of unoccupied bands. In mathematical terms, the problem of fragile and stable classification can be expressed as the computation of equivalence classes of vector bundles, with the equivalence being isomorphism \cite{denittis2014,denittis2015,kennedy2015homotopy} for the former case and isomorphism \emph{up to Whitney sums} for the latter, which falls under the purview of topological K-theory \cite{kitaev2009}. Given a choice of a reference band structure $H_\text{ref}$, a Bloch Hamiltonian $H$ is said to have ``stable topology'' if it is not equivalent to $H_\text{ref}$ under stable equivalence. If $H$ and $H_\text{ref}$ are  topologically equivalent under stable equivalence, but not under homotopic equivalence or vector bundle isomorphism, then $H$ is said to possess ``unstable'' topology. The space of stably equivalent band structures can be endowed with a group structure, the group operation corresponding to the direct sum of the Bloch Hamiltonians \footnote{
	Strictly speaking, defining the trivial element for the group operation does not require the identification of a trivial band structure, but, instead, proceeds via the Grothendieck construction, which considers topological equivalence of {\em pairs} $(H,H')$ of gapped band structures. Two pairs $(H_1,H_1')$ and $(H_2,H_2')$ are then considered topologically equivalent if $H_1 \oplus H_2'$ can be smoothly deformed into $H_2 \oplus H_1'$, without closing the excitation gap and without violating the symmetry constraints. 
}. 
As this group operation is incompatible with the restrictions imposed by fragile and delicate topology, they do not have a similar group structure. 

The rules of topological classification can be generalized by \emph{a priori} limiting the types of orbitals and Wyckoff positions allowed, \ie, restricting the band representations. Such \emph{representation-protected} phases have been previously investigated in Refs.~\cite{fu2011,song2020b,kobayashi2021}, whereby $\C_4$-symmetric band structures were shown to exhibit a topological phase for arbitrary number of bands, provided one only allows $p$-orbitals. As in the unrestricted case, one may have delicate, fragile, and stable versions of representation-protected topological classifications. All of these are sometimes referred to in the literature as ``fragile'' \cite{song2020b,kobayashi2021}; however, stable representation-protected topology exhibits many of the characteristics of conventional stable topology, such as a group structure and the existence of anomalous states at boundaries that respect the symmetry restrictions imposed on the bulk.

Our classification procedure starts by fixing the numbers $n_{\o}$ and $n_{\e}$ of occupied and unoccupied bands and thus naturally identifies all possible delicate topological phases. The fragile classification is obtained by taking the limit $n_\e\to\infty$ for a fixed $n_\o$, while the stable classification is obtained when we take $n_{\o,\e}\to\infty$. As our goal is to obtain a full homotopic classification, we respect symmetry-imposed obstructions between atomic-limit phases and do not follow the sometimes-used custom \cite{bradlyn2017,po2017,song2018quantitative} to refer to all band structures that are homotopically equivalent to an atomic-limit as ``trivial''. To infer whether a certain homotopic equivalence class corresponds to an atomic-limit phase, one has to compare the topological invariants to the topological invariants of atomic limit phases. An exhaustive listing of atomic-limit phases for all crystalline symmetry groups has been obtained in Ref.\ \cite{bradlyn2017} in the framework of ``topological quantum chemistry''.

\subsection{Lattice symmetries and the Bloch Hamiltonian} 
\label{sec:gen_bz}
We consider a $\nuc \times \nuc$ Bloch Hamiltonian $H(\vk)$ in the basis of Bloch states $\ket{\vk,\alpha}$, which are defined as 
\begin{equation}
  \ket{\vk,\alpha} 
    = \sum_{\vR} \e^{\i \vk \cdot (\vR+\vr_\alpha)} \ket{\vR,\alpha}.
  \label{eq:ketk}
\end{equation}
Here, $\ket{\vR,\alpha}$ denote the position-space basis states, with $\vR \in \ZZ^d$ the center of the unit cell and $\alpha=1,\ldots,\nuc$ an index labeling a set of $\nuc$ orbitals at positions $\vr_\alpha \in (-1/2,1/2]^d$ inside the unit cell. The lattice symmetries impose constraints on $H(\vk)$, which are encoded by a representation $C(g)$ of the \emph{reciprocal-space group} $G_\Lambda$ \cite{alexandradinata2014} as
\begin{equation}
  H(g \vk) = C^\dg(g) H(\vk) C(g),\ \ g \in G_{\Lambda}. \label{eq:Hsymm}
\end{equation}
The reciprocal-space group is generated by the point group and translations by reciprocal lattice vectors. The latter act nontrivially on $H(\vk)$, since our gauge choice for the Bloch states --- made to ensure that the matrices $C(g)$ representing the reciprocal-space group are independent of $\vk$ --- yields a Bloch Hamiltonian that is not periodic under reciprocal lattice translations. Instead, it satisfies
\begin{equation}
  H(\vk+2 \pi \ve_a) = \trans_a^\dg H(\vk) \trans_a^\pdg,
  \label{eq:Hsymm_ktrans}
\end{equation}
where $ a \in \{x,y,z\}$ and we have set $\trans_a \equiv C(g_a)$, where $g_a$ denotes translation by a reciprocal unit vector along the $a$-direction. 
Antiunitary symmetries, if present, are also included in $G_\Lambda$; however, for an antiunitary $\bar g \in G_{\Lambda}$, Eq.\ (\ref{eq:Hsymm}) must be replaced with
\begin{equation}
  H(\bar g \vk) = C^\dg(\bar g) H^*(\vk) C(\bar g), \quad  \bar g \in G_{\Lambda}.
  \label{eq:Hsymmb}
\end{equation}
The precise form of the representation matrices $C(g)$ depends on which orbitals are present. The relevant information is encoded in the Wyckoff positions of the orbitals and their representations under the relevant point group symmetries, which must be considered ``input data'' for the homotopic classification.

In view of the symmetry constraints (\ref{eq:Hsymm}) and (\ref{eq:Hsymmb}), the Bloch Hamiltonian $H(\vk)$ for all $\vk$ can be derived from its knowledge over a subset of the full reciprocal space. Such a subset is referred to as the ``irreducible Brillouin zone'', the ``effective Brillouin zone'' \cite{moore2007}, or the ``fundamental domain'' corresponding to the reciprocal-space group $G_\Lambda$
\footnote{
  Mathematically, the fundamental domain can be defined as the closure of the largest open set in the BZ on which the group action of $G$ is free (\ie, no point maps onto itself under the action of any non-identity element of the $G$).
}.
By construction, no two momenta in the fundamental domain are related by a symmetry operation in $G_\Lambda$, so that, within the fundamental domain, all symmetry restrictions on $H(\vk)$ are local in $\vk$. 

\subsubsection*{Fundamental domain as a CW complex}
The fundamental domain for a $d$-dimensional system with a given group of symmetries $G_\Lambda$ can be naturally constructed in terms of ``$p$-cells'', which are homotopically equivalent to the $p$-dimensional cube $(0,1)^p$ for $p \ge 1$ and points for $p = 0$. Starting from a disjoint union of $p$-cells with $0\leq p\leq d$, one recovers the structure of the fundamental domain by imposing a set of \emph{gluing maps}. 
For each $p$-cell, these maps assign to its boundary a closed loop formed by $q$-cells of lower dimensions, which may include multiple copies of a given $q$-cell related by symmetry operations in $G_\Lambda$. This construction endows the fundamental domain with the structure of a \emph{CW complex} \footnote{A similar construction, but in position space, appears in the construction of a ``topological crystal'' in Ref.\ \cite{song2019}.} --- a construct well known in the field of algebraic topology \cite{hatcher2002}.

\begin{figure}
	\includegraphics[width=0.98\columnwidth]{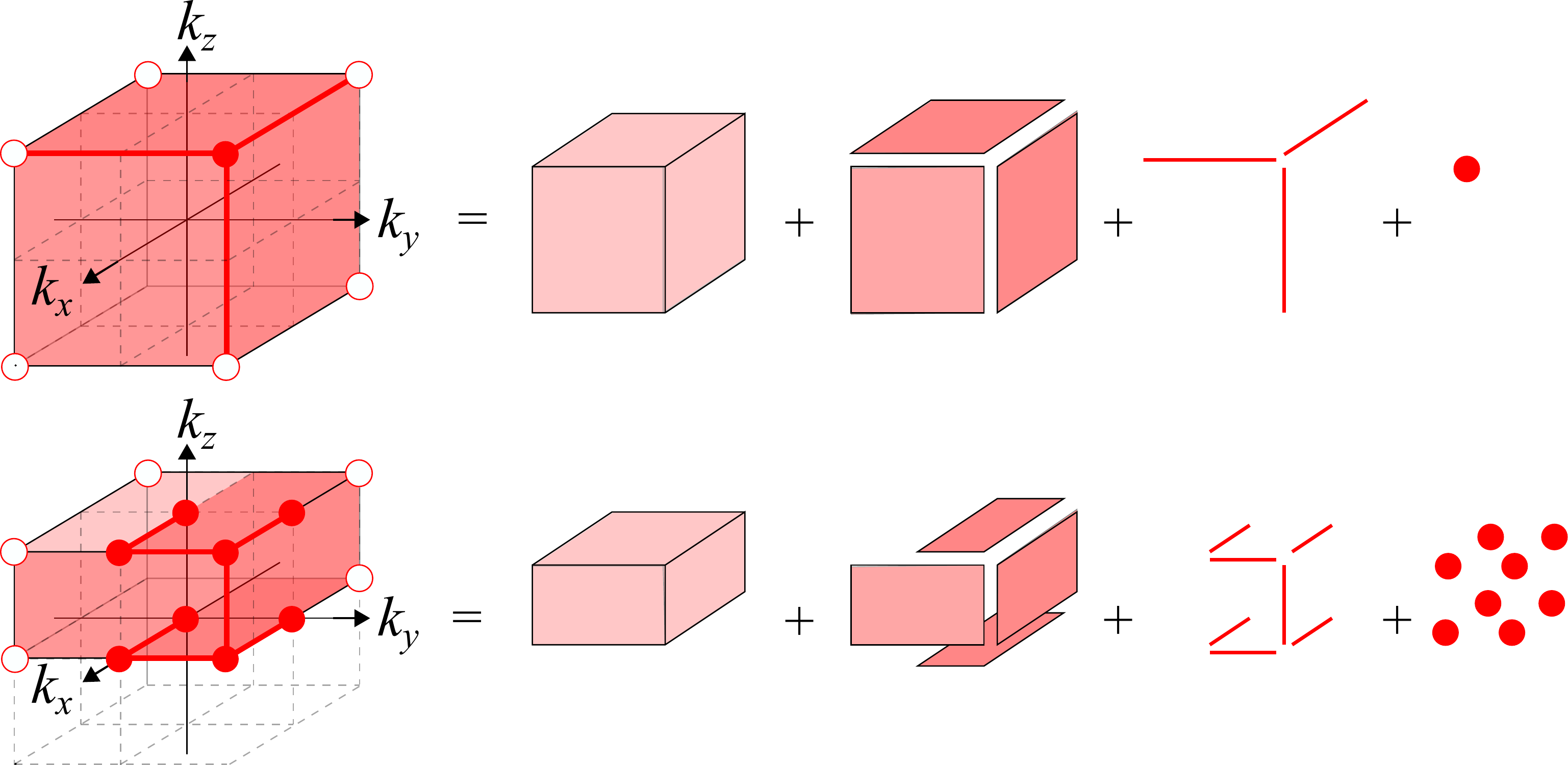}
	\caption{\label{fig:fundamental_general} Fundamental domain $\cwc$ for a three-dimensional band structure on a simple cubic lattice without (top) and with (bottom) time-reversal symmetry. Knowledge of the Bloch Hamiltonian on the fundamental domain is sufficient to determine it everywhere in the reciprocal space. Also shown is the decomposition of the fundamental domain into $p$-cells with $p = 0, 1,2,3$ (red dots, line segments, rectangles, and cuboid), endowing it with the structure of a ``CW complex''. Copies of 0-cells on the boundary $\partial \cwc$ that are obtained by a reciprocal lattice transformation are shown using open dots.}
\end{figure}

For example, in $d=1$ with inversion symmetry, we can take the set $[0,\pi]$ as the fundamental domain, which can be constructed by gluing the two 0-cells $\G$ and $\M$, which consists of the points $k=0$ and $k=\pi$, respectively, to the ends of the 1-cell $(0,\pi)$. Further examples of the fundamental domain and its decomposition into various $p$-cells are shown in Fig.~\ref{fig:fundamental_general} for a three-dimensional band structure without additional crystalline symmetries and without (top) and with (bottom) time-reversal symmetry.
For the fundamental domain shown in Fig.\ \ref{fig:fundamental_general} (top), the gluing condition identifies the boundary $\partial\cwc_3$ of the 3-cell with the three 2-cells, the three 1-cells, and the single 0-cell, with each of them occuring two, four and eight times on the boundary, respectively. Similarly, in the presence of time-reversal symmetry, the boundary $\partial\cwc_3$ of the 3-cell in Fig.~\ref{fig:fundamental_general} (bottom) is identified with the four 2-cells, seven 1-cells, and eight 0-cells and with their images under translation with a reciprocal wavevector and/or time-reversal, which sends $\vk \to -\vk$.

We hereafter denote the fundamental domain with this additional structure by $\cwc$ and its $p$-dimensional subset by $\cwc_p$. The action of the reciprocal-space group $G_\Lambda$ on $\cwc$ has two aspects:
\begin{enumerate}
  \item A given $p$-cell $\alpha$ may be left invariant by a subgroup $G_\alpha \subset G_\Lambda$, which is termed its \emph{little group}. These must be compatible with the gluing maps, in that the little groups of the $q$-cells $\beta\in\partial\alpha$ satisfy $G_\beta \supseteq G_\alpha$.
  Cells with $G_\alpha \neq 1$ are usually referred to as high-symmetry points/lines/planes.

  \item If multiple copies of a $q$-cell $\beta$ exist in the boundary $\partial\alpha$, then these copies are related by an element of the reciprocal-space group $G_\Lambda$, which is unique up to elements of the little group $G_\beta$.
\end{enumerate}
Both aspects are evident in Fig.~\ref{fig:fundamental_general} (bottom): All $0$-cells have a nontrivial little group isomorphic to $\ZZ_2$, generated by a combination of time-reversal and a translation by a reciprocal lattice vector, while the 1-cells have trivial little groups. Furthermore, in defining the boundary of the 2-cells, five of the seven 1-cells occur four times, whereas two of the 1-cells occur twice, with copies related by time-reversal and/or a reciprocal lattice translation. 

A $G_\Lambda$-symmetric Bloch Hamiltonian $H$ can be defined abstractly as a map $H\colon \cwc\to\Xsp$, where the target space $\Xsp$ consists of all gapped Hermitian $n \times n$ matrices. This map consists of a collection of maps $H_\alpha \colon \alpha \to \Xsp_\alpha$ for each $p$-cell $\alpha$, which are constrained by the CW-complex structure of $\cwc$. Explicitly, these maps must be compatible with the gluing map, \ie, if $\beta \in \partial \alpha$, then the restriction of $H_\alpha$ to $\beta$ is equal to $H_\beta$. Since $\alpha$ is an open set, we assume that the maps $H_\alpha$ can be continuously extended to its boundary. For instance, in Fig.~\ref{fig:fundamental_general} (top), the gluing condition implies that the maps $H_\beta$ for the three 2-cells $\beta$ as well as their images under a reciprocal lattice translation must coincide with $H_\alpha$ defined on the 3-cell $\alpha$, as one approaches the respective boundaries of $\alpha$. 

The maps constituting the Bloch Hamiltonian satisfy further constraints, which descend from the corresponding constraints on the $p$-cells:
\begin{enumerate}
  \item[1$'$.] The target space $\Xsp_\alpha \subseteq \Xsp$ consists of matrices that are invariant under the action of the little group $G_\alpha \subset G_\Lambda$, see Eqs.\ (\ref{eq:Hsymm}) and (\ref{eq:Hsymmb}). (This property assumes a representation of $G_{\Lambda}$ on $\Xsp$, see Subsec. \ref{sec:gen_bz}.) Consequently, the eigenstates of $H_\alpha$ can be labeled by the irreps of $G_\alpha$. For $\beta \in\partial\alpha$, the relation between little groups $G_\beta \supseteq G_\alpha$ implies that $\Xsp_\beta \subseteq \Xsp_\alpha$.

  \item[2$'$.] If two copies of $\beta$ in $\partial \alpha$ are related by $g \in G_\Lambda$, then the Hamiltonian on those two copies is related by the symmetry transformation of Eq.~(\ref{eq:Hsymm}) or (\ref{eq:Hsymmb}).
\end{enumerate}

\subsection{Homotopic classification}
\label{sec:gen_clfn}
We have reduced the classification problem of band structures with a given lattice symmetry to the classification of maps from a CW complex $\cwc$ to certain target spaces. The latter can be performed using the cellular decomposition of $\cwc$, starting at $0$-cells and inductively working our way up to the $d$-cells. We refer to the classification of $H_\alpha$ defined on $p$-cells $\alpha$ as the ``level-$p$ topological classification''. As shown below, classifications at different levels  are interlinked in both directions: 
\begin{enumerate}
	\item The precise value of the topological invariants of levels $<p$ determines the allowed types of level-$p$ topological invariants, and
	
	\item The consideration of level-$p$ topology yields compatibility relations constraining the topological invariants associated with all levels $\le p$.
\end{enumerate}

\subsubsection{Topology at level-0: Symmetry indicators}
We begin with a classification of zero-dimensional Hamiltonians, \viz, the restriction $H_S$ of the Bloch Hamiltonian $H$ to the 0-cell $S$. $H_S$ can be block diagonalized, with blocks corresponding to the irreps of the little group $G_S$. Since there are no further constraints within  each block, their topological class is completely determined by $n_{\o/\e,\rho}^S \geq 0$, the number of occupied/empty orbitals of irrep $\rho$.  This can be generalized to any $p$-cell $\alpha$, where the numbers of occupied/empty orbitals $n_{\o/\e,\rho}^\alpha \geq 0$ are well-defined for the restriction of $H$ to $\alpha$. If $G_{\alpha}$ has a single irrep, then the corresponding level-0 invariants are simply the total number of occupied/empty bands $n_{\o/\e}$.

Together, the natural numbers $n_{\o/\e,\rho}^\alpha$ constitute the level-0 invariants, which form the semigroup $\prod_\alpha \NN^{\nu_\alpha}$, where $\nu_\alpha$ is the number of irreps of $G_\alpha$. For a given $p$-cell $\alpha$, if two $q$-cells $\beta,\beta' \in \partial\alpha$, then $\Xsp_\beta, \Xsp_{\beta'} \subseteq \Xsp_\alpha$ and $H_\beta$ and $H_{\beta'}$ are homotopically equivalent within $\Xsp_\alpha$. Put differently, $H_\beta$ and $H_{\beta'}$ are homotopically equivalent, when we lift the symmetry restrictions on them to $G_\alpha$ for any cell $\alpha$ connecting them. This property constrains the values level-$0$ invariants can take \cite{kruthoff2017,po2017,bradlyn2017}.  In particular, since all $p$-cells $\alpha$ with $p<d$ lie on the boundary of $\cwc$, the corresponding level-0 invariants satisfy 
\begin{equation}
	n_{\o/\e} = \sum_{\rho=1}^{\nu_{\alpha}} \dim(\rho) \; n_{\o/\e, \rho}^\alpha, 
	\label{eq:lev0_constr}
\end{equation}
where $\nu_\alpha$ is the number of irreps of $G_\alpha$ and $\dim(\rho)$ is the dimension of the irrep $\rho$. Physically, this constraint expresses that the number of occupied/empty bands $n_{\o/\e}$ must be the same throughout reciprocal space for gapped band structures. This constraint is an example of a \emph{level-0 compatibility condition}. There may be further constraints on the level-0 invariants of a $q$-cell by their inclusion in a $p$-cells for $p<d$, which we discuss as required. 

For a given band, the set of representations of the little groups $G_\alpha$ define a \emph{band representation}. An arbitrary band representation can be decomposed into a direct sum of \emph{elementary band representations} (EBRs), which can be enumerated by considering all possible orbital types localized at high-symmetry points of the lattice. A systematic study of possible band representations has been central to the partial classifications based on ``symmetry-based indicators'' \cite{kruthoff2017,po2017,tang2019,tang2019comprehensive,zhang2019catalogue,tang2019efficient} and ``topological quantum chemistry'' \cite{bradlyn2017,cano2018,cano2021bandrep,vergniory2019complete}. The full homotopic classification is thus an extension of these partial classifications. 

\begin{figure}
	\includegraphics[width=0.75\columnwidth]{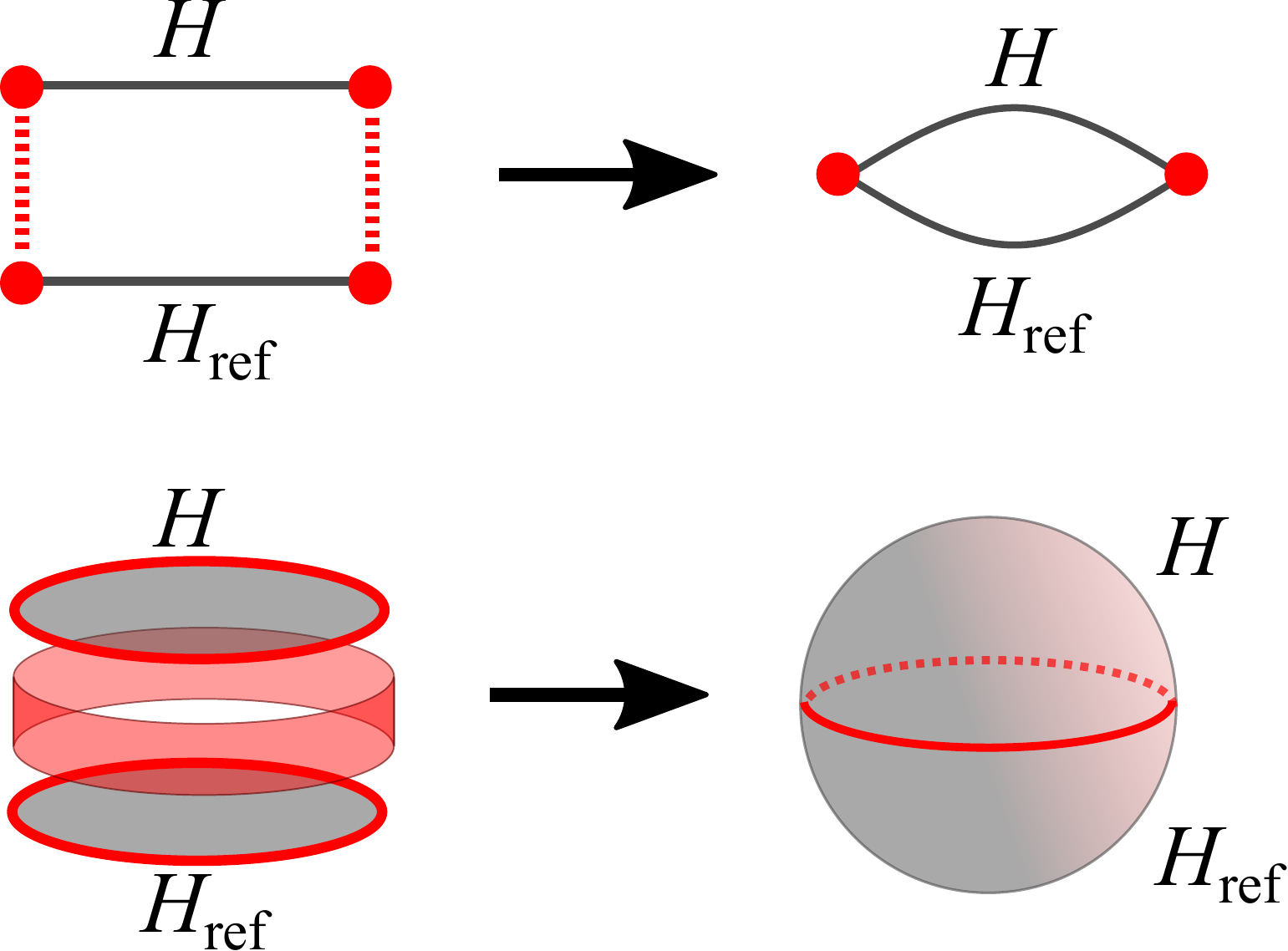}
	\caption{\label{fig:d1s1} 
		For two Hamiltonians $H$ and $H_\text{ref}$ whose topological invariants on the boundary of a $q$-cell are identical, we can define a boundary deformations (shown in gray) that renders the two Hamiltonians equal at the boundary. One thereby obtains a Bloch Hamiltonian defined on a $q$-sphere, which can be classified by a homotopy group. The figure shows this procedure for $q=1$ (top) and $q=2$ (bottom). 
	}
\end{figure}

\subsubsection{Topology at levels $p>0$}
We now turn to topology associated with the $p$-cells for $p>0$. We start with two observations:
\begin{itemize}
	\item The topological class of $H$ is always determined relative to a reference  $H_{\rm ref}$. The classification thus addresses \emph{pairs} of Bloch Hamiltonians $(H, H_\text{ref})$.
	
	\item If $H$ and $H_{\rm ref}$ have different invariants on level-$q$ for any $q<p$, then they are topologicaly inequivalent. Thus, for classification on $p$-cells, we may assume that $H$ and $H_\text{ref}$ have identical invariants up to level $p-1$.
\end{itemize}
The latter observation implies that for a $p$-cell $\alpha$, the restrictions of $H$ and $H_\text{ref}$ to $\partial \alpha$ can be smoothly deformed into each other, and, furthermore, that this deformation can be done simultaneously for all $p$-cells. Choosing such a smooth deformation, the ``difference'' between $H$ and $H_\text{ref}$ then defines a map from the $p$-sphere $S^p$ to $\Xsp_\alpha$ for each $p$-cell $\alpha$. This is illustrated in Fig.~\ref{fig:d1s1} for the cases $p=1$ and $p=2$. Such maps are classified by the homotopy spaces $K_{\alpha} \equiv \pi_{p} [\Xsp_\alpha]$, which we term the \emph{parent classification set} of the $p$-cell $\alpha$. The full level-$p$ parent classification set then reads
\begin{equation}	
	K_p = \prod_\alpha K_\alpha; \quad 
    K_\alpha = \pi_{p} \left[ \Xsp_\alpha \right].
\end{equation}

\begin{figure}
	\includegraphics[width=0.75\columnwidth]{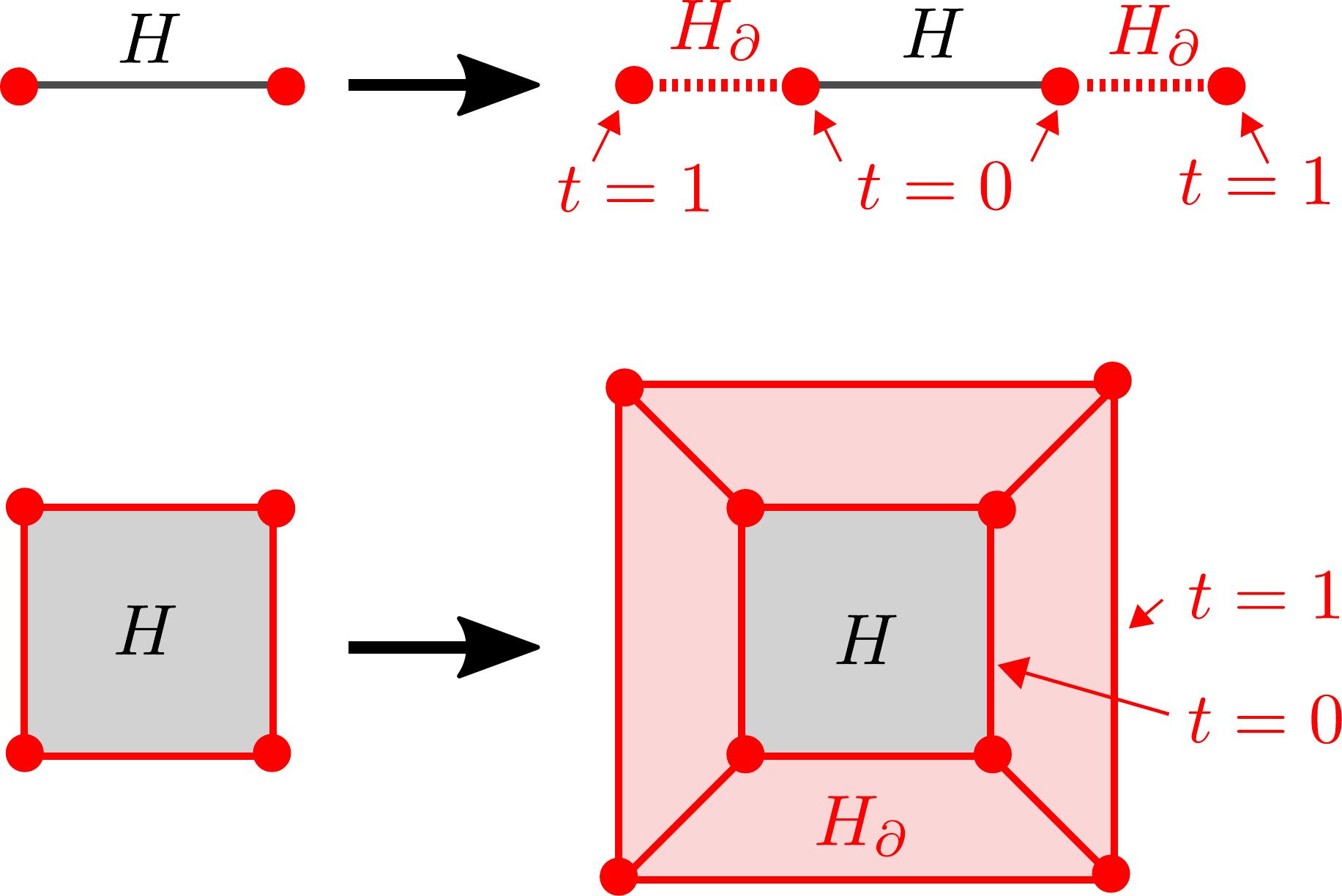}
	\caption{\label{fig:boundary_transform} Boundary deformation of a band structure $H(\vk)$ defined on a 1-cell (top) and on a 2-cell (bottom). A boundary deformation $H_{\partial}(\vk_{\partial},t)$, $0 \le t \le 1$, is continously added onto $H(\vk)$ at the boundary of the cell, such that $H(\vk)$ coincides with the reference band structure $H_{\rm ref}(\vk)$ on the cell boundary $\partial\cwc_m$ for $t=0$ and $t=1$ and $H(\vk)$ satisfies all symmetry constraints at each point in the transformation.}
\end{figure}

The parent classification set $K_p$ is, however, not the actual classifying set for equivalence classes of Bloch Hamiltonians $H$, because a given pair $(H,H_\text{ref})$ can yield different elements in $K_{p}$, depending on the deformation used to achieve equality of $H$ and $H_\text{ref}$ on $\partial\cwc_p$. Such elements of $K_{p}$ must be identified to get a true descriptor of homotopic equivalence classes. Explicitly, any deformation used to achieve equality of $H$ and $H_{\rm ref}$ on $\partial \cwc_p$ can be written as a $p$-dimensional ``Bloch Hamiltonian'' $H_{\partial}(\vk_{\partial},t)$, where $\vk_{\partial} \in \partial\cwc_p$, $t \in [0,1]$, and $H_\partial$ satisfies 
\begin{equation}
	H_{\partial}(\vk_{\partial},0) = H(\vk_\partial), \quad 
	H_{\partial}(\vk_{\partial},1) = H_\text{ref}(\vk_\partial), 
\end{equation}
as well as symmetry constraints for each $q$-cell in $\partial\cwc_p$ for any $t$. The ``difference'' of two boundary deformations transforming $H$ to $H_{\rm ref}$ on $\partial \cwc_p$ can be viewed as a boundary deformation that connects $H_{\rm ref}$ to itself, {\em i.e.}, with 
\begin{equation}
  H_{\partial}(\vk_{\partial},0) = H_{\partial}(\vk_{\partial},1) = H_{\rm ref}(\vk_{\partial}).
\end{equation}
These ``boundary deformations'' can be visualized as in Fig.~\ref{fig:boundary_transform}, whereby one imagines that $H_{\partial}(\vk_{\partial},t)$ is continuously ``grown'' onto the band structure defined by $H$ on $\alpha$ for each $p$-cell $\alpha \in \cwc_p$. The level-$p$ topological invariants are then obtained from $K_{p}$ by identification of elements related by deformations of this type.

To determine the action of boundary deformations $H_{\partial}$ on the elements of $K_p$, it is sufficient to consider at least one boundary deformation from each topological equivalence class of boundary deformations. Considering multiple boundary deformations from the same topological class is unproblematic, because they have the same action on $K_{p}$ and therefore lead to identification of the same elements. Since the domain $\partial \cwc_p \times [0,1]$ of a boundary deformation is again a CW complex, these topological equivalence classes can be determined in the same manner as described above. It is, in fact, sufficient to only determine the corresponding parent classification set 
\begin{equation}
	K_{p}^{\partial} = \prod_{\alpha\in\partial\cwc_p} \pi_{r_{\alpha}}[\Xsp_{\alpha}],
	\label{eq:Kboundaryparent}
\end{equation}
where $r_{\alpha} = \dim\alpha+1$. This is because the parent classification set $K_{p}^{\partial}$ is ``overcomplete'', so that considering one boundary deformation from each class in $K_{p}^{\partial}$ ensures that {\em at least} one boundary transformation from each (true) topological class is considered. Only having to determine $K_{p}^{\partial}$ leads to a substantial simplification by avoiding having to deal with ``boundary transformations of boundary transformations''. For the reader that is nevertheless interested in obtaining the exact boundary classification set, we review a recursive procedure in Appendix~\ref{app:CW}.

There are additional constraints on the (parent) topological invariants, which arise from the fact that the individual cells are contractible. For any $(p+1)$-cell $\alpha$ with $1 \le p \le d-1$, the boundary $\partial\alpha$ can thus be continuously deformed to a point within $\alpha$. Consequently, the topology of any Bloch Hamiltonian $H$ restricted to $\partial\alpha$, evaluated with respect the (larger) target space $\Xsp_\alpha \supseteq \Xsp_\beta$ for $\beta\in\partial\alpha$, must be trivial. This implies that the ``sum'' of level-$p$ invariants for $p$-cells $\beta\in\partial\alpha$ must vanish. We term these conditions --- one for each $(p+1)$-cell ---  the \emph{level-$p$ compatibility conditions}. Since this sum over level-$p$ invariants may also depend on topological invariants at levels $q < p$, the level-$p$ compatibility conditions may affect invariants at all levels $q \le p$. The compatibility condition equally apply to the boundary deformations and their parent classification $K_p^{\partial}$. 

To summarize, the topological invariants at all levels $0 \leq p \leq d$, modulo the ambiguities associated with boundary deformations, and constrained by the compatibility relations at all levels, constitute the complete classification of the $d$-dimensional systems. The compatibility relations ensure that for each set of topological invariants there is a gapped band structure with these invariants, while the robustness to boundary deformations ensures that the topological invariants can be uniquely associated with a gapped band structure. The classification naturally has a hierarchical reciprocal-space structure \cite{kennedy2015homotopy}, since the classification at level-$p$ depends on the invariants at all lower level, starting with the symmetry based indicators at level-0.

\subsection{Target spaces} 
\label{sec:gen_Gr}
We now discuss the various target spaces $\Xsp$ that arise in the classification for symmetry classess A, AI, and AII and describe their homotopy groups.

\subsubsection{Grassmannians and their topology}
We are interested in the space of $n\times n$ Hermitian matrices with a fixed number of occupied and empty bands (\ie, negative and positive eigenvalues) $n_{\o/\e}$ satisfying
\begin{equation}
  n = n_\o + n_\e.
\end{equation}
For a topological classification, $H(\vk)$ may be replaced by a Hermitian matrix with a ``flattened spectrum'' by deforming all positive/negative eigenvalues to $\pm1$, so that $H(k)$ is unitarily equivalent to $\diag \{\id_{n_\e}, -\id_{n_\o}\}$. The space of such Hamiltonians is the \emph{complex Grassmannian} $\Gr_{\CC}(n_\o,n_\e)$, which is a manifold that parametrizes $n_\o$-dimensional (or equivalently, $n_\e$-dimensional) complex planes in $\CC^n$ (see App.\ \ref{app:Gr} for more details). We also encounter Bloch Hamiltonians $H(\vk)$ with real $(\RR)$ or quaternion $(\HH$) entries, for which the target space is the corresponding Grassmannian $\Gr_{\RR}(n_\o,n_\e)$ or $\Gr_{\HH}(n_\o,n_\e)$, respectively. These three spaces constitute the top-level target spaces $\Xsp$ for the symmetries considered here. 

For a $p$-cell $\alpha$ with a nontrivial little group $G_\alpha$, the target spaces $\Xsp_\alpha \subseteq \Xsp$ can be written as a product of Grassmannians. This is because the eigenstates of $H$ restricted to $\alpha$ can be labeled with irreps $\rho$ of $G_\alpha$, so that after a spectral flattening, we get 
\begin{equation}
  \Xsp_\alpha = \prod_\rho \Gr_{{\FF}_\rho} \left( n_{\o,\rho}^\alpha, n_{\e,\rho}^\alpha \right),
\end{equation}
where $\rho$ runs over the irreps of $G_\alpha$ and $\FF_\rho$ is a number field, which, depending on $\rho$, may be $\CC$, $\RR$, or $\HH$. The latter two occur only if $G_\alpha$ contains antiunitary symmetries. Since we are interested in $\pi_p[\Xsp_\alpha]$ and the homotopy of a product of spaces satisfies 
\begin{equation}
  \pi_k \Big[ \prod_\alpha X_\alpha \Big] = \prod_\alpha \pi_k \left[ X_\alpha \right],
\end{equation}
we only need to compute $\pi_p[\Gr_\FF(n_{\o},n_{\e})]$.

Homotopy groups of Grassmannians have been extensively studied in the mathematical literature. Their computation is briefly outlined in Appendix~\ref{app:Gr}. We, however, make two general observations here: 
\begin{itemize}
	\item $\pi_p[\Gr_\FF(n_{\o},n_{\e})] = 0$ for all $p>0$ if $n_\o = 0$ or $n_\e = 0$, since in that case, $\Gr_\FF(n_{\o},n_{\e})$ degenerates to a single point.
	
	\item $\pi_p[\Gr_\FF(n_{\o},n_{\e})]$ becomes independent of $n_{\o}$ and $n_{\e}$ (``stabilizes'') once $n_{\o}$ and $n_{\e}$ are large enough, which typically means $n_{\o/\e} \gtrsim p$. 
\end{itemize}
As to the former observation, while $n_{\o,\e} > 0$ for gapped band structure, $\Gr_\FF(n_{\o},n_{\e})$ with $n_\o = 0$ or $n_\e = 0$ may occur in individual symmetry sectors at high-symmetry manifolds. The latter observation forms the basis for the classification of ``stable'' topology.

The homotopy invariants of Grassmannians can often be expressed in terms of Wilson lines or loops \cite{alexandradinata2014b, bouhon2019}. Given a path $\gamma$ in the reciprocal space, the Wilson line $\wilson^{\o/\e}(\gamma) \in \U(n)$ can be associated with the occupied/empty bands and encodes the evolution of the corresponding subspaces under parallel transport along $\gamma$. Wilson lines are not gauge invariant, since they depend on the choice of basis at the end points of $\gamma$. They are unambiguously defined only if there are additional constraints on the end points of $\gamma$ that fix the basis or if the path $\gamma$ is a closed loop. A local-in-$\vk$ symmetry on $H(\vk)$ implies a restriction on Wilson lines/loops; in particular, if the Hamiltonian is restricted by symmetries to be real- or quaternion-valued, then they are restricted to $\O(n), \, \Sp(n) \subset \U(n)$, respectively.

\subsubsection{Topological invariants for complex Grassmannians}
\label{sec:gen_Gr_cmplx}
We now discuss the nontrivial topological invariants associated with complex Grassmannians, which are summarized in Table.~\ref{tab:hom_cmplx}.  

\ \\ \para{$\pi_2[\Gr_{\CC}(n_\o,n_\e)]$} 
These stabilize to $\ZZ$ for all $n_{\o/\e} > 0$, with the corresponding invariant being the Chern number $\Chern$. 
For a complex Hamiltonian, $\det \wilson^{\o/\e}(\gamma)$ is a phase factor, \ie, it is defined on the unit circle. To define the Chern number, we take a family of Wilson loops $\wilson^{\o/\e}(\gamma_t)$ with $t\in[0,1]$ and $\gamma_0 = \gamma_1$ so that the $\gamma_t$ cover the two-sphere $S^2$ precisely once, in a manner compatible with the orientation of $S^2$. The Chern number $\Chern$ is then equal to the winding number of $\det(\wilson^{\o}({\gamma_t}))$ for $0 \le t \le 1$. As the total Chern number of a set of bands must vanish, the Chern number of the unoccupied bands equals $-\Chern$.

\ \\ \para{$\pi_3[\Gr_{\CC}(n_\o,n_\e)]$}
The only nontrivial case occurs for $n_\o = n_\e = 1$, with the topological invariant being the $\ZZ$-valued Hopf invariant. This arises from the fact that $\Gr_\CC(1,1) \cong  S^2$, so that its third homotopy group $\pi_3(S^2) \cong \ZZ$ is generated by the Hopf map. We refer to App.\ \ref{app:Hopf} for more details.

\begin{table}
	\centering
	\begin{tabular}{|r|l|l|l|}
		\hline
		$d$ & Bands 
		& \multicolumn{2}{c|}{Topological Invariant} \\
		\hline	
		1 & --- & \multicolumn{2}{c|}{---} \\
		\hline	
		2 & --- & $\ZZ$ & Chern number $\Chern$ \\
		\hline		
		\multirow{2}{*}{3}  
		& $n_\o,n_\e = 1$ & $\ZZ$ &  Hopf invariant $\Hopf$ \\ \cline{2-4}
		& $n_\o + n_\e > 2$ & \multicolumn{2}{c|}{---} \\ 
		\hline 
	\end{tabular}
	\caption{Topological invariants associated with complex Grassmannians $\Gr_\CC(n_\o, n_\e) \cong \Gr_\CC(n_\e, n_\o)$.}
	\label{tab:hom_cmplx}
\end{table}

\subsubsection{Topological invariants for real Grassmannians}
\label{sec:gen_Gr_real}
The nontrivial topological invariants associated with real Grassmannians are summarized in Table.~\ref{tab:hom_real}. These exhibit multiple low-dimensional exceptions.

\ \\ \para{$\pi_1[\Gr_{\RR}(n_\o,n_\e)]$}
These groups stabilize to $\ZZ_2$ for $n_{\o/\e} > 1$, with the corresponding topological invariant being the \emph{first Stiefel-Whitney invariant} $\SW_1$. To define this invariant, we note that the Wilson loops are orthogonal matrices if the Hamiltonian is real, so that $\det \wilson^{\o/\e}(\gamma) \in \{\pm1\}$ for any closed loop $\gamma$. The first Stiefel-Whitney invariant is defined as \cite{ahn2019stiefel}
\begin{equation}
  \SW_1 \equiv \sgn \left( \det \wilson^{\o/\e}(\gamma) \right).
  \label{eq:SW1_winding}
\end{equation}
As the total set of bands must be topologically trivial, $\SW_1$ may be calculated either from the occupied or from the empty bands. Since $\wilson^{\o/\e}(\gamma) \in \O(n_{\o/\e})$ encodes the basis transformation of the occupied/empty subspace under parallel transport along $\gamma$, a nontrivial $\SW_1$ represents an obstruction to defining an orientation of these subspaces continuously along $\gamma$.

For $n_\o = n_\e = 1$, this $\ZZ_2$ classification can be refined to a $\ZZ$ classification. To see this, we note that after flattening the spectrum, a gapped real $2 \times 2$ Hamiltonian can be written as 
\begin{equation}
  H(\vk) = \cos\theta_\vk \, \sigma_1 + \sin\theta_\vk \, \sigma_3, 
\end{equation}
where $\sigma_j$ are the Pauli matrices. Thus, for any closed path in the reciprocal space, $\vk\mapsto\theta_\vk$ defines a map from $S^1$ to $S^1$. Such maps are classified by a winding number $\Euler_1 \in \pi_1[S^1] \cong \ZZ$, termed the \emph{first Euler class}. The sign of the Euler class depends on a choice of orientation for $\Gr_\RR(1,1)$, since a unitary transformation can flip the sign of $\sigma_1$, thereby flipping the sign of $\Euler_1$ \cite{bouhon2020}. Thus, in absence of a preferred orientation, $\Euler_1$ is defined only up to a sign, so that the space of topological invariants is $\NN$ (instead of $\ZZ$). The first Euler invariant is related to the first Stiefel-Whitney invariant as
\begin{equation}
	\SW_1 = \Euler_1 \mod 2.
\end{equation}

\begin{table}
\setlength{\tabcolsep}{5pt}
\centering
\begin{tabular}{|r|l|l|l|}
	\hline
	$d$ & & \multicolumn{2}{c|}{Topological invariant(s)} \\
	\hline
	\multirow{2}{*}{1}  
	& $n_\o,n_\e = 1$  & $\ZZ$ & 1$^\text{st}$ Euler invariant $\Euler_1$  \\ \cline{2-4}
	& $n_\o, n_\e > 1$ & $\ZZ_2$ & 1$^\text{st}$ Stiefel-Whitney invariant $\SW_1$ \\
	\hline
	\multirow{12}{*}{2}  
	& $n_\o,n_\e = 1$ &  --- &  --- \\ \cline{2-4}
	& $n_\o = 1$, & \multirow{2}{*}{$2\ZZ$} & \multirow{2}{*}{2$^\text{nd}$ Euler invariant $\Euler_2^\o$} \\
	& $n_\e = 2$ && \\ \cline{2-4}
	& $n_\o = 2$, & \multirow{2}{*}{$2\ZZ$} & \multirow{2}{*}{2$^\text{nd}$ Euler invariant $\Euler_2^\e$} \\
	& $n_\e = 1$ && \\ \cline{2-4}     
	&  \multirow{2}{*}{$n_\o,n_\e = 2$} & \multirow{2}{*}{$\ZZ^2|_{\rm p}$} & 2$^\text{nd}$ Euler invariants $(\Euler_2^{\e},\Euler_2^{\e})$ \\
	& & & satisfying $\Euler_2^\o + \Euler_2^\e = 0 \mod 2$
	\\ \cline{2-4}
	& $n_\o = 2$, & \multirow{2}{*}{$\ZZ$}  & \multirow{2}{*}{2$^\text{nd}$ Euler invariant $\Euler_2^{\o}$} \\ 
	& $n_\e > 2$  && \\ \cline{2-4}
	& $n_\o > 2$, & \multirow{2}{*}{$\ZZ$}  & \multirow{2}{*}{2$^\text{nd}$ Euler invariant $\Euler_2^{\e}$} \\ 
	& $n_\e = 2$  && \\ \cline{2-4}
	& $n_\o, n_\e > 2$ &  $\ZZ_2$ & 2$^\text{nd}$ Stiefel-Whitney invariant $\SW_2$ \, \\ \hline
\end{tabular}
\caption{Topological invariants associated with real Grassmannians $\Gr_\RR(n_\o, n_\e) \cong \Gr_\RR(n_\e, n_\o)$. The $\ZZ$-valued invariants are defined only up to a sign ambiguity if the Grassmannian does not have a preferred orientation. The set $\ZZ^2|_{\rm p}$ consists of all ordered integer pairs $(p,q)$ with $p = q \!\! \mod 2$.}
\label{tab:hom_real}
\end{table}

\ \\ \para{$\pi_2[\Gr_{\RR}(n_\o,n_\e)]$}
These groups stabilize to $\ZZ_2$ for $n_{\o/\e} > 2$, with the corresponding topological invariant being the \emph{second Stiefel-Whitney invariant} $\SW_2$. If $n_{\o/\e} = 2$, the corresponding invariants are the \emph{second Euler invariants} $\Euler_2 ^{\o/\e}$. To define these invariants, we consider a one-parameter family of Wilson loops $\wilson^{\o/\e}({\gamma_t}) \in \O(n_{\o/\e})$, where $t \in [0,1]$ such that $\gamma_0 = \gamma_1$ and the $\gamma_t$ cover the two-sphere $S^2$ precisely once, in a matter compatible with its orientation. Such a family of Wilson loops defines a closed curve in $\O(n_{\o/\e})$. For $n_{\o/\e} = 2$, one associates an integer-valued winding number $\Euler_2^{\o/\e}$ with such a closed loop, since $2 \times 2$ orthogonal matrices are parameterized by a single phase variable. For $n_{\o/\e} > 2$ the winding number is only defined mod $2$, with the parity being the $\ZZ_2$-valued second Stiefel-Whitney number $\SW_{2}^{\o/\e}$ \cite{ahn2019stiefel}. 

Since the combination of all bands must be topologically trivial, one has $\SW_{2}^{\o} = \SW_{2}^{\e}$, so that one may write $\SW_2$ instead of  $\SW_{2}^{\o}$ and $\SW_{2}^{\e}$. The second Euler and Stiefel-Whitney invariants (if defined) further satisfy the parity constraint
\begin{equation}
  \Euler_2^{\o} = \Euler_2^{\e} = \SW_2 \mod 2,
\end{equation}
If $n_{\o} = 1$ or $n_{\e}=1$, continuity enforces that $\wilson^\o(\gamma_t) = 1$ or $\wilson^\e(\gamma_t) = 1$ for all $t \in [0,1]$, respectively, so that $\SW_{2}$ must be trivial and $\Euler_2^{\e}$ or $\Euler_2^{\o}$, if defined, must be even. The sign of $\Euler_2^{\o}$ and $\Euler_2^{\e}$ depends on the orientation of the basis of $\Gr_{\RR}(n_\o,n_\e)$. If this orientation is not fixed, then $\Euler_2^{\o}$ and $\Euler_2^{\e}$ have a sign ambiguity, which is a joint sign ambiguity if both $\Euler_2^{\o}$ and $\Euler_2^{\e}$ are defined. In this case the classifying space reduces to $\NN$ and $\NN \times \ZZ|_{\rm p}$, respectively, where $\NN \times \ZZ_{\rm p}$ consists of all ordered integer pairs $(p,q)$ with $p \ge 0$ and $p = q \mod 2$.

\subsubsection{Topological invariants for symplectic Grassmannians}
The first two homotopy groups of the symplectic Grassmannians are trivial,
\begin{equation}
  \pi_1[\Gr_\mathbb{H}(n_\o,n_\e)] = \pi_2[\Gr_\mathbb{H}(n_\o,n_\e)] = 0.
\end{equation}

\subsection{Topology and bulk-boundary correspondence}
Band structures with nontrivial stable topology in $d$ dimensions exhibit anomalous gapless boundary states on $(d-1)$-dimensional boundaries that respect all symmetries --- a phenomenon known as the  ``bulk-boudary correspondence'' \cite{hasan2010,qi2011,bernevig2013,ando2015}. In this context, a boundary mode is called ``anomalous'' if it cannot be realized by any  $(d-1)$-dimensional lattice model that respects all the symmetries. Of the three types of topological classification --- based on delicate, fragile, and stable equivalence ---  only the first allows for anomalous boundary states as a signature of the bulk topology. This, however, also includes the representation protected topological phases (defined in Sec.~\ref{sec:gen_top}), provided the boundary respects the symmetry as well as the representation constraints.

Our classification procedure can also be used to classify anomalous gapless states at a symmetry-preserving boundary. To this end, we first enumerate the possible gapless phases consistent with given symmetries by looking for all possible violations of the compatibility conditions in $d-1$ dimensions. Such violations indicate gapless phases, since a level-$p$ compatibility condition follows from the contractibility of a $(p+1)$-cell $\alpha$, so that its a violation implies that $H_\alpha: \alpha \to\Xsp_\alpha$ is defined only for a noncontractible subset of $\alpha$, \ie, the Bloch Hamiltonian is gapless somewhere on $\alpha$. A gapless ($d-1$)-dimensional Bloch Hamiltonian $H$ is anomalous if it necessarily violates a constraint imposed on the topological invariants by a gluing condition or by symmetry. We illustrate this classification scheme for anomalous boundary states by deriving some well known examples in the next section.

\section{Classification without crystalline symmetries}
\label{sec:C0}
We first apply our classification strategy to band structures without additional crystalline symmetries. Since these classification results are well known, this section mainly serves as an illustration of our classification scheme following the more abstract discussion of Sec.~\ref{sec:gen}.  

\begin{figure}
	\includegraphics[width=0.85\columnwidth]{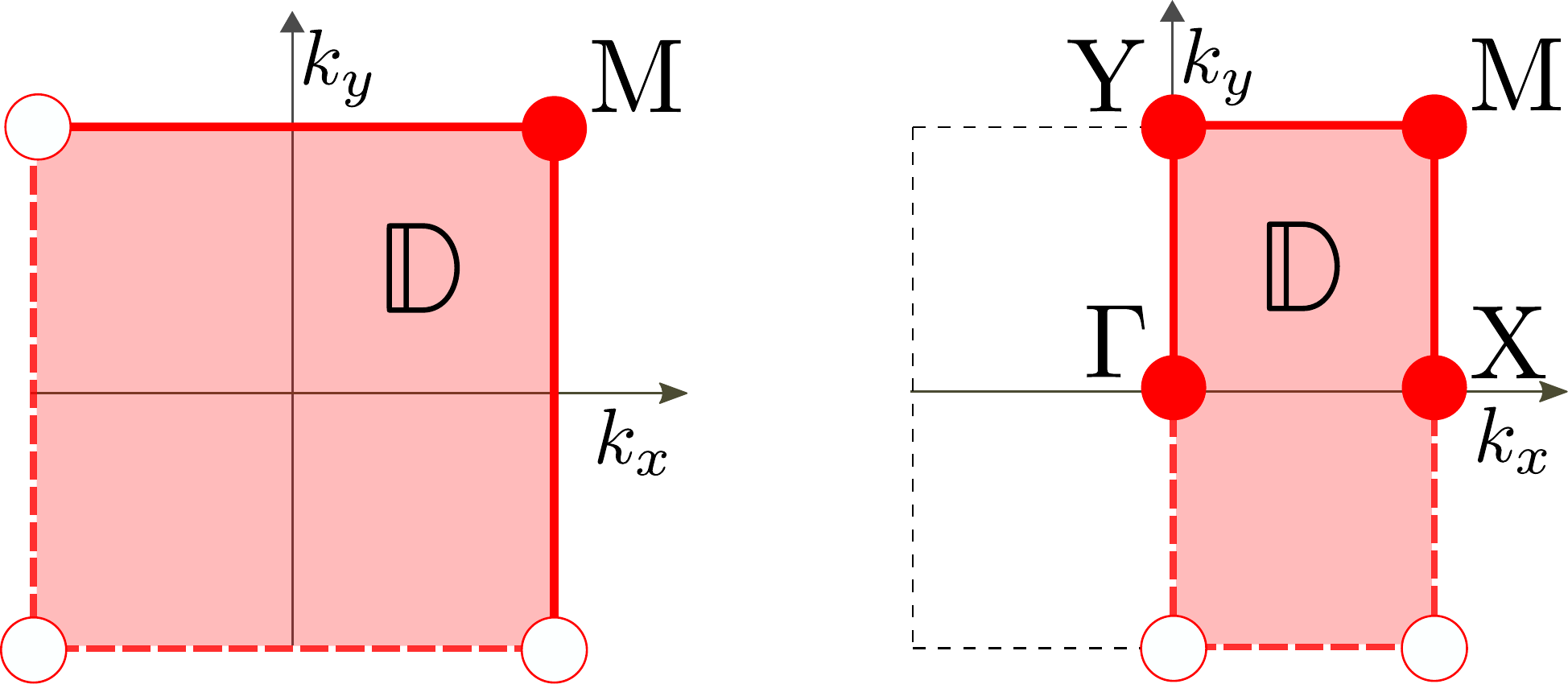}
	\caption{\label{fig:fundamental_general_d2} Fundamental domain of a band structure on a square lattice, without (left) or with (right) time-reversal symmetry, but without additional crystalline symmetries. In the latter case, two of the 0-cells and all 1-cells occur twice in $\partial\cwc$.
}
\end{figure}

\subsection{Classification without time-reversal symmetry}
\label{sec:C0_A}
In the absence of time-reversal symmetry (symmetry class A), the fundamental domain is the entire Brillouin zone, which can be decomposed into cells as shown in the left column of Fig.~\ref{fig:fundamental_general_d2} and Fig.~\ref{fig:fundamental_general} in two and three dimensions, respectively. The topological invariants for the unique 0-cell are the numbers of occupied and unoccupied bands $n_{\o/\e}$. The level-0 topological invariant is thus the pair $(n_\o, n_\e) \in \NN^2$. The target space $\Xsp$ for all cells is
\begin{equation}
  \Xsp = \Gr_{\CC}(n_\o,n_\e).
\end{equation}
We can now read off the parent topological invariants from Tab.~\ref{tab:hom_cmplx}. 

\begin{figure}
\includegraphics[width=.95\columnwidth]{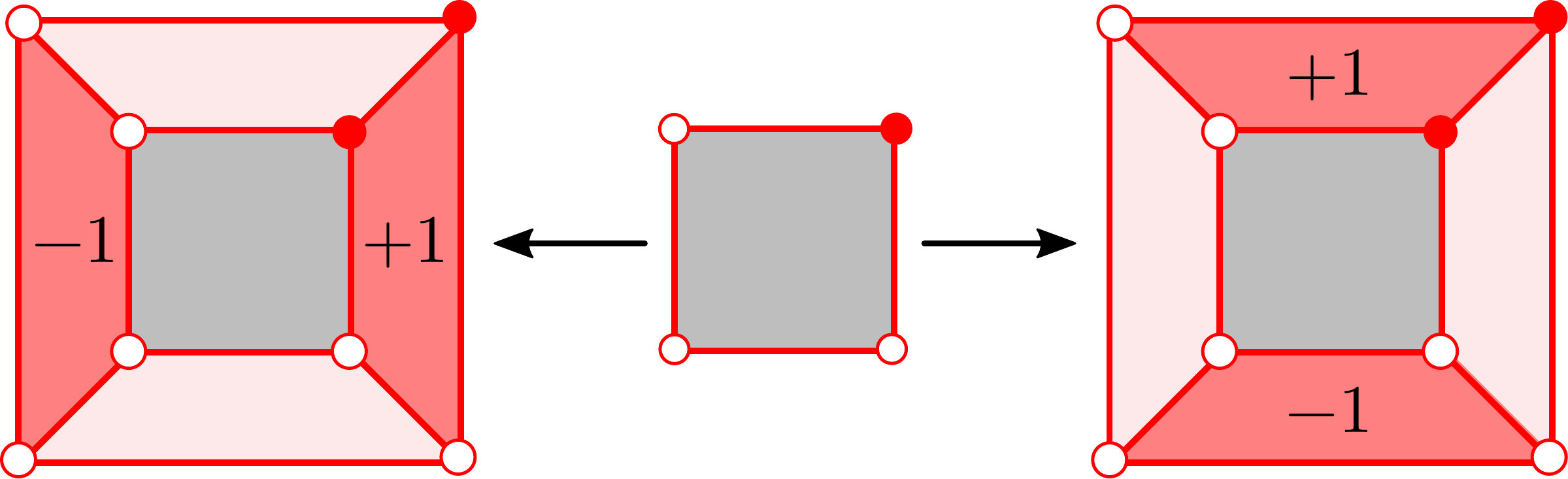}
\caption{\label{fig:boundary_Chern} Generators of the boundary deformations at the boundary of a 2-cell for a band structure without time-reversal symmetry and without additional crystalline symmetries. Periodicity of the Bloch Hamiltonian in reciprocal space forces that the net Chern number associated with the boundary deformation is zero. Hence, a boundary deformation cannot change the Chern number associated with the 2-cell.}
\end{figure}

There are no level-1 invariants, since the first homotopy group of complex Grassmannians is always trivial. At level-2, the parent classification set is $K_1 = \pi_2[\Xsp] \cong \ZZ$ and the corresponding topological invariant is the $\ZZ$-valued Chern number $\Chern$. To get the true topological invariant, however, we need to check if it is robust to deformations at the boundary of a 2-cell. As discussed in Sec.~\ref{sec:gen_clfn}, the parent classification set of boundary deformations $K_2^\partial$ is generated by deformations along the 0-cells and  the 1-cells, whose equivalence classes are parametrized by $\pi_1[\Xsp] = 0$ and $\pi_2[\Xsp] \cong K_2 \cong \ZZ$, respectively. Thus, boundary deformations may carry a nonzero Chern number along a 1-cell on $\partial\cwc$. However, since each 1-cell occurs twice in $\partial\cwc$ and the corresponding deformations carry opposite Chern numbers (see Fig.~\ref{fig:boundary_Chern}), the net Chern number of any boundary deformation is zero, rendering the parent Chern number robust. This argument applies for each 2-cell in the fundamental domain, so that the level-2 invariants are the single Chern number $\Chern$ in 2d and three Chern numbers $\Chern_{x,y,z}$ in 3d. 

\begin{table}
	\centering 	
	\setlength\tabcolsep{12pt}
	\begin{tabular}{|r|c|l|}
		\hline 
		& Bands & Topological invariant \\
		\hline
		Level 1 & --- & --- \\
		\hline 
		Level 2 &  ---  & $\Chern_{x}$, $\Chern_{y}$, $\Chern_{z} \in \ZZ$ \\ 
		\hline 
		\multirow{5}{*}{Level-3} 
		& \multirow{4}{*}{$n_\o = n_\e = 1$} & if $\Chern_x = \Chern_y = \Chern_z = 0$, \\ 
		&& then $\Hopf \in \ZZ$  \\ \cline{3-3}
		&& else: $\Hopf \in \ZZ_{2 \Chern}$, where \\ 
		&& $\Chern \equiv \textrm{gcd}\,(\Chern_x,\Chern_y,\Chern_z)$ \\ \cline{2-3}
		& $n_\o + n_\e > 2$ & --- \\ 
		\hline 
	\end{tabular}
	\caption{Homotopic classification of gapped band structures with broken time-reversal symmetry and no additional crystalline symmetries. }
	\label{tab:classification_A_3d}	
\end{table} 

At level-3, the parent invariant is the $\ZZ$-valued Hopf invariant $\Hopf$ if $n_\o = n_\e = 1$ \cite{moore2008}. The boundary deformations in $K_3^\partial$ depend on the three level-2 invariants $\Chern_{x,y,z}$ \cite{kennedy2016hopf}. There are six generators of the boundary deformations: Three carry unit Chern number associated with segments along one of the three 1-cells and its three images under translation by a reciprocal lattice vector and three carry unit Hopf number associated with segments along one of the three 2-cells and its image under translation by a reciprocal lattice vector. Because of the periodicity constraint in reciprocal space, the latter leave the Hopf number $\Hopf$ invariant, while the former change $\Hopf$ by $2 \Chern_{\alpha}$, $\alpha = x$, $y$, $z$ (see App.\ \ref{app:Hopf} for details). The minimum change to $\Hopf$ one may thus obtain is $2\, \mbox{gcd}(\Chern_x,\Chern_y,\Chern_z)$, where ``$\mbox{gcd}$'' denotes the greatest common divisor. Hence, $\Hopf$ is well defined up to multiples of $2\, \mbox{gcd}(\Chern_x,\Chern_y,\Chern_z)$, so that the level-3 invariant is $\ZZ$--valued if $\Chern_x = \Chern_y = \Chern_z = 0$ and $\ZZ_{2C}$--valued otherwise, where $C\equiv\mbox{gcd}(\Chern_x,\Chern_y,\Chern_z)$.

\ \\ \noindent These classification results are summarized in Tab.\ \ref{tab:classification_A_3d}.

\subsection{Classification with time-reversal symmetry}

\subsubsection{Symmetry class AI (spinless fermions)}
\label{sec:C0_AI}
In the presence of time-reversal and spin-rotation symmetry, the Bloch Hamiltonian satisfies
\begin{equation}
  H(\vk) = H^*(-\vk).
\end{equation}
The fundamental domain is half the Brillouin zone, which can be decomposed as shown in Fig.~\ref{fig:fundamental_general_d2} and Fig.~\ref{fig:fundamental_general} for two and three dimensions, respectively.  The level-0 invariants are given by the number of occupied/empty bands $n_{\o/\e}^S$ for each 0-cell $S$. Since $n_{\o/\e}$ must be constant throughout the Brillouin zone for a gapped Hamiltonian, one has $n_{\o/\e}^S = n_{\o/\e}$ for all $S$, leading to a single pair of level-0 invariants $(n_\o, n_\e) \in \NN^2$. At the 0-cells $S$, time-reversal symmetry constraints the Hamiltonian $H_S$ to be real-valued, so that the corresponding target spaces are 
\begin{align}
  \Xsp_S =&\, \Gr_{\RR}(n_\o,n_\e), 
\end{align}
For the $p$-cells with $p > 0$, we have $\Xsp = \Gr_{\CC}(n_\o,n_\e)$. Thus, beyond level-0, the \emph{parent topological invariant} for individual $p$-cells are identical to the symmtry class A, \ie, no invariants at level 1, a $\ZZ$-valued Chern number for each 2-cell at level-2 and a $\ZZ$-valued Hopf invariant at level-3 if $n_\o = n_\e = 1$. 

At level-2, the parent classification set of boundary deformations $K_2^\partial$ consists of a copy of $\pi_1[\Xsp_S]$ for each 0-cell and $\pi_2[\Xsp] = \ZZ$ for the 1-cells. The former homotopy set is nontrivial for each 0-cell, and since the 0-cells $\G$ and $\X$ occur only once in $\partial\cwc$, a boundary deformation that is topologically nontrivial at one of these can change the Chern number by an arbitrary integer. In App.\ \ref{app:deform}, we construct such a deformation explicitly as well as present another argument for the existence of such a boundary deformation. Thus, in contrast to symmetry class A, there are no level-2 topological invariants for symmetry class AI. Similarly, at level-3, there exists a boundary deformation that changes the Hopf number $\Hopf$ of the parent classification by one, as shown in Appendix~\ref{app:Hopf}, rendering the level-3 invariant trivial.

\subsubsection{Symmetry class AII (spinful fermions)}
\label{sec:C0_AII}
In the presence of time-reversal symmetry without spin-rotation symmetry, the Hamiltonian satisfies
\begin{equation}
  H(\vk) = \Sigma_2 H^\ast(-\vk) \Sigma_2,
\end{equation}
where $\Sigma_2 = \id_{n} \otimes \sigma_2$ is the time-reversal operator and $2n$ is the total number of bands. 
Kramers degeneracy implies that the number of occupied and unoccupied bands must be even, so that the level-0 invariants $(n_\o^S, n_\e^S)$ are \emph{half} the number of occupied/empty bands. The level-0 compatibility condition is ${n}_{\o/\e} = n_{\o/\e}^S$, where $2 {n}_{\o/\e}$ are the total number of filled/empty bands.

To derive the target space $\Xsp_S$, we note that at the high symmetry points, time-reversal implies that the Hamiltonian can be written as 
\begin{equation}
	H_S = h_0 \otimes \sigma^0 + h_j \otimes \i\sigma^j, 
	\label{eq:H_S_AII}
\end{equation} 
where $h_0$ and $h_j$ are real-valued $n \times n$ symmetric and antisymmetric matrices, respectively. Since the set of $2\times2$ conplex matrices $\{\sigma^0, \i\sigma^1, \i\sigma^2, \i\sigma^3 \}$ constitute a representation of the quaternion algebra, $H_S$ in Eq.~(\ref{eq:H_S_AII}) can be interpreted as a quaternion-valued matrix. The target space $\Xsp_S$ is thus a quaternionic Grassmannian
\begin{equation}
  \Xsp_S = \Gr_{\HH}(n_\o,n_\e).
\end{equation}
For $p$-cells with $p > 0$, we have $\Xsp = \Gr_{\CC}(2n_\o,2n_\e)$. 

The topological classification proceeds as for the symmetry class AI, with one important difference: While considering the boundary deformations to level-2 invariants, $\pi_1[\Xsp_S] = 0$ for the quaternionic Grassmannian, so that there are no nontrivial deformations at the 0-cells. The boundary deformations at 1-cells can carry a Chern number. These cancel for the two copies of $\Y\M$, but add up for the two copies of $\G\Y$ and $\M\X$, so that boundary deformations may add an even integer to the Chern number of the 2-cell. This renders the level-2 invariant $\ZZ_2$-valued. Thus, at level-2, we get one and four $\ZZ_2$-valued invariants for $d=2$ and $d=3$, respectively \cite{moore2007}. These are the well-known classifications of the quantum spin-Hall effect and of three-dimensional topological insulators \cite{kane2005qshegraphene,kane2005qshe,fu2007ti,moore2007,roy2009}.

\subsection{Anomalous boundary states}
\label{sec:C0_bdry}
In the symmetry class A, anomalous boundary states are associated with a nontrivial value of the only stable invariant, \viz, the Chern number. In two dimensions, these are the chiral edge states of the quantum Hall effect, while in three dimensions, these are chiral surface states of a weak Chern insulator. We now derive these invariants from a boundary perspective. 

\begin{figure}
\includegraphics[width=0.9\columnwidth]{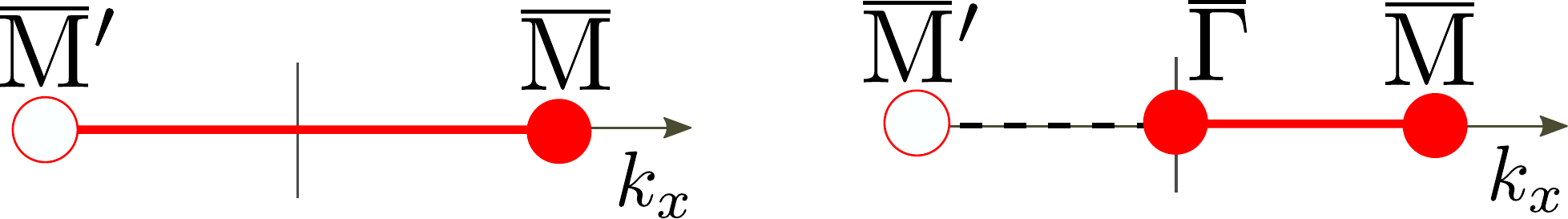}
\caption{\label{fig:fundamental_d1} Fundamental domain of a one-dimensional band structure, without (left) or with (right) time-reversal symmetry, and without additional crystalline symmetries.}
\end{figure}

Consider a 1d Bloch Hamiltonian in symmetry class A, whose fundamental domain is the entire 1d Brillouin zone, as shown in Fig.\ \ref{fig:fundamental_d1} (left).  We assume that $H$ is gapped at the 0-cells, so that the level-0 invariants $n_{\o/\e}^{\bar S}$ are well defined for $\bar{S} \in \{\sM, \sM'\}$. The presence of a chiral state crossing the Fermi level implies that the level-0 invariants $n_{\o/\e}^{\sM} \neq n_{\o/\e}^{\sM'}$, which violates both the compatibility condition and the gluing condition. The former violation signifies that a band structure with a chiral state crossing the Fermi level is necessarilty gapless, while the latter violation signals that such a band structure is anomalous, \ie, it can not be realized in a one-dimensional lattice model
\footnote{
	If $H$ has a zero energy mode at $\sM$, then we shift the Hamiltonian by an infinitesimal constant to ensure that $n_{\o/\e}^\sM$ is well-defined.	
}. 
An analogous argument holds for a chiral band crossing the Fermi energy in 2d. 

In symmetry class AI, there is no nontrivial bulk topology and, hence, there are no associated anomalous boundary states. The same conclusion can be obtained from a boundary perspective. To wit, the fundamental domain of a 1d Bloch Hamiltonian in symmetry class AI is half the 1d Brillouin zone, as shown in Fig.\ \ref{fig:fundamental_d1} (right). A band crossing between $\sG$ amd $\sM$ violates the level-0 compatibility condition for the $n_{\o/\e}^{\bar S}$ for $\bar{S} \in \{\sG,\sM\}$ --- signalling a gapless phase --- but it does not imply the violation of a gluing condition, so that such a band crossing is not anomalous. Instead, time-reveral symmetry imposes the presence of two opposite crossings of the Fermi energy in the full boundary Brillouin zone, which can be realized in a 1d lattice model and which may deformed to a gapped Hamiltonian. Thus, there are no anomalous boundary modes in symmetry class AI.

In symmetry class AII, anomalous boundary modes are associated with the $\ZZ_2$-valued stable invariant. In two dimensions, these are the helical edge states of the quantum spin Hall effect, while in three dimensions, these are the surface Dirac cones of a strong topological insulator. In this case, the fundamental domain is identical to that of symmetry class AI (Fig.\ \ref{fig:fundamental_d1}), but the level-0 invariants $n_{\o/\e}^{\bar S}$ are \emph{half} the number of occupied/unoccupied bands. Thus, if a chiral mode crosses the Fermi level between $\sG$ and $\sM$, then the total number of occupied/empty bands differ by an odd integer, which is incompatible with the level-0 invariants being integers, rendering such crossings anomalous. Application of a similar arguments to any pair of high-symmetry points in the surface Brillouin zone indicates the existence of anomalous surface states for 3d topological insulators \cite{fu2007ti,moore2007,roy2009}.

\section{Classification in three dimensions with crystalline symmetries: General remarks}
\label{sec:outline}

In the next sections, we apply the classification strategy of Sec.\ \ref{sec:gen} to band structures with $\C_2 \T$, $\C_2$, $\C_4$, and $\D_4$ symmetries. Since these symmetries are all two-dimensional and include a rotation, in two dimensions they do not exhibit a boundary that is invariant under the symmetries; however, they all exhibit invariant planes in three dimensions, which may host anomalous boundary states. We thus consider a simple-cubic lattice with unit lattice constant, with the rotation axis (and mirror planes for $\D_4$) parallel to the $z$-axis. For the crystalline symmetries considered in this article, this simplification does not affect our conclusions. Since $\rotn_2$ is an element of all of these groups, for the time-reversal-symmetric case, we choose a basis such that $\C_2 \T$ is represented by the identity matrix, so that 
\begin{equation}
	H(k_x,k_y,k_z) = H^*(k_x, k_y, -k_z). 
	\label{eq:tr_c2}
\end{equation}
For $k_z = 0,\pi$ (and for 2d systems), this reduces to a reality condition on $H(\vk)$. 

The fundamental domain in all such cases is a prism with the two-dimensional fundamental domain as the base and $k_z \in (-\pi,\pi]$ without and $k_z \in [0,\pi]$ with time-reversal symmetry. (See Fig.\ \ref{fig:fundamental_general} for the fundamental domains without crystalline symmetries.) The ``vertical'' faces are formed by extruding the 1-cells of the fundamental domain $\cwc_2$ along $z$, so that the vertical faces and edges inherit the little groups of the 1- and 0-cells of which they are an extrusion. The ``horizontal'' faces at $k_z = \pm\pi$ for the time-reversal broken case and $k_z = 0,\pi$ for the time-reversal-symmetric case are copies of $\cwc_2$, and thus satisfy the corresponding constraints. 
  
The classification on the horizontal faces depends on the details of the symmetry in question, but that on the vertical faces can be described more generally. Since time-reversal (if present) does not impose any additional constraints on the side faces for a two-dimensional symmetry, the target space for the vertical 1- and 2-cells is always the complex Grassmannian. Thus, there are no level-1 invariants for the vertical 1-cells, while the level-2 parent invariants are given by a set of Chern numbers, one for each irrep on the 1-cells of $\cwc_2$. The various Chern numbers satisfy a compatibility constraint arising from the fact that taken together, the 2-cells form the boundary of the 3-cell. 

The parent classification set on the 3-cell is identical to that obtained without additional symmetries, \ie, the $\ZZ$-valued Hopf invariant $\Hopf$ only if $n_\o = n_\e = 1$. Interestingly, it is robust even if the vertical 2-cells possess nonzero Chern numbers, as shown in App.\ \ref{app:Hopf}. This can alternatively be seen by extending the calculation of the Hopf invariant to the full Brillouin zone, for which the $\C_2$ or $\C_4$ symmetry rules out a nonzero Chern number associated with the vertical faces of the full Brillouin zone. However, if a cross section of the fundamental domain $\cwc$ at constant $k_z$ has a nonzero Chern number $\Chern$, then the Hopf invariant reduces to a $\ZZ_{2 \Chern}$-valued invariant, owing to a deformation along the vertical faces that changes $\Hopf$ by an even multiple of $\Chern$ (see App.\ \ref{app:Hopf}).

\section{$\C_2\T$ symmetry}
\label{sec:C2T}
As a first application with crystalline symmetries, we obtain the homotopic classification of two- and three-dimensional band structures with only a combined $\C_2 \T$ symmetry. In two dimensions, $\C_2\T$-symmetric Bloch Hamiltonians are real and the corresponding topological invariants are formulated in terms of the Euler class and the Stiefel-Whitney invariants (see Sec.\ \ref{sec:gen_Gr_real}). 

A stable classification of $\C_2 \T$-symmetric topological insulators was first given by Fang and Fu \cite{fang2015nonsymmorphic}. Delicate and fragile classifications for the two-dimensional case were obtained by Kennedy and Guggenheim \cite{kennedy2015homotopy} and by Bouhon, Bzdusek, and Slager \cite{bouhon2020}. These authors showed that level-2 classification spaces for delicate and fragile topology depends on the values of the level-1 invariants. Our discussion below shows that the topological classification may also depend on the number of orbitals present at each Wyckoff position.
To expose this feature, we begin with the analogous case in one dimension, where the combination of inversion ($\I$) and time-reversal imposes a reality condition on $H(k)$. 

\subsection{$\I\T$ symmetry in one-dimension}

\begin{figure}
	\includegraphics[width=0.9\columnwidth]{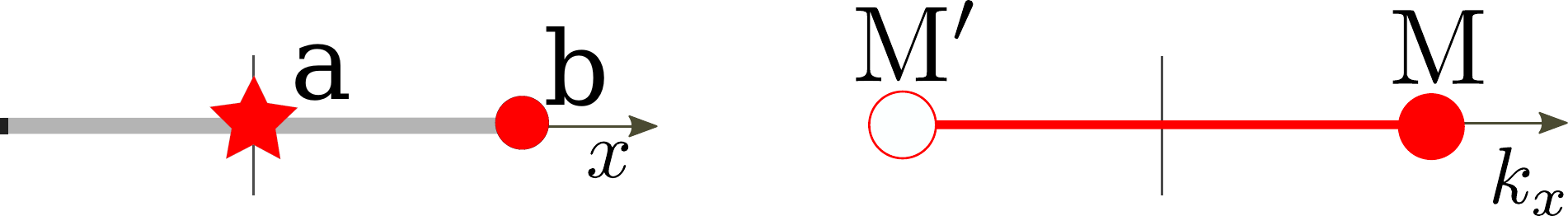}
	\caption{\label{fig:WyckoffC2T} Special Wyckoff positions (left) and fundamental domain of a one-dimensional $\I\T$ symmetric band structure.}
\end{figure}

\subsubsection{Wyckoff positions and the fundamental domain}
A lattice with $\I\T$ symmetry has two special Wyckoff positions ``a'' and ``b'' at $x=0$ and $x=1/2$ (shown in Fig.\ \ref{fig:WyckoffC2T}) as well as one generic Wyckoff position pair ``g'' at $x= \pm a$. Denoting the number of orbitals at Wyckoff position $\alpha$ by $n_\alpha$, the total number of orbitals $n$ is 
\begin{equation}
 n = n_a + n_b + 2 n_g.
\end{equation}
With a suitable choice of the basis orbitals at the Wyckoff positions a, b, and g, the constraint on $H(\vk)$ imposed by $\I \T$ symmetry can be cast in a form that is the same for all Wyckoff positions,
\begin{equation}
  H(k) = H^\ast(k).
\end{equation}
The symmetry relation (\ref{eq:Hsymm_ktrans}) for translations by reciprocal lattice vectors, however, bears reference to the Wyckoff positions:
\begin{equation}
  H(k+2 \pi) = \trans_x^\dg H(k) \trans_x,
  \label{eq:HboundaryC2T1d}
\end{equation}
with
\begin{equation}
  \trans_x = \diag \left\{ \id_{n_\a}, -\id_{n_\b}, \id_{n_g}\otimes\e^{2\pi\i a \sigma_2} \right\}.
  \label{eq:transC2T1d}
\end{equation}
Since a pair of orbitals at the generic Wyckoff position can always be moved to one of the special Wyckoff positions, we may without loss of generality assume that all orbitals are at the Wyckoff positions a and b.

In reciprocal space, the fundamental domain $\cwc$ consists of (two copies of) one 0-cell $\M$, corresponding to $k = \pi$, and one 1-cell $\M\M'$, corresponding to $-\pi < k < \pi$, as depicted in Fig.\ \ref{fig:WyckoffC2T}. The Hamiltonians at $\M$ and $\M'$ are related by Eq.~(\ref{eq:HboundaryC2T1d}).

\begin{table}
	\centering
	\setlength\tabcolsep{11pt}
	\begin{tabular}{|l|l|l|}
		\hline
		& Condition(s)  & Invariants \\ \hline
		\multirow{2}{*}{$n_\o = n_\e = 1$}
		&  Identical WPs & $\Euler_1 \in \ZZ$ \\ \cline{2-3}
		&  Distinct WPs & $\Euler_1 \! \mod 2 \in \ZZ_2$ \\ \hline
		$n_\o + n_\e > 2$ & --- & $\SW_1 \in \ZZ_2$ \\
		\hline
	\end{tabular}
	\caption{Homotopic classification of gapped band structures with $\I\T$ symmetry in one dimension. The classification for $n_\o = n_\e = 1$ depends on the Wyckoff positions (WPs) in the lattice model.
	\label{tab:classification_C2T_1d}}
\end{table}

\subsubsection{Homotopic classification}
The level-0 invariants are given by the number of occupied/empty bands $n_{\o/\e}$. For fixed level-0 invariants, the target space for all $p$-cells is 
\begin{equation}
  \Xsp = \Gr_{\RR}(n_\o,n_\e).
\end{equation}
At level-1, the parent classification set is $K_1 = \ZZ$ when $n_\o = n_\e = 1$ and $\ZZ_2$ otherwise, the corresponding topological invariants being the first Euler number $\Euler_1$ and the first Stiefel-Whitney invariant $\SW_1$, respectively (see Sec.\ \ref{sec:gen_Gr} for more details). 

The parent classification set for the boundary deformations $K_1^\partial$ consists of a copy of $\pi_1[\Xsp] \cong K_1$ for each end point $k = \pm\pi$. The deformations at these two points are related by Eq.~(\ref{eq:HboundaryC2T1d}) as 
\begin{equation}
	H_{\partial}(-\pi;t) = \trans_x H_{\partial}(\pi,t) \trans_x^\dagger. 
	\label{eq:IT_Hdeform}
\end{equation}
Since $\trans_x$ depends on the Wyckoff positions occupied, so does the effect of deformations on the 1d invariants. We now discuss the effect of boundary deformations on the Euler and Stiefel-Whitney invariant separately. 

\ \\ \para{First Euler invariant}
In this case, $n_\e = n_\o = 1$ and $\trans_x$ is proportional to $\sigma^0$ or $\sigma^3$, depending on whether the two Wyckoff positions involved are identical or distinct. In the former case, the windings of $H_\partial(\pm\pi; t)$ are equal and opposite and thus cancel each other, rendering $\Euler_1$ a robust topological invariant. In the latter case, the winding numbers associated with $H_\partial(\pm\pi; t)$ add up, so that the boundary deformation can add an even integer to $\Euler_1$. In this case, the true topological invariant is $\ZZ_2$ valued and is given by $\Euler_1\!\mod 2$. 

Interestingly, in the latter case, the two topological classes are both homotopic to atomic limits, \ie, Bloch Hamiltonians in either of these classes can be continuously deformed to $k$-independent Hamiltonians. To see how an atomic limit Hamiltonian can be assigned a nontrivial Euler invariant, we set $\Lambda_x = \sigma^3$ and (arbitrarily) take $H_\text{ref} = -\sigma^3$, corresponding to Wyckoff position ``a'' being occupied. With respect to this reference Hamiltonian, $H(k) = \sigma^3$ --- corresponding to Wyckoff position ``b'' being occupied --- carries a nontrivial level-1 invariant. This is precisely the invariant carried by the boundary deformation, which must rotate $\sigma^3$ to $-\sigma^3$ in opposite senses for $k = \pm\pi$. 

\begin{table*}
	\centering
	\setlength\tabcolsep{15pt}
	\begin{tabular}{|r|c|l|l|}
		\hline 
		& Bands & Additional condition(s) & Topological invariant \\
		\hline
		\multirow{4}{*}{Level 1} 
		& \multirow{2}{*}{$n_\o = n_\e = 1$} & Identical Wyckoff positions &  $\Euler_{1,x}, \Euler_{1,y} \in \ZZ$ \\ \cline{3-4}
		&& Distinct Wyckoff positions &  $\SW_1 \in \ZZ_2$  \\ \cline{2-4}
		& $n_\o \ge 1$, $n_\e > 1$ & \multirow{2}{*}{---} & \multirow{2}{*}{$\SW_{1,x}, \SW_{1,y} \in \ZZ$} \\ 
		& or $n_\o > 1$, $n_\e \ge 1$ &&  \\ 
		\hline
		\multirow{10}{*}{Level 2}   
		& $n_\o = n_\e = 1$ & \multicolumn{2}{c|}{------}
		\\ \cline{2-4}
		& \multirow{2}{*}{$n_\o = 1$, $n_\e = 2$} 
		& if $\Q^\e$ & $\Euler_2^\e \in 2\NN$ \\ \cline{3-4}
		&& if not $\Q^\e$  & $\SW_2^\e \in \ZZ_2$  
		\\ \cline{2-4}
		& $n_\o = 1$, $n_\e > 2$ & \multicolumn{2}{c|}{------}
		\\ \cline{2-4}
		& \multirow{4}{*}{$n_\o = n_\e = 2$} 
		& if $\Q^\o$ and $\Q^\e$  & $(\Euler_2^\o,\Euler_2^\e) \in \NN \times \ZZ|_{\rm p}$ \\ \cline{3-4}
		&& if $\Q^\o$, but not $\Q^\e$ & $\Euler_2^{\o} \in \NN$ \\ \cline{3-4}		
		&& if $\Q^\e$, but not $\Q^\o$ & $\Euler_2^{\e} \in \NN$ \\ \cline{3-4}		
		&& if not $\Q^\o$ and not $\Q^\e$ & $\SW_2 \in \ZZ_2$ 
		\\ \cline{2-4}
		& \multirow{2}{*}{$n_\o = 2$, $n_\e > 2$} 
		& if $\Q^\o$ & $\Euler_2^\o \in \NN$ \\ \cline{3-4}
		&& if not $\Q^\o$ &  $\SW_2 \in \ZZ_2$
		\\ \cline{2-4}
		& $n_\o$, $n_\e > 2$ & --- &  $\SW_2 \in \ZZ_2$ \\	
		\hline
	\end{tabular}	
	\caption{Homotopic classification of gapped band structures with $\C_2 \T$ symmetry in two dimensions. The classification for $n_\o = n_\e = 1$ depends on the Wyckoff positions of the two sites. The symbol $\Q^{\o/\e}$ refers to the condition $\Q^{a,\o/\e}_2$ of Eq.\ (\ref{eq:cond1c2td2}) being met for $a=x$ and for $a=y$. The set $\NN \times \ZZ_{\rm p}$ consists of all ordered integer pairs $(p,q)$ with $p \ge 0$ and $p = q \mod 2$.
	\label{tab:classification_C2T_2d}}
\end{table*}

\ \\ \para{First Stiefel-Whitney invariant} 
This invariant is given by $\SW_1 = \sgn \large( \det \wilson^{\o/\e} \large)$, see Eq~(\ref{eq:SW1_winding}). The Wilson lines associated with the deformations at $\pm\pi$ are related as 
\begin{equation}
	\wilson^{\o/\e}_{\partial}(\pi) = \trans_x^{\o/\e} \wilson_{\partial}^{\o/\e}(-\pi)
	\trans_x^{\o/\e \dagger}.
\end{equation}
where $\trans_x^{\o/\e}$ denotes the projection of $\trans_x$ onto the occupied/empty subspaces. Thus, 
\[
\det \wilson_{\partial}^{\o/\e}(\pi) =  \det \wilson_{\partial}^{\o/\e}(-\pi) \in \{\pm1\},
\]
so that $\det W^{\o/\e}$ for the entire loop remains unchanged under boundary deformation. The first Stiefel-Whitney invariant is thus robust under boundary deformations.

\ \\ \noindent These classification results are summarized in Tab.\ \ref{tab:classification_C2T_1d}.

\subsection{$\C_2\T$ symmetry in two dimensions}

\subsubsection{Wyckoff positions and the fundamental domain}
In two dimensions, there are four special Wyckoff positions ``a'', ``b'', ``c'', and ``d'', as shown in Fig.~\ref{fig:C2}, and one pair of generic Wyckoff positions ``g'' at coordinates $(a_x,a_y)$. Denoting the number of orbitals at the Wyckoff position $\alpha$ by $n_{\alpha}$, the total number of orbitals is
\begin{equation}
  n = n_\a + n_\b + n_\c + n_\d + 2 n_\g.
\end{equation}
We choose the basis vectors that are invariant under $\C_2\T$ (see Appendix~\ref{app:symm_tr}), so that the Bloch Hamiltonian satisfies $H^\ast(\vk) = H(\vk)$. The reciprocal-space group $G_\Lambda$ is generated by the two reciprocal lattice translations. We denote the corresponding representation matrices (see Eq.\ (\ref{eq:Hsymm})) by $\trans_{x/y}$, which are explicitly given by 
\begin{align}
  \trans_x &= \diag \left\{ \id_{n_\a}, \id_{n_\b}, -\id_{n_\c}, -\id_{n_\d}, \id_{n_g}\e^{2\pi\i a_{x} \sigma_2} \right\}, \nonumber \\
  \trans_y &= \diag \left\{ \id_{n_\a}, -\id_{n_\b}, -\id_{n_\c}, \id_{n_\d}, \id_{n_g}\e^{2\pi\i a_{y} \sigma_2} \right\}.
\end{align}
For the purpose of classifying topological phases, we may continuously shift any orbitals at the Wyckoff position ``g'' to the special Wyckoff position ``a'' and use the corresponding expressions for $\trans_{x/y}$. The fundamental domain is the same as symmetry class A, as depicted in Fig.~\ref{fig:fundamental_general_d2}.

\subsubsection{Homotopic classification}
The level-0 invariants are given by the number of occupied/empty bands $n_{\o/\e}$. The target space for all cells is again the real Grassmannian $\Xsp = \Gr_{\RR}(n_\o,n_\e)$. The level-1 classification is similar to the one-dimensional case discussed above: each of the two 1-cells contribute the $\ZZ$-valued first Euler classes $\Euler_{1,x/y}$ if $n_\o = n_\e = 1$ and $\ZZ_2$-valued Stiefel-Whitney invariants $\SW_{1,x/y}$ otherwise. The latter are robust to boundary deformations, while the former are robust only if both orbitals lie at Wyckoff position with the same $x/y$ coordinate, and reduce to a $\ZZ_2$ invariant otherwise. 

The level-1 invariants are further constrained by compatibility relations. Since the boundary of the 2-cell consists of two copies of each 1-cell, the $\ZZ_2$-valued $\SW_1$ always add up to $0$, so that there are no additional restrictions on $\SW_{1, x/y}$. There are nontrivial compatibility constraints on $\Euler_{1,x/y}$, however, that depend on the reciprocal-space translation operators $\trans_{x/y}$. If the two orbitals are at the same Wyckoff position, such that $\trans_{x/y} \propto \sigma^0$, then there are no additional constraints on $\Euler_{1,x/y}$, so that the level-1 classification set is $\ZZ^2$. If only one of the translation operators ($\Lambda_{x}$, say) is proportional to $\sigma^3$, then $\Euler_{1,x}$ is only defined modulo 2, while $\Euler_{1,y}$ adds up for the two copies of the 1-cell along $y$, so that the compatibility constraint requires $2\, \Euler_{1,y}=0$. This leaves only one $\ZZ_2$-valued invariant, \emph{viz}, $\SW_{1} \equiv \Euler_{1,x} \mod 2$. On the other hand, if both $\trans_{x/y} \propto \sigma^3$ (such as when the two orbitals are at Wyckoff positions ``a'' and ``d''), then the compatibility condition reads 
\[ 
	2\, \Euler_{1,x} - 2\, \Euler_{1,y} = 0 \implies \Euler_{1,x} = \Euler_{1,y} \equiv \Euler_1. 
\]
Furthermore, the boundary deformations (described for the 1d case) simultaneously changes both $\Euler_{1,x}$ and $\Euler_{1,y}$ by an even integer. Thus, we are again left with only one $\ZZ_2$-valued invariant $\Euler_1$.

At level-2, $K_2 = \pi_2[\Xsp]$ depends on $n_{\o/\e}$ (see Tab.~\ref{tab:hom_real}), and the corresponding invariants can be derived from the winding number $\Euler_2^{\o/\e}$ of a family of Wilson loops. These are the $\ZZ$-valued second Euler invariant (up to a parity constraint, see Tab.~\ref{tab:hom_real}) when $n_{\o/\e}=2$, while for $n_{\o/\e}>2$, we get the $\ZZ_2$-valued second Stiefel-Whitney invariant $\SW_2 \equiv \Euler_2^{\o/\e} \mod 2$. For the boundary deformations, $K_2^\partial$ consists of a copy of $\pi_1[\Xsp]$ for the 0-cell and $\pi_2[\Xsp] \cong K_2$ for each of the 1-cells. We now discuss their effect on the Euler and Stiefel-Whitney invariant separately. 

\ \\ \para{Second Euler invariant}
Since these invariants are defined only when $n_\o=2$ or $n_\e=2$, the contribution of the 0-cell to $K_2^\partial$ is $\pi_1[\Xsp]=\ZZ_2$. We first consider deformations that are trivial on the 0-cell. Nontrivial deformation along each 1-cell can carry a second Euler invariant given by the winding number of the Wilson loops in $\SO(2)$ (see Sec.~\ref{sec:gen_Gr_real}). However, each 1-cell occurs twice in  $\partial\cwc$, and the corresponding windings cancel each other, unless the orientation of the basis states is changed while going from one copy of the 1-cell to the other. 
This change of orientation can originate from either a reciprocal space translation operator or a  nontrivial first Stiefel-Whitney invariant. Explicitly, denoting the 1-cells along $x$ and $y$ by $\alpha_x$ and $\alpha_y$, respectively, the windings on two copies of $\alpha_x$ cancel if either $\det\trans_y = -1$ or $\SW_{1,y} = 1$. Thus, $\Euler_2$ is robust to boundary deformations only if
\begin{equation}
	\Q_{2,a}^{\o/\e}:\ (-1)^{{\rm SW}_{1,a}} \det \trans_a^{\o/\e} = 1. \label{eq:cond1c2td2}
\end{equation}
for $a=x,y$. Note that while $\SW_{1,a}$ and $\det \trans_{x/y}^{\o/\e}$ both depend on the choice of the reference Hamiltonian $H_{\rm ref}$, this condition does not. If this condition is obeyed, then $\Euler_2^{\o/\e}$ is robust to boundary deformations, while if it is violated, the boundary deformations can change it by even integers, rendering the level-2 invariant $\ZZ_2$-valued. 

A boundary deformation that is nontrivial at a 0-cell changes the sign of $\Euler_2^\o$ and $\Euler_2^\e$ simultaneously, since it changes the orientation of the bases for both the occupied and the unoccupied states \cite{bouhon2020}. 

\noindent \emph{Second Stiefel-Whitney invariant --- }
These are robust to boundary deformations, since the effect of the latter always cancels out modulo 2. 

\ \\ \noindent The classification results are summarized in Tab.\ \ref{tab:classification_C2T_2d}. They are consistent with the conclusions of Ref.\ \cite{bouhon2020}, which considered the case $\trans_{x/y} = \openone$, corresponding to all orbitals being at the same Wyckoff position. In the stable limit, there are two $\ZZ_2$-valued level-1 invariants and one $\ZZ_2$-valued level-2 invariant.

\subsection{$\C_2\T$ symmetry in three dimensions}
\label{sec:C2T3d}
In three dimensions, $\C_2\T$ symmetric Hamiltonians satisfy Eq.~(\ref{eq:tr_c2}). The fundamental domain is a cuboid with the top/bottom faces (at $k_z = 0,\pi$) formed by copies of the 2d fundamental domain, and the side faces formed by copies of two distinct 2-cells.

\subsubsection{Homotopic classification}
The level-1 invariants are two independent copies of the invariants obtained for the 2d classification. As discussed in Sec.\ \ref{sec:outline}, there are no topological invariants at the 1-cells parallel to $k_z$.

The level-2 invariants for the $k_z = 0,\pi$ planes are copies of those obtained for the 2d classification. For the two distinct faces parallel to the $k_z$-axis, the parent topological invariants are the Chern numbers $\Chern_{x/y}$. There are two kinds of boundary deformations for these invariants: those along the 1-cells parallel to the $k_z$-axis and those perpendicular to it. The former cannot change the Chern number, since the two vertical boundaries of each face are copies of the the same 1-cell, so that a deformation with Chern number $\Chern$ on one copy of the 1-cell is exactly canceled by the other. The latter deformations cannot carry a Chern number, since the Hamiltonian is constrained to be real at the boundary. Thus, the Chern numbers $\Chern_{x/y}$ are robust topological invariants. There are no additional constraints on them from level-3 topology, since each of these 2-cells occurs twice in the boundary of the 3-cell, so that the total Chern number always vanishes. 

The $\ZZ$-valued $\Chern_{x/y}$ can be combined with the $\ZZ_2$ valued difference between $\SW_{1,x/y}$ computed at $k_z = 0$ and $\pi$ to get a $\ZZ$-valued invariant $\Chern^\text{full}_{x/y}$, which is simply the Chern number computed on the $k_x/k_y$ plane for the full range $k_z \in (-\pi, \pi)$. This is because the symmetry condition $H(k_x,k_y,k_z) = H^\ast(k_x,k_y,-k_z)$ implies that the Chern numbers for $k_z \in (-\pi,0)$ and $(0,\pi)$ add up to give even integers. Furthermore, a difference between $\SW_{1,x/y}$ at $k_z = 0$ and $\pi$ implies that the Wilson loop along $k_x/k_y$ changes from $0$ to $\pi$ along $k_z$, which, combined with the symmetry of Wilson loops under time reversal, implies that the Chern number computed for $k_z \in (-\pi,\pi)$ is unity (see Appendix~\ref{app:deform_Chern} for a similar argument). Combining this with the even total Chern number yields two $\ZZ$-valued invariants  $\Chern^\text{full}_{x/y}$, which represent weak Chern insulator phases protected by discrete translation symmetry in the $x$ and $y$ direction, respectively. 

At level-3, we get the Hopf invariant when $n_\o = n_\e = 1$. This is not affected by deformations, since the Bloch Hamiltonian is real for $k_z = 0$, $\pi$ and thus cannot carry a Chern number (see Sec.~\ref{sec:outline} for details).

\ \\ \noindent The classification results are summarized in Tab.\ \ref{tab:classification_C2T_3d}.

\begin{table}
	\centering 	
	\setlength\tabcolsep{8pt}
	\begin{tabular}{|r|c|l|}
		\hline 
		& Bands & Topological invariant \\
		\hline
		\multirow{2}{*}{Level 1} & \multirow{2}{*}{---} & Two copies of invariants \\
		&&  from Tab.\ \ref{tab:classification_C2T_2d} \\ 
		\hline 
		\multirow{2}{*}{Level 2} & \multirow{2}{*}{---} & $\Chern_x$, $\Chern_y \in \ZZ$ and two copies \\
		&& of invariants from Tab.\ \ref{tab:classification_C2T_2d} \\ 
		\hline 
		\multirow{2}{*}{Level-3} 
		& $n_\o = n_\e = 1$ & $\Hopf \in \ZZ$  \\ \cline{3-3}
		& $n_\o + n_\e > 2$ & --- \\ 
		\hline 
	\end{tabular}
	\caption{Homotopic classification of gapped band structures with $\C_2 \T$ symmetry in three dimensions. 
	\label{tab:classification_C2T_3d}}
\end{table}

\subsubsection{Anomalous boundary states}
\label{sec:C2T_bdry}
The presence of stable topological indices for a 3d band structure with $\C_2 \T$ symmetry implies the existence of anomalous boundary states on boundaries invariant under $\C_2$, \ie, a surface parallel to the $xy$ plane \cite{fang2015nonsymmorphic}. The stable topological invariants are the $\ZZ$-valued $\Chern_{x,y}^\text{full}$ associated with the vertical 2-cells and the $\ZZ_2$-valued difference between $\SW_2$ associated with the planes $k_z = 0, \pi$ \footnote{
	A band structure with identical invariants for $k_z = 0, \pi$ and $\Chern_x^\text{full} = \Chern_y^\text{full} = 0$ is a weak topological phase that is protected by the discrete translation symmetry in the $z$ direction. As a surface parallel to the $xy$ plane breaks this symmetry, we do not generically get surface modes. 
}.
The former lead to anomalous chiral states at the surface that propagate along $x$ and $y$, respectively, while the latter leads to a surface Dirac cone. 

We now derive these anomalous states from the boundary perspective. Following the argument for symmetry class A (Sec.~\ref{sec:C0_bdry}), a chiral state violates the level-0 compatibility condition as well as the gluing condition and is thus anomalous. To see that a surface Dirac cone is anomalous, one first observes that a closed path $\gamma$ around a single Dirac cone has Berry phase $\pi$, so that $\det \wilson^{\o/\e}(\gamma) = -1$ \cite{fang2015nonsymmorphic}. On the other hand, for any set of $\ZZ_2$-valued level-1 invariants $\SW_{1,x/y}$, the level-1 compatibility condition arising from the triviality of the Wilson loop $\det \wilson^{\o/\e}(\partial\cwc) = 1$ is trivially satisfied, since each 1-cell occurs twice in $\partial\cwc$ and the corresponding invariants add up to $0 \mod 2$. The two copies of a 1-cell have identical $\SW_1$, since the Hamiltonians at them are related by a unitary transformation given by $\trans_{x,y}$. As a Dirac cone manifestly violates the level-1 compatibility condition, it can be realized only if $\SW_1$ is different for the two copies of a 1-cell, \ie, if the the gluing condition is violated. This implies that a single surface Dirac cone must be anomalous.

\section{\texorpdfstring{$\C_2$}{C2} symmetry}
\label{sec:C2}
A classification of $\C_2$-symmetric topological insulators in the stable limit was previously obtained in Refs.\ \cite{turner2012,chiu2013,morimoto2013,shiozaki2014,trifunovic2019}. A partial fragile classification, based on level-0 invariants only, was obtaied using the method of topological quantum chemistry \cite{bradlyn2017}. Delicate topological invariants at levels 1 and 2 were constructed by Kobayashi and Furusaki \cite{kobayashi2021}, who also identified representation-protected phases. The full homotopic classification we present below rederives and completes these results from the literature in a consistent mathematical framework.

\subsection{Wyckoff positions, basis orbitals, and symmetries}  
\label{sec:C2_symm}
\begin{figure}
	\includegraphics[width=0.75\columnwidth]{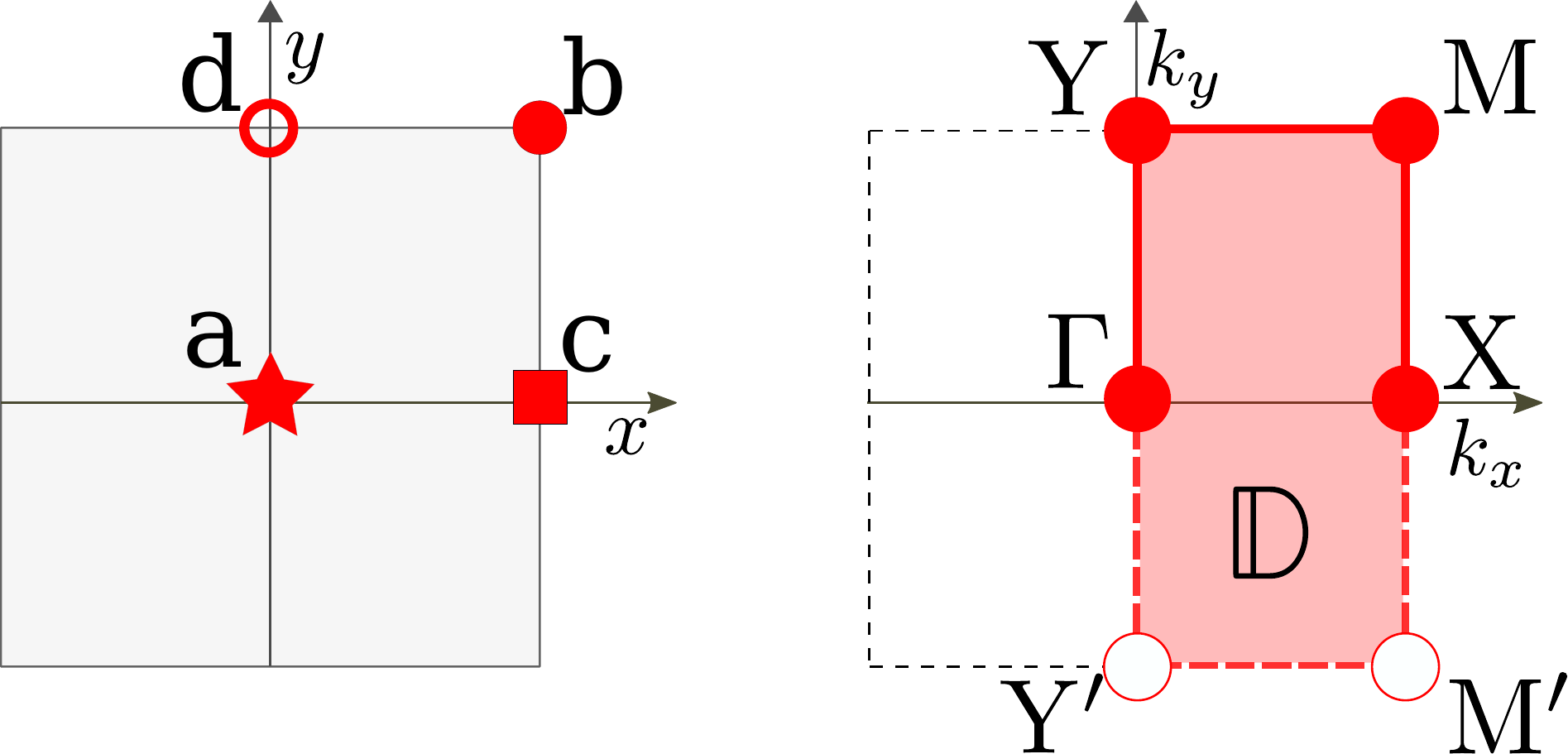}
	\caption{\label{fig:C2} 
		(Left) The Wyckoff positions for a square lattice with $\C_2$ symmetry. 
		(Right) The fundamental domain $\cwc$ for two-dimensional band structures with $\C_2$ symmetry. 
	}
\end{figure}

With twofold rotation symmetry and in two dimensions, there are four special Wyckoff positions ``a'', ``b'', ``c'', and ``d'' (shown in Fig.~\ref{fig:C2}) and one pair of generic Wyckoff positions ``g''. At each special Wyckoff position, electron orbitals are even ($+$) or odd ($-$) under twofold rotation. Combining orbitals at the two generic Wyckoff positions, one even ($+$) and one odd ($-$) orbital may be obtained. Denoting the number of orbitals of parity $\pm$ by $n_{\pm}$, the total number of orbitals is 
\begin{align}
	n =&\, n_\a + n_\b + n_\c + n_\d + 2 n_\g = n_+ + n_-.
\end{align}
The reciprocal-space group $G_\Lambda$ is generated by twofold rotation $g_2$ and two reciprocal-space translations $g_{x}$ and $g_{y}$. We denote the corresponding representation matrices (see Eq.\ (\ref{eq:Hsymm})) by $\rotn_2$, $\trans_x$, and $\trans_y$, respectively. The Bloch Hamiltonian satisfies
\begin{align}
	H(k_x, k_y) = \rotn_2 H(-k_x,-k_y) \rotn_2^\dg
\end{align}
as well as Eq.~(\ref{eq:Hsymm_ktrans}) for the reciprocal-space translations $g_{x,y}$. Since the matrices $C_2$, $\trans_x$, and $\trans_y$ generate a representation of $G_\Lambda$, they satisfy
\begin{equation}
	C_2 \trans_{x,y} = \trans_{x,y}^\dg C_2^\pdg.
	\label{eq:CDrelationC2}
\end{equation}
We choose the over-all phase factor of the twofold rotation operator $C_2$ such that $C_2^2 = \id$. Hence, in its eigenbasis, it can be written explicitly as
\begin{equation}
	C_2 = \diag\{\openone_{n_+},-\openone_{n_-}\}.
\end{equation}
Explicit expressions for $\trans_{x,y}$ are derived in Appendix~\ref{app:symm_C2}.

We take the fundamental domain $\cwc$ as the right half of the Brillouin zone, as shown in Fig.\ \ref{fig:C2}. It consists of four 0-cells, three 1-cells and one 2-cell. The 0-cells all have a nontrivial little group $G_S = \{0, g_S \}$, with $g_\Gamma = g_2$ a twofold rotation, whereas $g_{\X}$, $g_{\Y}$, and $g_{\M}$ are combinations of a twofold rotation and a reciprocal lattice translation. The corresponding representation matrices $\rotn_S \equiv C(g_S)$ are
\begin{align}
	\rotn_\G &= C_2,  \nonumber \\ 
	\rotn_{\X,\Y} &= \trans_{x,y}^\dg \rotn_2 = \rotn_2 \trans_{x,y},   \nonumber \\
	\rotn_\M &= \trans_x^\dg \trans_y^\dg \rotn_2 = \rotn_2 \trans_x \trans_y.
	\label{eq:CS}
\end{align}
These matrices all square to $\id$ and thus have eigenvalues $\pm1$. The eigenstates of $H_S$ are thus labeled by the $\rotn_S$-eigenvalue $\pm1$. The four matrices $C_S$ are not independent, since they satisfy 
\begin{equation}
  C_{\G} C_{\X} C_{\M} C_{\Y} = \openone.
  \label{eq:Cmatrixrelation}
\end{equation}
The eigenvalues of $C_S$ for band structures with a minimal set of orbitals at a single Wyckoff position constitute the EBRs \cite{bradlyn2017,cano2018ebr,po2017}. These are derived in Appendix \ref{app:symm_C4} and listed in Tab.~\ref{tab:C2_ebr}. 

The three 1-cells $\G\Y$, $\Y\M$, and $\M\X$ each have two copies, which are related as 
\begin{align}
	H(0,-k) =&\, \rotn_{\Gamma}^\dg H(0,k) \rotn_{\Gamma}, \nonumber \\
	H(\pi,-k) =&\, \rotn_{X}^\dg H(\pi,k) \rotn_{X}, \nonumber \\
	H(k,\pi) =&\, \trans_y^\dg H(k,-\pi) \trans_y.
	\label{eq:C2_Cmat_1cell}
\end{align}
The last expression is independent of $\rotn_2$, since the two copies of $\Y\M$ are related by a translation $2\pi\ve_y$. There are no local-in-$\vk$ constraints on the Hamiltonian at the 1-cells or the 2-cell. If time-reversal symmetry is present, we choose a basis where $\rotn_2 T = \id$, so that the Hamiltonian satisfies the reality condition $H(\vk) = H^*(\vk)$.

\subsection{Homotopic classification in two dimensions}
\label{sec:C2_2d}

\subsubsection{Without time-reversal (symmetry class A)}
The level-$0$ invariants are given by the number of occupied/empty bands $n_{\o/\e,\pm}^S$ for each 0-cell $S$ with even/odd parity under $\C_2$. These satisfy a  ``level-0 compatibility relation''  (see Eq.~(\ref{eq:lev0_constr}))
\begin{equation}
	n_{\o/\e} = \sum_{x = \pm} n_{\o/\e,x}^S.  
	\label{eq:C2_lev0_comp}
\end{equation}
Additionally, there are constraints on the level-0 invariants that follow from the restrictions imposed on the representation matrices $C_S$. In particular, from Eqn.~(\ref{eq:Cmatrixrelation}),it follows that $\prod_S \det C_S = 1$, so that
\begin{align}
	\sum_{S} \left( n_{\o,-}^S + n_{\e,-}^S \right) = 0 \mod 2.
	\label{eq:C2_ncond}
\end{align}
This is sufficient to ensure that the level-0 invariants, summed over occupied and unoccupied bands, constitute a linear combination of the EBRs listed in Tab. \ref{tab:C2_ebr}. 

For fixed level-0 invariants, the target spaces are 
\begin{equation}
	\Xsp_S = \Gr_\CC(n_{\o,+}^S, n_{\e,+}^S) \times \Gr_\CC(n_{\o,-}^S, n_{\e,-}^S).
\end{equation}
for the 0-cells and $\Xsp = \Gr_\CC(n_\o, n_\e)$ for the 1- and the 2-cell. The corresponding parent invariants are identical to those obtained for class A (Sec.~\ref{sec:C0}), \ie, trivial at level-1 and a $\ZZ$-valued Chern invariant $\Chern$ at level-2. Proceeding as in Sec.\ \ref{sec:C0}, one sees that the Chern invariant is robust to boundary deformations, because the Chern number of a boundary deformation along a 1-cell enters through both its copies at the boundary of the 2-cell, which precisely cancel each other. Thus, we obtain the Chern number $\Chern\in\ZZ$ as the level-2 topological invariant.

\ \\ \noindent These results are summarized in Tab.\ \ref{tab:classification_C2_AI_2d}.

\subsubsection{With time-reversal (symmetry class AI)}
The level-$0$ topological invariants are identical to the time-reversal-broken case, but now subject to an additional constraint
\begin{align}
	\label{eq:constraintC2}
	\sum_{S} n_{\o,-}^S = 0 \mod 2,
\end{align}
which arises from the level-1 topology (as we show below) and thus constitutes a level-1 compatibility relation. Following Eq.~(\ref{eq:C2_ncond}), an identical constraint also applies to the unoccupied bands. The target spaces now consist of real Grassmannians, and are explicitly given for the 0-cells by
\begin{align}
	\Xsp_S =&\, \Gr_\RR(n_{\o,+}^S, n_{\e,+}^S) \times \Gr_\RR(n_{\o,-}^S, n_{\e,-}^S)
	\label{eq:C2_Xsp0}
\end{align}
and for the 1-cells and the 2-cell by $\Xsp =\Gr_\RR(n_\o, n_\e)$. Thus, the parent classification is similar to that of $\C_2\T$ discussed in Sec.~\ref{sec:C2T}. We summarize our classification results in Tab.\ \ref{tab:classification_C2_AI_2d} and briefly discuss their derivation here. 

\subsubsection*{Classification on 1-cells}
At level-1, the  parent invariants for the 1-cell $SS'$ are the $\ZZ$-valued first Euler invariant $\Euler_1^{SS'} = -\Euler_1^{S'S}$ if $n_\o = n_\e = 1$ and the $\ZZ_2$-valued first Stiefel-Whitney number $\SW_{1}^{SS'}$ otherwise. The parent classification set for boundary deformations $K_1^\partial$ consists of a factor $\pi_1[\Xsp_S]$ for each 0-cell $S$. The boundary deformations at $S$ do not have an effect on the level-1 invariants if $\pi_1[\Xsp_S] = 0$, which happens if $\Gr_\RR(n_{\o,a}^S, n_{\e,a}^S)$ degenerates to a point for both $a = \pm$. In terms of the level-0 invariants, this happens when 
\begin{equation}
	\Q_1^S:\ n_{\o,+}^S n_{\e,+}^S = n_{\o,-}^S n_{\e,-}^S = 0.
	\label{eq:condC2}
\end{equation}
We now discuss the effect of boundary deformations on the Euler and Stiefel-Whitney invariant separately. 

\ \\ \para{First Euler invariant} 
A boundary deformation at the 0-cell $S$ may add an integer simultaneously to $\Euler_1^{SS'}$ for all 1-cells with $S$ as an end point if $\pi_1[\Xsp_S] \neq 0$. Thus, $\Euler_1^{SS'}$ is robust to boundary deformations only if the conditions $\Q^S_1$ and $\Q^{S'}_1$ are satisfied, see Eq.~(\ref{eq:condC2}). If $\Q^S_1$ is obeyed at all four 0-cells, then all three first Euler numbers $\Euler_1^{\X\M}$, $\Euler_1^{\M\Y}$, and $\Euler_1^{\Y\G}$ are well defined. If this condition is violated, then using $n_{\o/\e,\pm}^S \in \{0,1\}$ and Eq.~(\ref{eq:constraintC2}), we conclude that it must be violated at either two or all four 0-cells. In these cases, upon taking into account the ambiguity from boundary deformations, we are left with one and no first Euler invariant, respectively.  

The number of first Euler invariants are further reduced by the level-1 compatibility conditions. If $\Q_1^S$ is obeyed at all four 0-cells, then one has the symmetry relations $\Euler_1^{\G \Y'} = - \Euler_1^{\G\Y}$, $\Euler_1^{\Y'\M'} = -\Euler_1^{\Y\M}$, and $\Euler_1^{\M'\X} = -\Euler_1^{\X\M}$. Triviality of the first Euler invariant computed for the boundary of the 2-cell then gives the constraint
\begin{equation}
	\Euler_1^{\X\M} + \Euler_1^{\M\Y} + \Euler_1^{\Y\G} = 0,
\end{equation} 
which effectively removes one level-1 invariant. There is a similar constraint if $\Q_1^S$ is obeyed only at two of the 0-cells, which removes the only level-1 invariant present in that case. Thus, there are two independent first Euler invariants if $\Q_1^S$ is obeyed at all 0-cells $S$, while there are none otherwise.

\ \\ \para{First Stiefel-Whitney invariant} 
If a boundary deformation flips the orientation at a 0-cell $S$, then it flips the sign of $\det \wilson^{\o/\e}_{SS'}$ for all 1-cells having $S$ as their end point. Such a boundary deformation is allowed if $\Q_1^S$ is violated, thereby effectively ``removing'' one level-1 invariant. 

The level-1 compatibility condition follows from triviality of the Wilson loop around the boundary of the 2-cell
\begin{equation}
	\wilson_{\partial\cwc} = \wilson_{\G\Y'} \wilson_{\Y'\M'} \wilson_{\M'\X} \wilson_{\X\M} \wilson_{\M\Y} \wilson_{\Y\G} = \id,
\end{equation}
where we have skipped the superscript $\o/\e$ to avoid notational clutter. The Wilson loops at the symmetry-related 1-cells satisfy
\begin{align}
	\wilson^{\o/\e}_{\G\Y'} &= \rotn^{\o/\e \dagger}_{\G} \wilson^{\o/\e}_{\G\Y} \rotn^{\o/\e}_{\Y}, \nonumber \\ 
	\wilson^{\o/\e}_{\Y'\M'} &= \wilson^{\o/\e}_{\Y\M}, \nonumber \\ 
	\wilson^{\o/\e}_{\M'\X} &= \rotn^{\o/\e \dagger}_{\M} \wilson^{\o/\e}_{\M\X} \rotn^{\o/\e}_{\X}. 
	\label{eq:C2_wilson_symm}
\end{align}
Using these and $\det \wilson_{SS'} \in \{ \pm1 \}$, we get 
\begin{align}
	\det \wilson^{\o/\e}_{\partial\cwc} 
	&= \det\left( \rotn^{\o/\e\dagger}_{\G} \rotn^{\o/\e}_{\Y} \rotn^{\o/\e\dagger}_{\M} \rotn^{\o/\e}_{\X} \right), 
	\label{eq:WCdet}
\end{align}
so that the compatibility condition becomes 
\begin{equation}
	1 = \prod_{S} \det \rotn^{\o/\e}_{S} = \prod_{S} (-1)^{n_{\o/\e,-}^S},
	\label{eq:constraintC2anomalous}
\end{equation}
which is equivalent to  Eq.\ (\ref{eq:constraintC2}). Thus, the compatibility condition does not constrain the level-1 invariants, but imposes a constraint on the level-0 invariants.

\begin{table*}
	\centering
	\setlength{\tabcolsep}{15pt}
	\begin{tabular}{|r|c|l|l|}
		\hline 
		& Bands & Additional condition(s) & Topological invariant \\
		\hline
		A, Level 1
		& $n_\o$, $n_\e \ge 1$ & \multicolumn{2}{c|}{------} \\
		\hline 
		A, Level 2
		& $n_\o$, $n_\e \ge 1$ & --- & \highlight{$\Chern\in\ZZ$ \ [S]}  \\ 
		\hline \hline
		\multirow{4}{*}{AI, Level 1} 
		& \multirow{2}{*}{$n_\o = n_\e = 1$} & if $\Q_1^{\land}$ &  $\Euler_1^{\Gamma \Y}$, $\Euler_1^{\Y\M} \in \ZZ$ \\ \cline{3-4}
		&& if not $\Q_1^{\land}$ & ---   \\ \cline{2-4}
		& $n_\o \ge 1$, $n_\e > 1$ & \multirow{2}{*}{---} & $\SW_1^{\Gamma \Y}$, $\SW_1^{\Y\M}$, $\SW_1^{\M\X}\in \ZZ_2$ \\ 
		& or $n_\o > 1$, $n_\e \ge 1$ &  &  one less for every violation of $\Q_1^S$ \\ 
		\hline
		\multirow{17}{*}{AI, Level 2}   
		& $n_\o = n_\e = 1$ & \multicolumn{2}{c|}{------}
		\\ \cline{2-4}
		& \multirow{3}{*}{$n_\o = 1$, $n_\e = 2$} 
		& if $\Q_1^{\lor}$ and $\Q_2^\e$ & $\Euler_2^\e \in 2\ZZ$ \\ \cline{3-4}
		&& if $\Q_2^\e$, but not $\Q_1^{\lor}$ & $\Euler_2^\e \in 2\NN$ \\ \cline{3-4}
		&& if not $\Q_2^{\rm e}$  & $\Euler_2^\e \in \ZZ_2$  
		\\ \cline{2-4}
		& $n_\o = 1$, $n_\e > 2$ & \multicolumn{2}{c|}{------}
		\\ \cline{2-4}
		& \multirow{8}{*}{$n_\o = n_\e = 2$} 
		& if $\Q_1^{\lor}$ and $\Q_2^\e$ and $\Q_2^\o$ & $(\Euler_2^\o,\Euler_2^\e) \in \ZZ^2|_{\rm p}$  \\ \cline{3-4}
		&& if $\Q_2^\e$ and $\Q_2^\o$, but not $\Q_1^{\lor}$ & $(\Euler_2^\o,\Euler_2^\e) \in \NN \times \ZZ|_{\rm p}$ \\ \cline{3-4}
		&& if $\Q_1^{\lor}$ and $\Q_2^\o$, but not $\Q_2^\e$ & $\Euler_2^{\o} \in \ZZ$ \\ \cline{3-4}		
		&& if $\Q_1^{\lor}$ and $\Q_2^\e$, but not $\Q_2^\o$ & $\Euler_2^{\e} \in \ZZ$ \\ \cline{3-4}		
		&& if $\Q_2^\o$, but not $\Q_1^{\lor}$ and not $\Q_2^\e$ & $\Euler_2^{\o} \in \NN$ \\ \cline{3-4}		
		&& if $\Q_2^\e$, but not $\Q_1^{\lor}$ and not $\Q_2^\o$ & $\Euler_2^{\e} \in \NN$ \\ \cline{3-4}		
		&& if $\Q_2^{\land}$, but not $\Q_2^\e$ and not $\Q_2^\o$ & $\SW_2 \in \ZZ_2$ \\ \cline{3-4}		
		&& if not $\Q_2^{\land}$ & ---	
		\\ \cline{2-4}
		& \multirow{4}{*}{$n_\o = 2$, $n_\e > 2$} 
		& if $\Q_1^{\lor}$ and $\Q_2^\o$ & $\Euler_2^\o \in \ZZ$ \\ \cline{3-4}
		&& 	\highlight{if $\Q_2^\o$, but not $\Q_1^{\lor}$} & 	\highlight{$\Euler_2^\o \in \NN$ \ [F] }\\ \cline{3-4}	
		&& if $\Q_2^{\land}$, but not $\Q_2^\o$ & $\SW_2 \in \ZZ_2$ \\ \cline{3-4}	
		&& if not $\Q_2^{\land}$ &  --- 
		\\ \cline{2-4}
		& \multirow{2}{*}{$n_\o$, $n_\e > 2$} 
		& \highlight{if $\Q_2^{\land}$} & \highlight{$\SW_2 \in \ZZ_2$ [F]} \\ \cline{3-4} 
		&& if not $\Q_2^{\land}$ & --- \\	
		\hline
	\end{tabular}	
	\caption{Homotopic classification of gapped band structures with $\C_2$ symmetry in two dimensions for various possible combinations of number of bands and level 0 invariants, the latter encoded in the $\Q$'s. In particular, $\Q_1^S$, $\Q_2^{\o/\e}$, $\Q_2^S$, and $\Q_2^\land$ are defined by Eqs.\ (\ref{eq:condC2}), (\ref{eq:condC2b}), (\ref{eq:condC2c}), and (\ref{eq:condC2cand}), respectively. The condition $\Q_1^{\land}$ holds if $\Q_1^S \, \forall S$, while $\Q_1^{\lor}$ holds if $\exists S$ where $\Q_1^S$ is true. The set $\ZZ^2|_{\rm p}$ consists of all ordered integer pairs $(p,q)$ with $p = q \mod 2$, while the set $\NN \times \ZZ_{\rm p}$ consists of all ordered integer pairs $(p,q)$ with $p \ge 0$ and $p = q \mod 2$. The table lists the cases $n_\o \leq n_\e$ only. The remaining classifications with $n_\o > n_\e$ can be obtained from the entries in the Table by exchanging the labels ``$\o$'' and ``$\e$''. The fragile and stable topological phases are marked by \highlight{[F]} and \highlight{[S]}, respectively. 
	\label{tab:classification_C2_AI_2d}}
\end{table*}

\subsubsection*{Classification on the 2-cell}
At level-2, the the parent classification set is given by $K_2 = \pi_2[\Xsp]$, with the corresponding invariants being the $\ZZ$-valued second Euler invariant or the $\ZZ_2$-valued second Stiefel-Whitney invariant, depending on $n_{\o/\e}$ (see Table~\ref{tab:hom_real}). The parent classification set for boundary deformations $K_2^\partial$ consists of a factor of $\pi_2[\Xsp] \cong K_2$ for each 1-cell and of $\pi_1[\Xsp_S]$ for each 0-cell, which consists of two factors corresponding to the two parity sectors (see Eq.~(\ref{eq:C2_Xsp0})). We now discuss the effect of boundary deformations on the Euler and Stiefel-Whitney invariant separately. 

\ \\ \para{Second Euler invariant}
For the occupied/empty subspaces, $\Euler_2^{\o/\e}$ is defined if $n_{\o/\e} = 2$, whereby it is $\ZZ$-valued if $n_{\e/\o} \geq 2$ and $2\ZZ$-valued if $n_{\e/\o} = 1$. For the former case, boundary deformations that are nontrivial along a 1-cell (\ie, in $\pi_2[\Xsp] \cong K_2$) can add an arbitrary integer to $\Euler_2^{\o/\e}$. However, as each 1-cell occurs twice in $\partial\cwc$, following the argument for $\C_2\T$ (Sec.~\ref{sec:C2T}), these individual contributions cancel out unless the orientation of the occupied/empty subspace is flipped between the two copies of the 1-cell. From Eq.~(\ref{eq:C2_wilson_symm}), this can happen if $\det C_{S}^{\o/\e} = -1$ for any of the 0-cells. Thus, the two boundary contributions cancel if $\det C_{S}^{\o/\e} = 1$ for all $S$, or, equivalently, if  
\begin{equation}
	\Q_2^{\o/\e}:\
	n^S_{\o/\e,+} n^S_{\o/\e,-} = 0 \ \mbox{for all $S$}.
	\label{eq:condC2b}
\end{equation}
If this condition is violated, then the boundary deformations at the two copies of the 1-cell add up. For $n_{\o/\e} = 2$ and $n_{\e/\o} \geq 2$, such a boundary deformation may add an even integer to $\Euler^{\o/\e}$, rendering the classification group $\ZZ/2\ZZ \cong \ZZ_2$. On the other hand, for $n_{\o/\e} = 2$ and $n_{\e/\o} = 1$, such boundary deformations may add an arbitrary multiple of four to $\Euler_2^{\o/\e}$, so that the classification group becomes $2 \ZZ/4 \ZZ = \ZZ_2$.

Next, we consider boundary deformations that are nontrivial at a 0-cell $S$, which may be nontrivial in either one or both factors of $\pi_1[\Xsp_S]$. In the former case, boundary deformations can flip the orientation of the basis of the occupied/empty subspace for one of the parity sectors at $S$, and hence of that subspace also away from $S$. By continuity, such a deformation must be nontrivial at each 0-cell, so that it cannot exist if $\pi_1[\Xsp_S] = 0$ for at least one 0-cell, \ie, if $\exists S$ for which $\Q_1^S$ holds. If that is not the case, then such a boundary deformation changes the sign of both $\Euler_2^{\o}$ and $\Euler_2^{\e}$, replacing the level-2 classifying spaces $\ZZ$ by $\NN$ and $\ZZ^2|_{\rm p}$ by $\NN \times \ZZ|_{\rm p}$.

Boundary deformations with nontrivial topology in both factors of $\pi_1[\Xsp_S]$ cannot change the orientation of the occupied/empty subspaces, so that they are trivial away from $S$. Such deformations may, however, change both $\Euler_2^{\o}$ and $\Euler_2^{\e}$ by arbitrary integers. (See App.\ \ref{app:deform} for an explicit construction of such a boundary deformation.) These deformations do not exist if either Grassmannian in $\Xsp_S$ degenerates to a point, \ie, if 
\begin{equation}
	\Q_2^S:\  n^S_{\o,+} n^S_{\o,-} n^S_{\e,+} n^S_{\e,-} = 0. 
	\label{eq:condC2c}
\end{equation}
We further define 
\begin{equation}
	\Q_2^{\land}:\ \Q_2^S\ \mbox{for all}\ S.
	\label{eq:condC2cand}
\end{equation}
Thus, if $\Q_2^{\land}$ is violated, \ie, if $\Q_2^S$ is violated at any $S$, then the boundary deformation around $S$ can change both $\Euler_2^{\o}$ and $\Euler_2^{\e}$ by arbitrary integers, so that there are no topological invariants at level-2.  

\ \\ \para{Second Stiefel-Whitney invariant}
The boundary deformations at 1-cells do not affect $\ZZ_2$-valued $\SW_2$, since the contributions from its two copies always cancel out. On the other hand, boundary deformations at the 0-cells with nontrivial topology in both factors of $\pi_1[\Xsp_S]$, if they exist, can add arbitrary integers to $\SW_2$, rendering the topology at level-2 trivial.

\begin{table*}
	\centering
	\begin{tabular}{|r|c|l|l|l|}
		\hline
		&  & Level 1 & Level 2 & Level 3 
	\\ \hline 
		\multirow{4}{*}{A} & \multirow{2}{*}{$n_\o = n_\e = 1$}
		& \multirow{2}{*}{---} &
		$\Chern_z$, $\Chern^{\G\Y}$, $\Chern^{\Y\M}$, $\Chern^{\M\X} \in \ZZ$, & 
		if $\Chern_z = 0$: $\Hopf \in \ZZ$ \\ \cline{5-5} &
		&& one less for every violation of $\Q_1^S$ & 
		if $\Chern_z \neq 0$: $\Hopf \in \ZZ_{2 \Chern_z}$ 
	\\ \cline{2-5}
		& \multirow{1}{*}{$n_\o$, $n_\e \ge 1$,} &
		\multirow{2}{*}{---}
		& {\color{blue}{$\Chern \in \ZZ$ [S]}}; 
		$\Chern^{\G\Y}$, $\Chern^{\Y\M}$, $\Chern^{\M\X} \in \ZZ$, &
		\multirow{2}{*}{---} \\ &
		\multirow{1}{*}{$n_\o + n_\e > 2$} &&
		\multirow{1}{*}{one less for every violation of $\Q_1^S$} & 
	\\ \hline\hline  
		\multirow{5}{*}{AI} & \multirow{2}{*}{$n_\o = n_\e = 1$}
		& \multirow{1}{*}{Two copies of invariants} &
		$\Chern^{\G\Y}$, $\Chern^{\Y\M}$, $\Chern^{\M\X} \in \ZZ$,
		& \multirow{2}{*}{$\Hopf \in \ZZ$} \\ & & 
		\multirow{1}{*}{from Tab.\ \ref{tab:classification_C2_AI_2d}} &
		one less for every violation of $\Q_1^S$ & 
	\\ \cline{2-5}
		& \multirow{2}{*}{$n_\o$, $n_\e \ge 1$,} & \multirow{2}{*}{Two copies of invariants} & \color{blue}{two copies of invariants from} & 	\multirow{3}{*}{---} \\ &
		\multirow{2}{*}{$n_\o + n_\e > 2$} &
		\multirow{2}{*}{from Tab.\ \ref{tab:classification_C2_AI_2d}} & 
		{\color{blue}{Tab.\ \ref{tab:classification_C2_AI_2d}} [F]}; $\Chern^{\G\Y}$, $\Chern^{\Y\M}$, $\Chern^{\M\X} \in \ZZ$, & \\ 
		&&& one less for every violation of $\Q_1^S$ & \\
		\hline 
	\end{tabular}
	\caption{Homotopic classification of gapped band structures with $\C_2$ symmetry in three dimensions. The condition $\Q_1^S$ is given in Eq.\ (\ref{eq:condC2}). Nontrivial fragile and stable invariants are denoted with \highlight{[F]} and \highlight{[S]}, respectively.
	\label{tab:classification_C2_A_3d}}
\end{table*}

\subsubsection*{Summary}
Table \ref{tab:classification_C2_AI_2d} summarizes the classification for $\C_2$-symmetric gapped band structures with time-reversal symmetry. To obtain the fragile and stable classifications, besides taking the limits $n_\e \gg 1$ or $n_\o$, $n_\e \gg 1$, one must also verify compatibility of the various conditions for the existence of these invariants with the rules of fragile and stable topology. In particular, the conditions $\Q_1^S$ and $\Q_2^{\e}$ are incompatible with fragile classification rules, since the latter allows for addition of unoccupied bands with arbitrary band representations. Furthermore, $\Q_2^{\land}$ and $\Q_2^\o$ become equivalent, since $n_{\e,\pm} > 0$ after a possible addition of empty bands. Finally, all of these conditions are incompatible with the rules of stable classification, which allows for adding occupied/empty bands with arbitrary band representations. Thus, there are no $\C_2$-symmetric stable topological phases. For $n_\e \gg 1$, we get the classification of fragile phases consistent with the fragile band structures identified via topological quantum chemistry \cite{bradlyn2017}. The delicate topological invariants agree with those obtained by Kobayashi and Furusaki \cite{kobayashi2021}. 

We also encounter our first example of \emph{representation-protected stable topology}, whereby a stable topological phase exists if we constrain the band representations such that the condition $\Q_2^{\land}$ is always satisfied, without limiting the number of occupied and unoccupied bands. This can be ensured by demanding that all bands at each 0-cell must have the same parity. To ensure the representation constraint in real space, we turn to the EBRs \cite{bradlyn2017,cano2018ebr,po2017} of a $\C_2$-symmetric band structure, listed in Tab.~\ref{tab:C2_ebr}. Since each EBR, labelled by a special Wyckoff position and a parity, corresponds to a fixed set of parities at each 0-cell, we can ensure a representation-protected stable phase only by allowing a \emph{single} orbital type at a single Wyckoff position. Inspection of Tab.~\ref{tab:C2_ebr} further shows that the presence of multiple EBRs always leads to the existence of different parity bands at at least one of the high-symmetry points, thereby violating the condition for representation-protected stable topology.

\subsection{Classification in three dimensions}
\label{sec:C2_3d}

\begin{figure}
	\includegraphics[width=0.9\columnwidth]{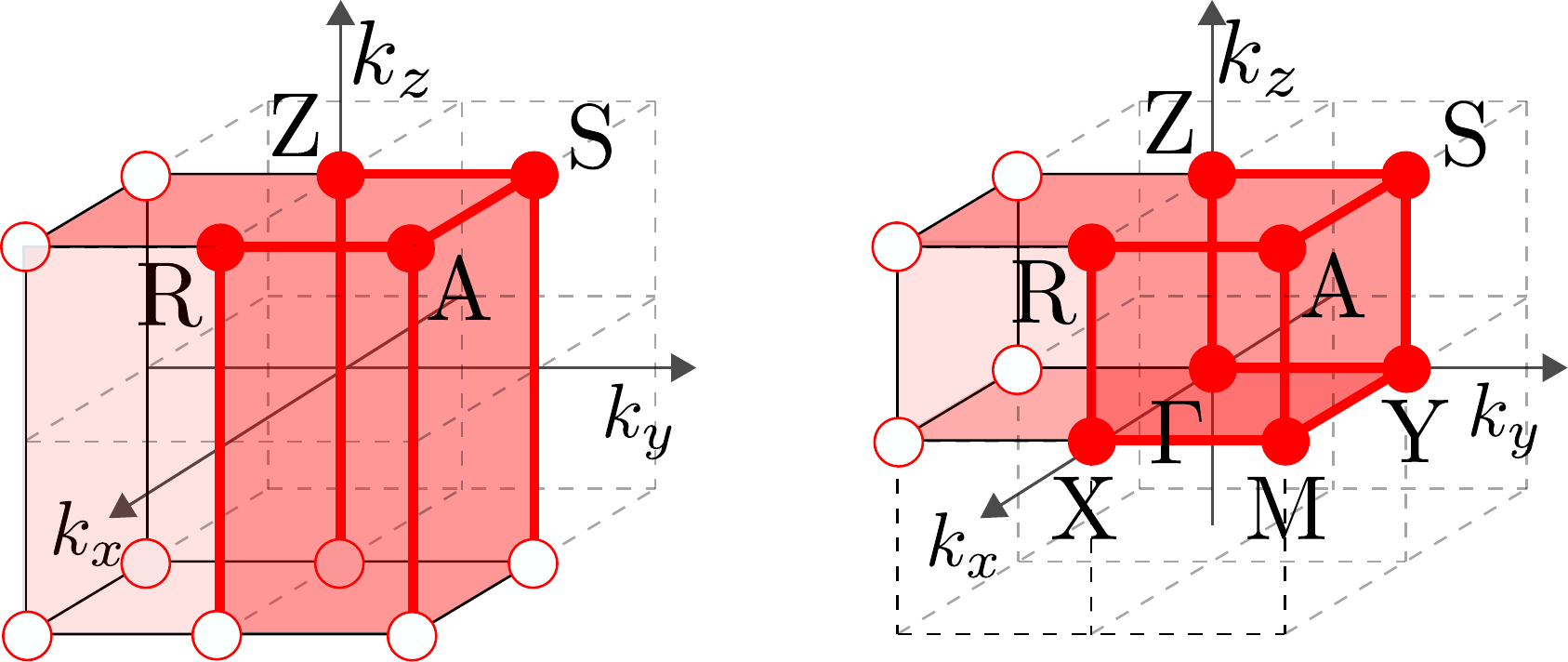}
	\caption{\label{fig:fundamentalC2_3d} Fundamental domains $\cwc$ for a three-dimensional band structure with twofold rotation symmetry without (left) and with (right) time-reversal symmetry.}
\end{figure}

\subsubsection{Without time-reversal symmetry}
The fundamental domain is a cuboid as shown in Fig.~\ref{fig:fundamentalC2_3d}(left). There are no level-1 invariants, as follows for horizontal and vertical 1-cells from the 2d classification and Sec.~\ref{sec:outline}, respectively. At level-2, the parent invariants are Chern numbers for each 2-cell. For the horizontal 2-cell, $\Chern_z$ is robust to boundary deformations and can be computed for any fixed-$k_z$ slice of $\cwc$. For the vertical 2-cells,  boundary deformations at the vertical line through a 0-cell $S$ can add arbitrary integers to all $\Chern^{SS'} = -\Chern^{S'S}$. Such deformations are allowed if $\pi_1[\Xsp_S] \neq 0$, \ie, if  $\Q_1^S$  (Eq.~(\ref{eq:constraintC2})) is violated. Thus, the number of level-2 invariants associated with vertical faces is one less than the number of 0-cells satisfying $\Q_1^S$. These invariants were previously identified by Nelson {\em et al.} \cite{nelson2022}, who refer to them as ``returning Thouless pump''. At level-3, we get the $\ZZ$-valued Hopf invariant when $n_\o = n_\e = 1$. This reduces to a $\ZZ_{2 \Chern_z}$-valued invariant if $\Chern_z \neq 0$.

\subsubsection{With time-reversal symmetry}
The fundamental domain is a cuboid as shown in Fig.~\ref{fig:fundamentalC2_3d}(right). At level-0, level-1, and level-2, we get two independent sets of topological invariants as for the two-dimensional case, associated with the planes $k_z = 0$ and $k_z = \pi$. At level-2, we additionally get three Chern numbers as parent invariants, associated with the vertical faces of $\cwc$. The conditions for their robustness with respect to boundary deformations is the same as in the case of broken time-reversal symmetry. At level-3, we get the Hopf invariant if $n_\o = n_\e = 1$, which is always $\ZZ$-valued, since the Chern number associated to a cross section of $\cwc$ at constant $k_z$ vanishes. 

\ \\ \noindent These classification results are summarized in Table \ref{tab:classification_C2_A_3d}. Fragile and stable invariants in 3d derive entirely from those in 2d (see Sec.\ \ref{sec:C2_2d}).

\subsubsection{Anomalous boundary states}
In the absence of time-reversal symmetry, a nontrivial Chern number $\Chern_z$ for the horizontal 2-cell indicates a weak Chern-insulator phase, which is a stable topological phase with topologically protected chiral surface states on surfaces parallel to the rotation axis.

In the presence of time-reversal symmetry, we get a pair of surface Dirac cones related by $\C_2$ \cite{kobayashi2021}, which are associated with a nontrivial representation-protected stable topology if the level-2 invariants at $k_z = 0,\pi$ are different. From the boundary perspective, following the $\C_2\T$-symmetric case (Sec.~\ref{sec:C2T_bdry}), a single Dirac node in the fundamental domain implies that $\det \wilson^{\o/\e}(\partial\cwc) = -1$, which violates the level-1 compatibility constraint of Eq.~(\ref{eq:constraintC2}). This condition cannot be satisfied for a lattice model with only orbitals of a single parity at a single Wyckoff position, as evident from the EBRs of Tab.\ \ref{tab:C2_ebr}. Hence, a surface band structure with a pair of Dirac cones (or an odd number of such pairs) is anomalous under the rules of representation-protected stable topology.

\section{\texorpdfstring{$\C_4$}{C4} symmetry}
\label{sec:C4}
A four-band lattice model of spinless electrons in three dimensions with $\C_4$ and $\T$ symmetries was introduced by Fu, as an early example of topology protected by a crystalline symmetry \cite{fu2011}. Song, Elcoro, and Bernevig discuss a model of spinless electrons with $C_4$ symmetry in two dimensions and show that it is topologically inequivalent to a insulator with localized orbitals under the rules of fragile topology. Both of these are particularly instructive, because, depending on the precise constraints imposed, they exhibit nontrivial stable, fragile, or delicate topology. Previous theoretical analyses of these models have identified multiple topological invariants specific to delicate or fragile classifications \cite{alexandradinata2014,alexandradinata2016,alexandradinata2020,song2020,song2020b,kobayashi2021}. We now show how these invariants arise in the framework of a systematic homotopic classification.

\begin{figure}
	\includegraphics[width=0.85\columnwidth]{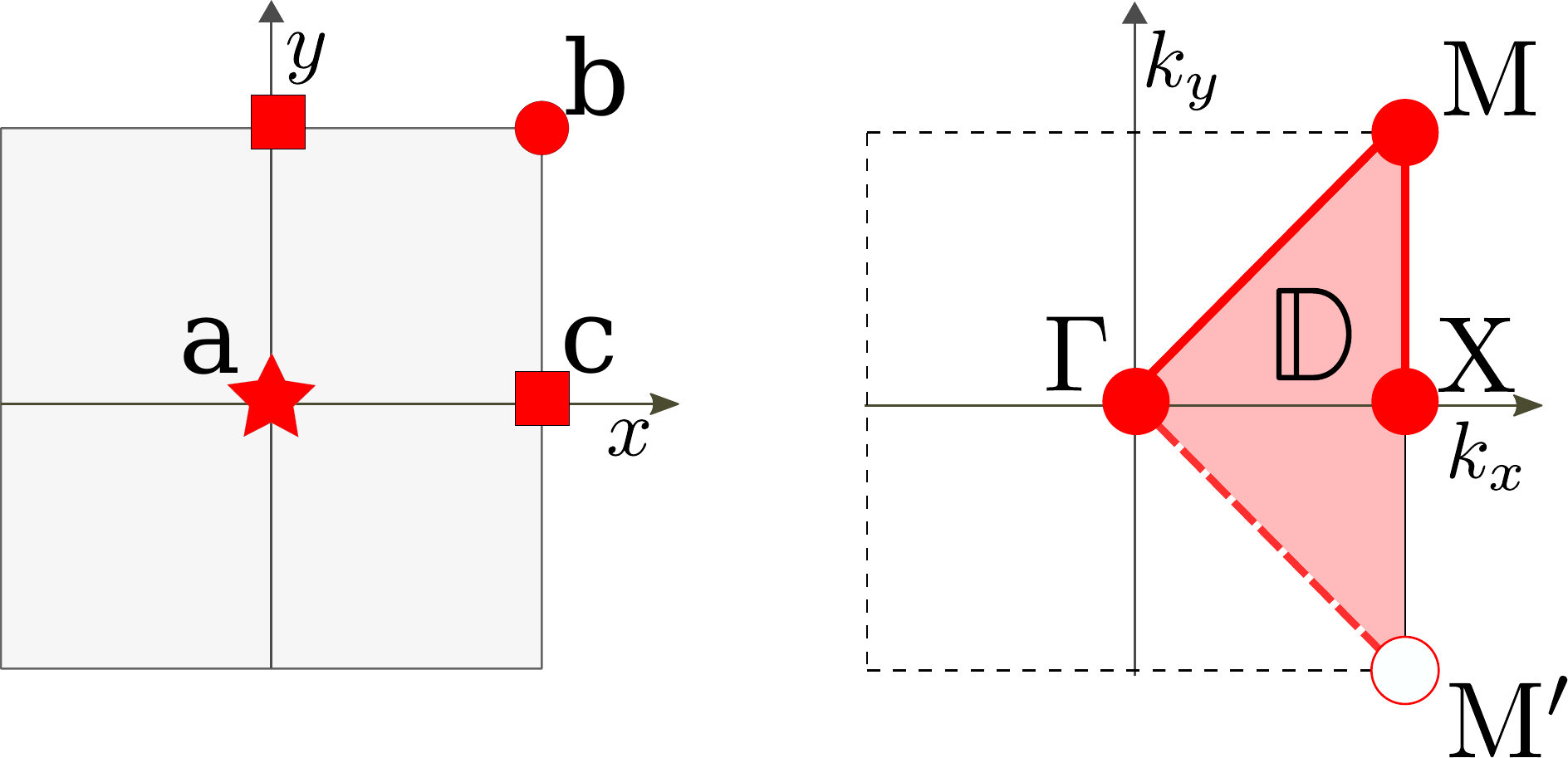}
	\caption{\label{fig:C4} 
		(Left) The Wyckoff positions for a square lattice with $\C_4$ symmetry. 
		(Right) The fundamental domains $\cwc$ for two-dimensional band structures with $\C_4$ symmetry. 
	}
\end{figure} 

\subsection{Wyckoff positions and basis orbitals}
\label{sec:C4_symm}
With fourfold rotation symmetry, there are three special Wyckoff positions ``a'', ``b'', and ``c'', as well as one pair of generic Wyckoff positions ``g'', as shown in Fig.~\ref{fig:C4}. The special Wyckoff positions ``a'' and ``b'' have fourfold rotation symmetry, whereas the special Wyckoff position ``c'' has twofold rotation symmetry only. Correspondingly, one may choose a basis of orbitals at the Wyckoff positions that transform under one of the irreducible representations of the fourfold or twofold rotation operation, respectively. For ``a'' and ``b'', this means that orbitals can be of $s$, $d$, or $p\pm$ type, corresponding to eigenvalue 1, $-1$, or $\pm i$ under fourfold rotation. With time-reversal symmetry the $p\pm$ orbitals must be combined into a doublet of $p$ orbitals that transforms under a two-dimensional representation of a fourfold rotation. For the two ``c'' positions, orbitals are even ($+$) or odd ($-$) under twofold rotation. Combining orbitals at the two ``c'' positions, a pair of even-parity orbitals may be rearranged as one $s$ orbital and one $d$ orbital, whereas the pair of odd-parity orbitals may be arranged as a pair of $p$ orbitals. Finally, at the generic Wyckoff position orbitals appear in quartets that can be arranged as one $s$ orbital, one $d$ orbital, and a pair of $p$ orbitals. The parity of $s$ and $d$ orbitals under twofold rotation is even ($+$); the parity of $p$ orbitals under twofold rotation is odd ($-$).

The reciprocal-space group $G_\Lambda$ is generated by fourfold rotation $g_4$ and two reciprocal-space translations $g_{x}$ and $g_{y}$. The corresponding representation matrices $C(g)$ (cf.\ Eq.~(\ref{eq:Hsymm})) are denoted $C_4$, $\trans_x$, and $\trans_y$, respectively. The Bloch Hamiltonian satisfies
\begin{align}
  H(k_x, k_y) = C_4^\pdg H(-k_y, k_x) C_4^\dg
\end{align}
for a fourfold rotation, as well as Eq.~(\ref{eq:Hsymm_ktrans}) under $g_{x,y}$. Since the matrices $C_4$, $\trans_x$, and $\trans_y$ generate a representation of $G_\Lambda$, they satisfy
\begin{equation}
  C_4 \trans_x = \trans_y C_4, \qquad
  C_4 \trans_y = \trans_x^\dg C_4^\pdg.
  \label{eq:CDrelationC4}
\end{equation}
We choose the overall phase factor of $C_4$ such that $C_4^4 = \id$. In its eigenbasis, it can be written explicitly as
\begin{equation}
  C_4 = \diag\{\id_{n_s},-\id_{n_\d},\i \id_{n_{p+}},-\i \id_{n_{p-}}\}.
\end{equation}
In the presence of time-reversal symmetry, the $p_\pm$ orbitals always come in pairs, and changing to a basis where $C_4$ is manifestly real, we can write
\begin{align}
  C_4 = \diag\{\id_{n_s},-\id_{n_\d},\i \sigma_2 \id_{n_p}\}.
\end{align}
Explicit expresions for $\trans_{x,y}$ are derived in Appendix~\ref{app:symm_C4}.

We take the fundamental domain $\cwc$ as the right quadrant of the Brillouin zone enclosed between the lines $k_y = \pm k_x$, as shown in Fig.\ \ref{fig:C4}. It consists of three 0-cells, two 1-cells and one 2-cell. The little groups of the 0-cells are $G_\G \cong G_\M \cong \C_4$ and $G_\X \cong \ZZ_2$, with $G_\G$ generated by the fourfold rotation $g_\G = g_4$, $G_\M$ generated by $g_\M = g_x^{-1}g_4^\pdg$ and $G_\X$ by the twofold rotation followed by a reciprocal lattice translation $g_\X = g_x^{-1}g_4^2$. The corresponding matrices $\rotn_S \equiv C(g_S)$ are thus given by
\begin{align}
  \rotn_\G &= \rotn_4, \nonumber  \\
  \rotn_\M &= \trans_x^\dagger \rotn_4^\pdg = \rotn_4 \trans_y,  \nonumber \\
  \rotn_\X &= \trans_x^\dagger \rotn_4^2 = \rotn_4^2 \trans_x^\pdg
  \label{eq:Cmatrices}
\end{align}
and satisfy
\begin{equation}
  C_\X C_\M C_\G = \openone. \label{eq:C4matrixrelation}
\end{equation}
We can derive the spectrum of these matrices without knowing their explicit form. By definition, $\rotn_\G^4 = \id$, and furthermore, using Eq.~(\ref{eq:CDrelationC4}), we get
\begin{align}
  \rotn_\M^4 
  = \left (\trans^\dagger_x \rotn_4^\pdg \right)^4
  = \trans^\dagger_x  \trans^\dagger_y \rotn^4 \trans_x \trans_y
  = \id.
  \label{eq:C4_M4}
\end{align}
Thus, the spectrum of $\rotn_{\G,\M}$ is $\{1,-1, \i, -\i\}$, corresponding to $\{s,d,p_+,p_-\}$-type orbitals. in the presence of time-reversal symmetry, the $p_\pm$-type orbitals always come in pairs. Finally, 
\begin{align}
  \rotn_\X^2 
  = (\trans_x^\dagger \rotn_4^2) (\rotn_4^2 \trans_x^\pdg)
  = \trans_x^\dagger \trans_x^\pdg = \id,
  \label{eq:C4_X2}
\end{align}
so that the spectrum of $\rotn_\X$ consists of $\pm1$, \ie, an even and an odd orbital. The EBRs for a $\C_4$-symmetric band structure are listed in the appendix, see Table~\ref{tab:C4_ebr}.

\subsection{Homotopic classification in two dimensions}
\label{sec:C4_2d}
The homotopic classification for this case closely follows that for $\C_2$-symmetric band structures described in Sec.\ \ref{sec:C2_2d}. We summarize these results in Tab.\ \ref{tab:clfn_C4_2d}. 

\subsubsection{Without time-reversal symmetry}
The level-0 invariants are given by $n_{\o/\e,\rho}^S$, the number of occupied/empty bands at $S$, with $\rho \in \{s,p_\pm,d\}$ for $S = \G,\M$ and $\rho \in \{\pm\}$ for $S = \X$. The ``level-0 compatibility relation'' (Eq.~(\ref{eq:lev0_constr})) reads
\begin{equation}
  n_{\o/\e} = \sum_{\rho \in \{s, p_\pm, d\}} n_{\o/\e,x}^S
            = \sum_{\rho = \pm} n_{\o/\e,\rho}^\X,     
  \label{eq:C4_lev0_comp}
\end{equation}
where $S = \G,\M$. Additionally, there are constraints on the level-0 invariants that follow from Eq.~(\ref{eq:C4matrixrelation}). The target spaces for the 0-cells are given by 
\begin{align}
  \Xsp_S = \prod_\rho \Gr_{\CC}\left( n_{\o,\rho}^S,n_{\o,\rho}^S \right),
\end{align}
where the direct product is over $\rho = s$, $d$, $p+$, and $p-$ for $S = \Gamma$, $\M$ and $\rho = \pm$ for $S = \X$. For the 1-cells and the 2-cell, the target space is $\Xsp = \Gr_\CC(n_\o, n_\e)$.

At level-1, there are no topological invariants, while at level-2, we get a $\ZZ$-valued topological invariant, \emph{viz}, the Chern number, irrespective of the level-0 invariants. The latter is robust to boundary deformations following an argument identical to Sec.~\ref{sec:C2}.

\subsubsection{With time-reversal symmetry}
In the presence of time-reversal symmetry, the $p$ bands always come in pairs at the high-symmetry points $S = \G$, $\M$. Denoting the number of these doublets by $n_{\o/\e,p}^S$, the level-0 invariants for the $\G$ and $\M$ point are $\large(n_{\o/\e,s}^S,n_{\o/\e,d}^S,n_{\o/\e,p}^S \large)$, while those at $\X$ are identical to the time-reversal broken case. The level-0 compatibility relation (\ref{eq:lev0_constr}) takes the form
\begin{align}
  n_{\o/\e} =&\, n_{\o/\e,+}^\X + n_{\o/\e,-}^\X \nonumber \\ =&\,
  n_{\o/\e,s}^S + n_{\o/\e,d}^S + 2 n_{\o/\e,p}^S,
  \label{eq:constraintC0}
\end{align}
where $S = \G$, $M$. The level-0 invariants also satisfy the level-1 compatibility condition 
\begin{align}
  \label{eq:constraintC}
  n_{\o/\e,-}^\X = n_{\o/e,d}^\G + n_{\o/\e,d}^\M \mod 2,
\end{align}
which will be derived below. The target spaces for the high-symmetry points are given by 
\begin{align}
  \Xsp_S &= \Gr_{\RR}(n_{\o,s}^S,n_{\e,s}^S) \times \Gr_{\RR}(n_{\o,d}^S,n_{\e,d}^S) \times \Gr_{\CC}(n_{\o,p}^S,n_{\e,p}^S), \nonumber \\
  \Xsp_\X &= \Gr_{\RR}\left(n_{\o,+}^\X,n_{\e,-}^\X\right) \times \Gr_{\RR}\left(n_{\o,-}^\X,n_{\e,-}^\X\right),
\end{align}
where $S = \G,\M$. To see that the target space for $p$ orbitals is a complex Grassmannian, we note that for $p$-orbitals, we can choose a basis where $C_S = \id_{n_p^S} \otimes \i\sigma^2$, where $n^S_p = n^S_{\o,p} + n^S_{\e,p}$. Any $H_S$ that commutes with $C_S$ can thus be written as 
\[
	H_S = h_S \otimes \sigma^0 + \i h_S' \otimes \sigma^2,
\]
where $h_S$ amd $h_{S'}$ are $n^S_p$-dimensional real-valued symmetric and antisymmetric matrices, respectively. Hence, the $2 n^S_p$-dimensional real symmetic matrix $H_S$ can equivalently be represented by the complex $n^S_p$-dimensional matrix $h_S + \i h_S'$. 

\begin{table*}
	\centering
	\begin{tabular}{|r|c|l|l|}
		\hline 
		& Bands & Additional condition(s) & Topological invariant \\
		\hline
		A, Level 1
		& $n_\o$, $n_\e \ge 1$ & \multicolumn{2}{c|}{------} \\
		\hline 
		A, Level 2
		& $n_\o$, $n_\e \ge 1$ & --- & \highlight{$\Chern\in\ZZ$ \ [S]}  \\ 
		\hline \hline
		\multirow{3}{*}{AI, Level 1} 
		& $n_\o = n_\e = 1$ &  \multicolumn{2}{c|}{------}
		\\ \cline{2-4}
		& $n_\o \ge 1$, $n_\e > 1$ & \multirow{2}{*}{---} & $\SW_1^{\Gamma \M}$, $\SW_1^{\M\X} \in \ZZ_2$ \\ 
		& or $n_\o > 1$, $n_\e \ge 1$ &  &  one less for every violation of $\Q_1^S$ \\ 
		\hline 
		\multirow{17}{*}{AI, Level 2} %
		& $n_\o = n_\e = 1$ & \multicolumn{2}{c|}{------}
		\\ \cline{2-4}
		& \multirow{3}{*}{$n_\o = 1$, $n_\e = 2$} 
		& if $\Q_1^{\lor}$ and $\Q_2^\e$ & $\Euler_2^\e \in 2\ZZ$ \\ \cline{3-4}
		&& if $\Q_2^\e$, but not $\Q_1^{\lor}$ & $\Euler_2^\e \in 2\NN$ \\ \cline{3-4}
		&& if not $\Q_2^{\rm e}$  & $\Euler_2^\e \in \ZZ_2$  
		\\ \cline{2-4}
		& $n_\o = 1$, $n_\e > 2$ &  \multicolumn{2}{c|}{------}
		\\ \cline{2-4}
		& \multirow{8}{*}{$n_\o = n_\e = 2$} 
		& if $\Q_1^{\lor}$ and $\Q_2^\e$ and $\Q_2^\o$ & $(\Euler_2^\o,\Euler_2^\e) \in \ZZ^2|_{\rm p}$  \\ \cline{3-4}
		&& if $\Q_2^\e$ and $\Q_2^\o$, but not $\Q_1^{\lor}$ & $(\Euler_2^\o,\Euler_2^\e) \in \NN \times \ZZ|_{\rm p}$ \\ \cline{3-4}
		&& if $\Q_1^{\lor}$ and $\Q_2^\o$, but not $\Q_2^\e$ & $\Euler_2^{\o} \in \ZZ$ \\ \cline{3-4}		
		&& if $\Q_1^{\lor}$ and $\Q_2^\e$, but not $\Q_2^\o$ & $\Euler_2^{\e} \in \ZZ$ \\ \cline{3-4}		
		&& if $\Q_2^\o$, but not $\Q_1^{\lor}$ and not $\Q_2^\e$ & $\Euler_2^{\o} \in \NN$ \\ \cline{3-4}		
		&& if $\Q_2^\e$, but not $\Q_1^{\lor}$ and not $\Q_2^\o$ & $\Euler_2^{\e} \in \NN$ \\ \cline{3-4}		
		&& if $\Q_2^{\land}$, but not $\Q_2^\e$ and not $\Q_2^\o$ & $\SW_2 \in \ZZ_2$ \\ \cline{3-4}		
		&& if not $\Q_2^{\land}$ & ---	
		\\ \cline{2-4}
		& \multirow{4}{*}{$n_\o = 2$, $n_\e > 2$} 
		& if $\Q_1^{\lor}$ and $\Q_2^\o$ & $\Euler_2^\o \in \ZZ$ \\ \cline{3-4}
		&& 	\highlight{if $\Q_2^\o$, but not $\Q_1^{\lor}$} & 	\highlight{$\Euler_2^\o \in \NN$ \ [F] }\\ \cline{3-4}	
		&& if $\Q_2^{\land}$, but not $\Q_2^\o$ & $\SW_2 \in \ZZ_2$ \\ \cline{3-4}	
		&& if not $\Q_2^{\land}$ &  --- 
		\\ \cline{2-4}
		& \multirow{2}{*}{$n_\o$, $n_\e > 2$} 
		& \highlight{if $\Q_2^{\land}$} & \highlight{$\SW_2 \in \ZZ_2$ [F]} \\ \cline{3-4} 
		&& if not $\Q_2^{\land}$ & --- 
		\\	\hline
	\end{tabular}
	\caption{Homotopic classification of spinless gapped band structures with $\C_4$ symmetry in two dimensions for various possible combinations of number of bands and level 0 invariants, the latter encoded in the $\Q$'s. In particular, $\Q_1^S$, $\Q_2^{\o/\e}$, and $\Q_2^S$ are defined by Eqs.\ (\ref{eq:condC4}), (\ref{eq:condC4b}), and (\ref{eq:condC4c}), respectively.  The condition $\Q_1^{\land}$ holds if $\Q_1^S \, \forall S$, $\Q_1^{\lor}$ holds if $\exists S$ where $\Q_1^S$ is true, and $\Q_2^{\land}$ holds if $\Q_2^S \, \forall S$. The set $\ZZ^2|_{\rm p}$ consists of all ordered integer pairs $(p,q)$ with $p = q \mod 2$; the set $\NN \times \ZZ_{\rm p}$ consists of all ordered integer pairs $(p,q)$ with $p \ge 0$ and $p = q \mod 2$. The invariants are only specified for $n_\o \leq n_\e$, since the classification is symmetric under $n_\o \leftrightarrow n_\e$. The fragile and stable topological phases are marked by \highlight{[F]} and \highlight{[S]}, respectively. 		
		\label{tab:clfn_C4_2d}}
\end{table*}

\begin{table}
	\centering 
	\setlength{\tabcolsep}{15pt}
	\begin{tabular}{|c|c|c|}
		\hline
		Allowed EBRs & Level 1 & Level 2 \\ \hline
		$p_\a$, $p_\b$ & $\ZZ_2$ & $\ZZ_2$ \\
		$p_\a \oplus p_\b$ & $\ZZ_2$ & 0 \\
		$s_\a$,	$s_\b$, $d_\a$, $d_\b$ & 0 & $\ZZ_2$ \\ 
		\hline
	\end{tabular}
	\caption{\label{tab:stable2} Representation-protected stable topology: Restrictions to certain EBRs (combinations of orbital types and Wyckoff positions) allow for a nontrivial stable representation-protected classification in the presence of time-reversal symmetry. The first row indicates that we can have either $p_\a$ or $p_\b$ \emph{but not both}, while $p_\a \oplus p_\b$ means that $p$-orbitals are allowed simultaneously at Wyckoff positions $\a$ and $\b$.
	}
\end{table}

\subsubsection*{Classification on 1- and 2-cells}
The parent invariants at level-1 and -2 are identical to that of $\C_2$-symmetric band structures described in Sec.~\ref{sec:C2}. The effect of boundary deformations is also quite similar, so that we only describe the difference from the $\C_2$-symmetric case. The condition of Eq.~(\ref{eq:condC2}) for the triviality of $\pi_1[\Xsp_S]$ now reads 
\begin{align}
	\label{eq:condC4}
	\Q_1^S:\ 
	\begin{cases}
		n^S_{\o s} n^S_{\e s} =\, n^S_{\o d} n^S_{\e d} = 0, & S = \G,\M, \\[5pt]
		n^S_{\o +} n^S_{\e +} =\, n^S_{\o -} n^S_{\e -} = 0, & S=\X.
	\end{cases} 
\end{align}
No condition is imposed on the level-0 invariants associated with the $p$-orbital, since the corresponding factor in $\Xsp_S$ is a complex Grassmannian. 

At level-1, the band structure is always trivial when $n_\o = n_\e = 1$, since the existence of a first Euler invariant requires that $\Q_1^S$ be satisfied $\forall S$, which is incompatible with Eq.~(\ref{eq:constraintC}). In the stable limit, there are up to two first Stiefel-Whitney invariants and every violation of $\Q_1^S$ effectively removes one of these. The level-1 compatibility condition, following from the triviality of the Wilson loop around the boundary of the 2-cell, reads  
\begin{equation}
	1 = \prod_{S} \det \rotn^{\o/\e}_{S} = (-1)^{n_{\o/\e,d}^\G + n_{\o/\e,d}^\M + n_{\o/\e,-}^\X},
	\label{eq:constraintC4anomalous}
\end{equation}
which is equivalent to Eq.\ (\ref{eq:constraintC}).

\begin{table*}
	\centering
	\begin{tabular}{|r|c|l|l|l|}
		\hline
		&& level 1 & level 2 & level 3 
		\\ \hline
		\multirow{4}{*}{A} & \multirow{2}{*}{$n_\o = n_\e = 1$}
		& \multirow{2}{*}{---} &
		$\Chern_z$, $\Chern^{\G\M}$, $\Chern^{\M\X} \in \ZZ$, & 
		if $\Chern_z = 0$: $\Hopf \in \ZZ$ \\ 
		&&& one less for every violation of $\Q_1^{Sz}$ & 
		if $\Chern_z \neq 0$: $\Hopf \in \ZZ_{2 \Chern_z}$ 
		\\ \cline{2-5}
		& \multirow{1}{*}{$n_\o$, $n_\e \ge 1$,} &
		\multirow{2}{*}{---}
		& {\color{blue}$\Chern_z \in \ZZ$ [S];} 
		$\Chern^{\G\M}$, $\Chern^{\M\X} \in \ZZ$, &
		\multirow{2}{*}{---} \\
		& \multirow{1}{*}{$n_\o + n_\e > 2$} &&
		\multirow{1}{*}{one less for every violation of $\Q_1^{Sz}$} & 
		\\ \hline\hline  
		\multirow{5}{*}{AI} & \multirow{2}{*}{$n_\o = n_\e = 1$}
		& two copies of invariants 
		& $\Chern^{\G\M}$, $\Chern^{\M\X} \in \ZZ$, & 
		\multirow{2}{*}{$\Hopf \in \ZZ$} \\
		&& from Tab.\ \ref{tab:clfn_C4_2d}
		& one less for every violation of $\Q_1^{Sz}$ & 
		\\ \cline{2-5}
		& \multirow{2}{*}{$n_\o$, $n_\e \ge 1$,} &
		\multirow{2}{*}{two copies of invariants}
		& {\color{blue} two copies of invariants from} &
		\multirow{3}{*}{---} \\
		& \multirow{2}{*}{$n_\o + n_\e > 2$} &
		\multirow{2}{*}{from Tab.\ \ref{tab:classification_C2_AI_2d}} & 
		{\color{blue} Tab.\ \ref{tab:clfn_C4_2d} [F];} $\Chern^{\G\M}$, $\Chern^{\M\X} \in \ZZ$, & \\
		&&& one less for every violation of $\Q_1^{Sz}$ & \\
		\hline
	\end{tabular}
	\caption{Homotopic classification of spinless gapped band structures with $\C_4$ in three dimensions. The condition $\Q_1^{Sz}$ is given in Eq.\ (\ref{eq:condC4Q2zz}). Nontrivial fragile and stable invariants are marked with ``[F]'' and ``[S]'', respectively.
		\label{tab:classification_C4_AI_3d}}
\end{table*}

\begin{figure}
	\includegraphics[width=0.95\columnwidth]{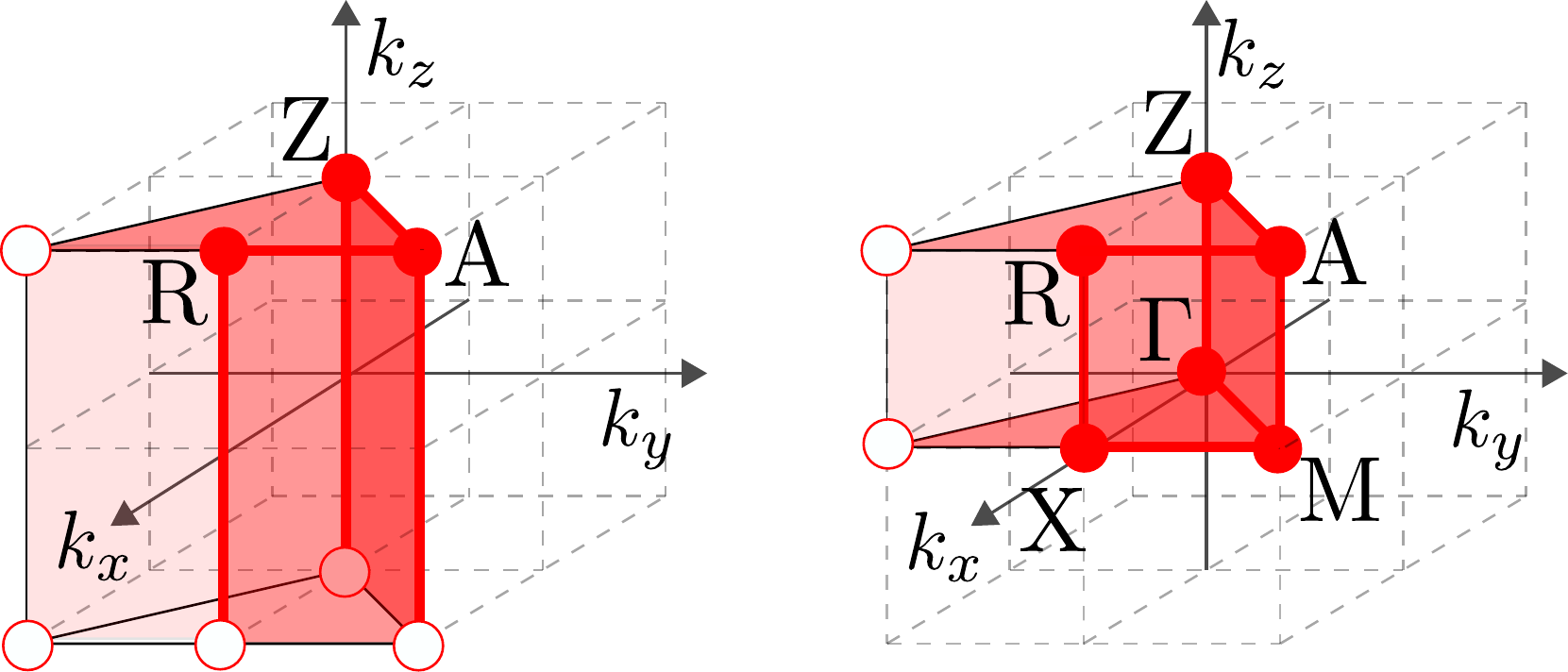}
	\caption{\label{fig:fundamental3d} Fundamental domains $\cwc$ for a three-dimensional band structure with fourfold rotation symmetry without (left) and with (right) time-reversal symmetry.}
\end{figure}

At level-2, the analogue of condition $\Q_2^{\o/\e}$ (Eq.~(\ref{eq:condC2b})) now reads 
\begin{align}
	\Q_2^{\o/\e}:\ 
	n^S_{\o/\e, s} n^S_{\o/\e, d} = n^X_{\o/\e, +} n^X_{\o/\e, -} = 0,
	\label{eq:condC4b} 
\end{align}
with $S \in \{\G,\M\}$, and $\Q_2^S$ (Eq.~(\ref{eq:condC2b})) becomes 
\begin{align}
	\Q_2^S:\ 
	\begin{cases}
		n^S_{\o s} n^S_{\e s} n^S_{\o d} n^S_{\e d} = 0, & S = \G,\M, \\[5pt]
		n^S_{\o +} n^S_{\e +} n^S_{\o -} n^S_{\e -} = 0, & S=\X,
	\end{cases} 
	\label{eq:condC4c}
\end{align}
and we again define $\Q_2^\land$ as $\Q_2^S \, \forall S$. Thus, $\Euler_2$ is $\NN$-valued if $\Q_1^S$ is violated $\forall S$, $\ZZ_2$-valued if $\Q_2^{\o/\e}$ are violated, and trivial if $\Q_2^S$ is violated for any $S$, while $\SW_2$ is is $\ZZ_2$-valued if $\Q_2^\land$ holds and trivial otherwise. 

\ \\ \noindent These results are summarized in Table~\ref{tab:clfn_C4_2d}. 
\subsubsection*{Summary}
To obtain the fragile and stable classification from the results listed in Table~\ref{tab:clfn_C4_2d}, we again need to check whether the various conditions for the existence of topologically nontrivial phases are compatible with the rules of the corresponding topology. For $n_{\o,\e} \gg 1$, there are no stable topological phases, since the required conditions $\Q_1^S$ or $\Q_2^\land$ are both incompatible with the rules of stable topology. However, imposing either of these conditions, we get representation-protected stable topological phases. Unlike the case of $\C_2$, these phases can have a nontrivial level-1 invariant, since $\Q_1^S$ for $S = \G,\M$ does not include the $p$-type orbitals, so that they are satisfied for arbitrary number of $p$-orbitals. In Table~\ref{tab:stable2}, we list all possible EBRs (\ie, combinations of Wyckoff positions and allowed orbital types) for which such a phase exists. 

\subsection{Classification in three dimensions}
\label{sec:C4_3d}
The 3d classification is again analogous to those for $\C_2$ symmetry discussed in Sec.~\ref{sec:C2_3d}. 

\subsubsection{Without time-reversal symmetry}
The fundamental domain is a triangular prism depicted in Fig.~\ref{fig:fundamental3d} (left). There are no level-1 invariants. At level-2, we get Chern numbers for each face of $\cwc$, where $C_z$ for the horizontal face is robust. For the vertical faces, boundary deformations at vertical line through a 0-cell $S$ can add arbitrary integers to all $\Chern^{SS'}$ if $\pi_1[\Xsp_S] \neq 0$, which happens if 
\begin{equation}
	\Q_1^{S,z}:\ n^S_{\o,\rho} n^S_{\e,\rho} = 0 \;\; \forall\rho
	\label{eq:condC4Q2z}
\end{equation}
is violated, where $\rho$ runs over the irreps of $G_S$, \ie, $\rho \in \{s,d,p_+, p_-\}$ for $S = \G,\M$ and $\rho \in \{+,-\}$ for $S = \X$. At level-3, we get the $\ZZ$-valued Hopf invariant if $n_\o = n_\e = 1$, which reduces to a $\ZZ_{2 \Chern_z}$-valued if $\Chern_z \neq 0$.

\subsubsection{With time-reversal symmetry}
The fundamental domain is a triangular prism based depicted in Fig.~\ref{fig:fundamental3d}(right). We get two independent copies of the 2d classification problem (Sec.~\ref{sec:C4_2d}) at levels 0, 1, and 2, as well as Chern numbers $\Chern^{SS'}$ for each of the vertical faces of $\cwc$. Boundary deformations can add arbitrary integers to $\Chern^{SS'}$ if 
\begin{equation}
	\Q_1^{S,z}:\ n^S_{\o,\rho} n^S_{\e,\rho} = 0 \;\; \forall\rho
	\label{eq:condC4Q2zz} 
\end{equation}
is violated, where $\rho$ now runs over the \emph{real} irreps of $G_S$, \ie, $\rho \in \{s,d,p \}$ for $S = \G,\M$ and $\rho \in \{+,-\}$ for $S = \X$. At level-3, we get the $\ZZ$-valued Hopf invariant. 

\ \\ \noindent These classification results are summarized in Tab.\ \ref{tab:classification_C4_AI_3d}.

\subsubsection{Anomalous boundary states}
Without time-reversal symmetry, a nonzero Chern number at fixed $k_z$ signals a weak Chern-insulator phase. 
With time-reversal symmetry, we get anomalous surface modes associated with the nontrivial representation-protected stable topology. A difference in the level-1 topological invariants between $k_z = 0$ and $\pi$ leads to a protected quadratic band touching at the $\sG$ and $\sM$ points in the surface Brillouin zone, while a difference in level-2 invariants is associated with the presence of a set of four surface Dirac cones related by $\C_4$ \cite{kobayashi2021}. 

From the boundary perspective, an odd number of Dirac cones in the fundamental domain is anomalous when we restrict to a single orbital type at a single Wyckoff position. This follow from the violation of the level-1 compatibility condition (Eq.~(\ref{eq:constraintC})), as in the case of the surface Dirac cones for $\C_2$. On the other hand, a single quadratic band touching at $\sG$ or $\sM$ comes with an {\em odd} number of levels crossing the Fermi energy between $\overline \G$ and $\overline M$. However, for band structures with only $p$-orbitals, the number of occupied bands at $\sG$ and $\sM$ must be even, rendering the quadratic band touching anomalous.

\newcommand\invminus{\phantom{-}}
\begin{table}[t]
	\centering
	\setlength\tabcolsep{5pt}
	\begin{tabular}{|c|ccccc|}
		\hline
		Irrep.    & $\{\id\}$   & $\{\rotn_4, \rotn_4^{-1}\}$   & $\{\rotn_2\}$    & $\{\mirror_x, \mirror_y\}$  & $\{\mirror_\d, \mirror_{\bar{d}} \}$    \\
		\hline     
		$A_1$   & 1   & $\invminus 1$  & $\invminus 1$  & $\invminus 1$  & $\invminus 1$  \\ 
		$A_2$   & 1   & $\invminus 1$  & $\invminus 1$  & $-1$  & $-1$  \\ 
		$A_3$   & 1   & $-1$  & $\invminus 1$  & $\invminus 1$  & $-1$  \\ 
		$A_4$   & 1   & $-1$  & $\invminus 1$  & $-1$  & $\invminus 1$  \\ 
		$B$   & 2   & $\invminus 0$  & $-2$  & $\invminus 0$  & $\invminus 0$  \\ 
		\hline
	\end{tabular}
	\caption{The character table of $\D_4$. The character corresponding to $\{\id\}$ is equal to the dimensionality of the representation.} 
	\label{tab:D4_chr}
\end{table}

\section{\texorpdfstring{$\D_4$}{D4} symmetry}
\label{sec:D4}
Dihedral groups $\D_n$ are the groups of symmetries of regular polygons, which includes rotation as well as reflections. The homotopic classification of $\D_4$-symmetric band structures is instructive for two reasons: 
(i) $\D_4$-symmetry in two dimensions allows for high-symmetry lines, in contrast to $\C_n$, which only has high-symmetry points, and 
(ii) The group $\D_4$ is nonabelian, so that not all of its irreps are one-dimensional. Explicitly, the group $\D_4$ can be defined as 
\begin{equation}
  \D_4 = \{ \id, \rotn_4, \rotn_2, \rotn_4^{-1}, \mirror_x, \mirror_y, \mirror_{d}, \mirror_{\bar{d}} \},
\end{equation}
where $\rotn_N$ denotes the $N$-fold rotation and $\mirror_{x,y,d,\bar{d}}$ denote reflection in the $x$-axis, $y$-axis, the lines $y=\pm x$, respectively. The non-commutativity of $\D_4$ can be seen, for instance, in $\rotn_4 \mirror_x = \mirror_\d$ while $\mirror_x \rotn_4 = \mirror_{\bar{d}}$. The irreps of $\D_4$ are listed in the its character table, which is shown in Table.~\ref{tab:D4_chr}. For the one-dimensional irreps, the characters are the eigenvalues of the corresponding symmetry operator. Thus, irreps $A_1$, $A_3$ and $A_4$ correspond to $s$, $d_{x^2-y^2}$, and $d_{xy}$ orbitals, while the irrep $A_2$ requires angular momentum $\ell = 4$. For the two-dimensional irrep $B$, we can choose the representation matrices as 
\begin{equation}
  \rotn_4 = -\i\sigma^2, \qquad 
  \mirror_x = \sigma^3,
  \label{eq:irrepB}
\end{equation}
which act on a doublet formed by the $p_x$ and $p_y$ orbitals. These symmetry operators are also real-valued and thus compatible with time-reversal symmetry.

\begin{figure}
	\includegraphics[width=0.85\columnwidth]{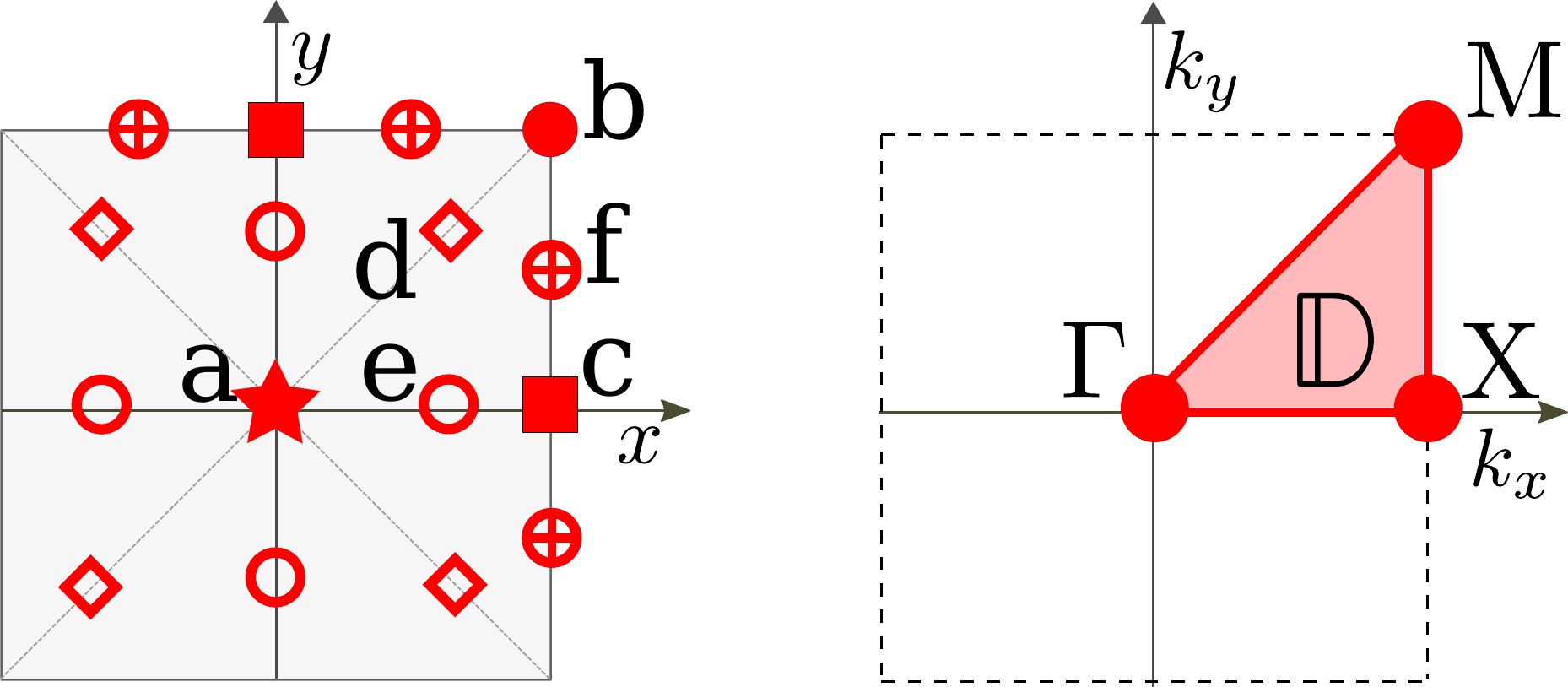}
	\caption{\label{fig:D4} 
		(Left) The Wyckoff positions for a square lattice with $\D_4$ symmetry. 
		(Right) The fundamental domains $\cwc$ for two-dimensional band structures with $\D_4$ symmetry. 
	}
\end{figure} 

\subsection{Wyckoff positions, basis orbitals and symmetries}
\label{sec:D4_symm}
A lattice with $\D_4$ symmetry has six special Wyckoff positions (labelled ``a''--``f'') as well as the generic Wyckoff position(``g''), as shown in Fig.~\ref{fig:D4}. The Wyckoff position ``a'' and ``b'' are invariant under the full symmetry group $\D_4$ and thus have a multiplicity of one. The Wyckoff position ``c'' with multiplicity two is invariant under $M_x$, $M_y$, and a $\C_2$ rotation, so that its little group is isomorphic to $\ZZ_2^2$. The Wyckoff positions ``d'', ``e'' and ``f'' with multiplicity four are invariant under various mirror symmetries, making their little groups isomorphic to $\ZZ_2$. 

The reciprocal space group $G_\Lambda$ is generated by fourfold rotation $g_4$, reflection in the $k_x$-axis $m_x$, and the reciprocal space translations $g_x$ and $g_y$. The Bloch Hamiltonian satisfies 
\begin{align}
  H(k_x, k_y) 
  &= \rotn_4^\dagger H(-k_y, k_x) \rotn_4^\pdg,  \nonumber \\
  &= \mirror_x^\dagger H(k_x, -k_y) \mirror_x^\pdg,
\end{align}
as well as Eq.~(\ref{eq:Hsymm_ktrans}) for reciprocal space translations. Since the unitary matrices $\rotn_4$ and $\mirror_x$ generate a representation of $\D_4$, they satisfy 
\begin{equation}
	\rotn_4^4 = \mirror_x^2 = \id, \quad
	\rotn_4 \mirror_x \rotn_4 = \mirror_x. 
\end{equation} 
The matrices $\rotn_4$ and $\mirror_x$ cannot be diagonalized simultaneously. Together with $\trans_x$ and $\trans_y$, they form a representation of $G_\Lambda$, so that 
\begin{align}
	\rotn_4 \trans_x =&\, \trans_y \rotn_4,  &
	\rotn_4 \trans_y =&\, \trans_x^\dg \rotn_4,
	\label{eq:CDrelationD4} \\
	\mirror_x \trans_x =&\, \trans_x \mirror_x,  &
	\mirror_x \trans_y =&\, \trans_y^\dagger \mirror_x. 
	\label{eq:MDrelationD4}
\end{align}
Explicit expressions for the representation matrices for various Wyckoff positions are derived in Appendix~\ref{app:symm_D4}.

In two dimensions, we take the fundamental domain $\cwc$ as the triangle $\G\X\M$, as shown in Fig.~\ref{fig:D4}. It consists of three 0-cells, three 1-cells and one 2-cell. The 0-cells $\G$ and $\M$ are invariant under $\D_4$, while $\X$ is invariant under its subgroup $\D_2 \cong \ZZ_2\times\ZZ_2$ consisting of the mirror symmetries $\mirror_{x,y}$ and their product $\mirror_x\mirror_y = \rotn_2$. Thus, $H_S$ commutes with two matrices $\rotn_S$ and $\mirror_S$, which, using Eqs.~\eqref{eq:CDrelationD4} and \eqref{eq:MDrelationD4}, are given by
\begin{align}
   \rotn_\G &= \rotn_4, 
  &\mirror_\G &= \mirror_x, \nonumber  \\
   \rotn_\M &= \trans_x^\dagger \rotn_4 = \rotn_4 \trans_y,
  &\mirror_\M &= \trans_y^\dagger \mirror_x = \mirror_x \trans_y, \nonumber \\
   \rotn_\X &= \trans_x^\dagger \rotn_4^2 = \rotn_4^2 \trans_x,
  &\mirror_\X &= \mirror_x.
\end{align}
The matrices $\rotn_\M$ and $\mirror_\M$ generate a representation of $\D_4$, since $\rotn_\M^4 = 1$, $\mirror_\M^2 = \trans_y^\dagger \mirror_x^2 \trans_y = \id$, and
\begin{equation}
  \rotn_\M \mirror_\M \rotn_\M
  = \rotn_4 \mirror_x \rotn_4 \trans_y
  = \mirror_x \trans_y
  = \mirror_\M.
\end{equation}
Similarly, $\rotn_\X$ and $\mirror_\X$ satisfy $\rotn_\X^2 = \mirror_\X^2 = 1$. Thus, the eigenstates of $H_S$ at $S = \G,\M$ can be labeled by an irrep of $\D_4$, while those of $H_\X$ can be labeled by $(\pm,\pm)$, the eigenvalues of $\rotn_\X$ and $\mirror_\X$, respectively. 

The 1-cells $SS'$ are invariant under a reflection, so that $H_{SS'}$ satisfies 
\begin{equation}
  H(\vk) = \mirror_{SS'}^\dagger H(\vk) \mirror_{SS'},\ \ 
  \vk \in \cwc_\alpha
\end{equation}
with 
\begin{align}
	\mirror_{\G\X} = \mirror_x, \quad 
	\mirror_{\G\M} = \mirror_d, \quad 
	\mirror_{\M\X} = \trans_x^\dagger \mirror_y. 
	\label{eq:D4_M_SS'}
\end{align}
The bands can thus be labeled by a mirror parity at each 1-cell. There are no additional constraints on the 2-cell.

\subsection{Homotopic classification in two dimensions}
\label{sec:D4_2d}
The homotopic classification for $\D_4$-symmetric band structures exhibit two features that are distinct from the previously discussed cases: 
\begin{itemize}
	\item The 1-cells have nontrivial little groups, leading to additional level-0 constraints from the gluing of the 0-cells at the ends of the 1-cells. 
	
	\item All 1-cells occur only once in the fundamental domain, so that we do not get the symmetry-based cancellations of boundary deformations, in contrast to the symmetries discussed previously. 
\end{itemize}

\begin{figure}
	\includegraphics[width=0.75\columnwidth]{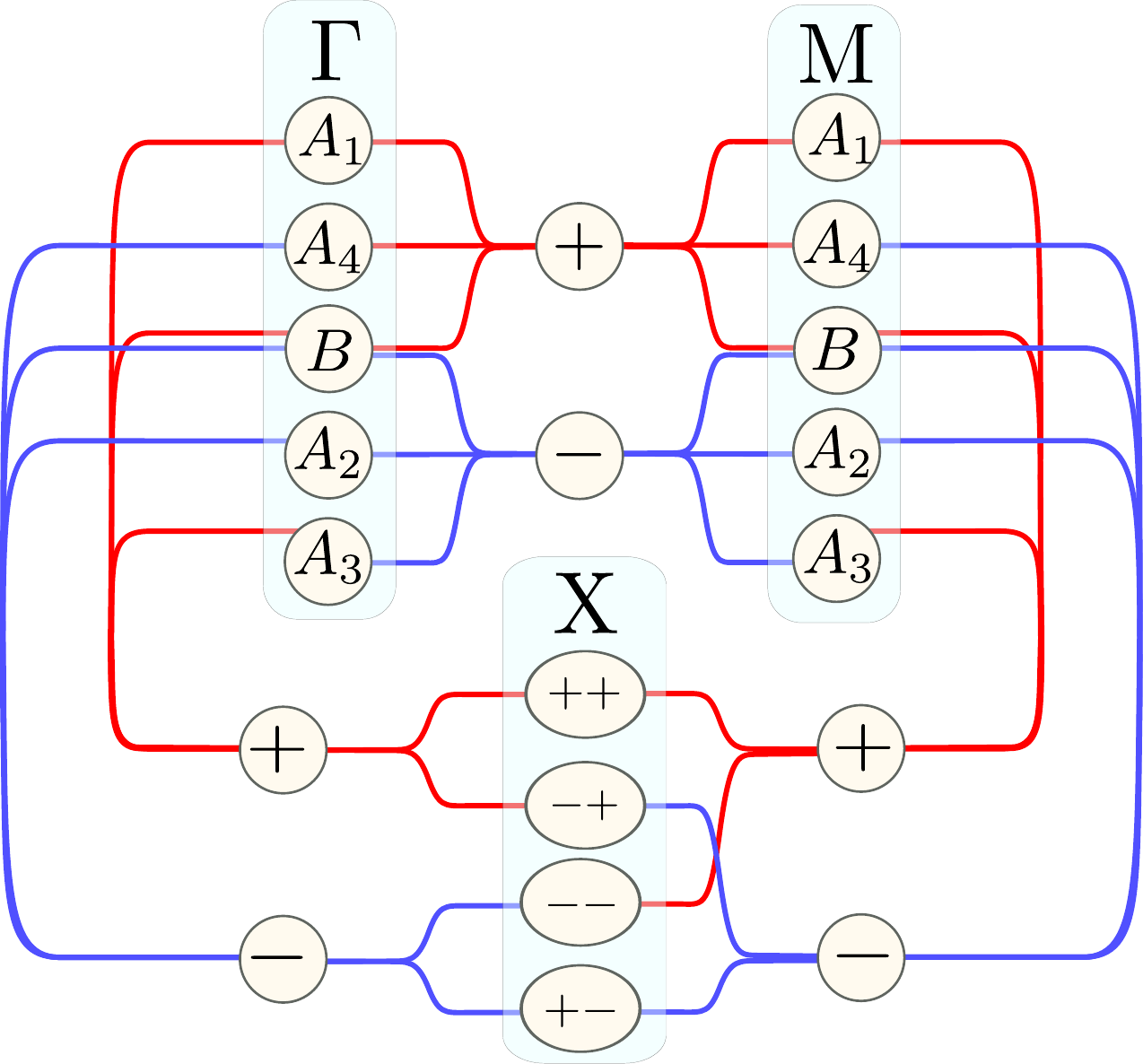}
	\caption{\label{fig:D4_bandreps} Splitting rules for the band representations of $\D_4$-symmetric lattices at high-symmetry points/lines. The irreps $A_i$ for $i = 1,2,3,4$ at the $\G$ and $\M$ point are one-dimensional, while the irrep $B$ is two-dimensional. }
\end{figure}

\subsubsection{Without time-reversal symmetry}
\label{sec:D4_2d_A}
The level-0 invariants $n_{\o/\e, \rho}^S$ count the number of occupied/empty copies of each irrep $\rho$ of the little group $G_S$ of the 0-cells $S$. Explicitly, for $S\in\{\G,\M\}$, $n_{\o/\e, A_j}^S$ is the number of orbitals transforming under irrep $A_j$, while $n^S_{\o/\e,B}$ is the number of \emph{doublets} transforming under the irrep $B$. Similarly, numbers $n_{\o/\e, a}^{SS'}$ of occupied/empty copies of the irreps can also be defined for the 1-cells $SS'$. The level-0 compatibility relations (\ref{eq:lev0_constr}) then read
\begin{align}
  n_{\o/\e} =&\, \sum_{j=1}^{4} n^{S}_{\o/\e,A_j} + 2 n^S_{\o/\e,B} \nonumber \\
  =&\, n^{\X}_{\o/\e,++} + n^{\X}_{\o/\e,+-} + n^{\X}_{\o/\e,-+} + n^{\X}_{\o/\e,--} \nonumber \\[5pt]
  =&\, n^{SS'}_{\o/\e,+} + n^{SS'}_{\o/\e,-},
\end{align}
where $S = \G$, $\M$. Additional constraints on the level-0 invariants arise from the \emph{splitting rules}, which describe how the two parities at the 1-cell $SS'$ connect to various irreps of $G_S$ as one approaches the end point. For the 1d irreps, the splitting rules can be read off directly from the character table of $\D_4$, while the 2d irrep $B$ connects to one odd and one even state on the 1-cells (see Fig.~\ref{fig:D4_bandreps}). The resulting constraints are identical at the 0-cells $S = \G,\M$, and read 
\begin{align}  
	n_{\o/\e,+}^{\G\M} &= n_{\o/\e, A_1}^S +  n_{\o/\e, A_4}^S +  n_{\o/\e, B}^S,  \nonumber \\ 
	n_{\o/\e,-}^{\G\M} &= n_{\o/\e, A_2}^S +  n_{\o/\e, A_3}^S +  n_{\o/\e, B}^S,  \nonumber \\ 	
	n_{\o/\e,+}^{S\X} &= n_{\o/\e, A_1}^S +  n_{\o/\e, A_3}^S +  n_{\o/\e, B}^S,   \nonumber \\ 
	n_{\o/\e,-}^{S\X} &= n_{\o/\e, A_2}^S +  n_{\o/\e, A_4}^S +  n_{\o/\e, B}^S	
	\label{eq:D4constr_GM}
\end{align}
For the 0-cell $\X$, we get 
\begin{align}
	n_{\o/\e,+}^{\G\X} &= n_{\o/\e,++}^\X + n_{\o/\e,-+}^\X,  \nonumber \\  
	n_{\o/\e,-}^{\G\X} &= n_{\o/\e,+-}^\X + n_{\o/\e,--}^\X,  \nonumber \\ 
	n_{\o/\e,+}^{\M\X} &= n_{\o/\e,++}^\X + n_{\o/\e,--}^\X,  \nonumber \\ 
	n_{\o/\e,-}^{\M\X} &= n_{\o/\e,+-}^\X + n_{\o/\e,-+}^\X.
	\label{eq:D4constr_X}
\end{align}
For fixed level-0 invariants, the target spaces for the $p$-cell $\alpha$ with $p = 0,1$ are given by 
\begin{equation}
	\Xsp_\alpha = \prod_\rho \Gr_\CC(n_{\o,\rho}^\alpha, n_{\o,\rho}^\alpha),
\end{equation}
where $\rho$ runs over the irreps of the little group $G_\alpha$. For the 2-cell, we have $\Xsp = \Gr_\CC(n_\o, n_\e)$.

At level-1, there are no topological invariants, while at level-2, the parent invariant is the Chern number. The parent classification set for boundary deformations $K_2^\partial$ consists of a copy of $\pi_1[\Xsp_S]$ for each 0-cell $S$ and of $\pi_2[\Xsp_{SS'}]$ for each 1-cell $SS'$. The former are always trivial, while the latter are trivial only if $\Gr_\CC(n_{\o,\pm}^{SS'}, n_{\e,\pm}^{SS'})$ degenerates to a point, \ie, if 
\begin{equation}
	\Q_{1,a}^{SS'}:\ n_{\o,a}^{SS'} n_{\e,a}^{SS'} = 0 
	\label{eq:condD4_SS'}
\end{equation}
for $a = \pm$. If this condition is violated for either parity at any 1-cell, then boundary deformations can add an arbitrary integer to the level-2 invariant. Thus, there is a $\ZZ$-valued level-2 invariant if $\Q_{1,a}^{SS'}$ is obeyed for both parities at each 1-cell \footnote{
	A similar phase for $\C_6$-symmetry has previously been studied by Nelson {\em et al.} \cite{nelson2021multicellularity}.	
}. 
This condition also indirectly constraints the level-0 invariants because of the splitting rules. For instance, the presence of irrep $B$ at $S = \G, \M$ violates $\Q_{1,\pm}^{SS'}$, since $B$ splits into an even and an odd orbitals on the adjoining 1-cell.

\begin{table*}
	\centering
	\setlength{\tabcolsep}{15pt}
	\begin{tabular}{|r|c|l|l|}
		\hline 
		& Bands & Additional condition(s) & Topological invariant \\ 
		\hline
		A, Level 1
		& $n_\o$, $n_\e \ge 1$ & \multicolumn{2}{c|}{See main text,  Sec.~\ref{sec:D4_2d_A}} \\
		\hline 
		\multirow{2}{*}{A, Level 2} 
		& \multirow{2}{*}{$n_\o$, $n_\e \ge 1$}
		& if $\Q_1^{\land}$	 & $\Chern \in \ZZ$
		\\ \cline{3-4}
		&& if not $\Q_1^{\land}$ & --- \\
		\hline \hline
		AI, Level 1
		& $n_\o$, $n_\e \ge 1$ & \multicolumn{2}{c|}{See main text, Sec.~\ref{sec:D4_2d_AI}} \\
		\hline 
		\multirow{17}{*}{AI, Level 2} 
		& $n_\o = n_\e = 1$ & --- & --- 	\\ \cline{2-4}
		& \multirow{3}{*}{$n_\o = 1$, $n_\e = 2$} 
		&  if $\Q_1^{\lor}$ and $\Q_2^{\e}$   & $\Euler_2^\e \in 2\ZZ$ 	\\ \cline{3-4}
		&& if $\Q_2^{\e}$, but not $\Q_1^{\lor}$	&  $\Euler_2^\e \in 2\NN$ 		\\ \cline{3-4}
		&& if not $\Q_2^{\e}$ &  --- 
		\\ \cline{2-4}
		& $n_\o = 1$, $n_\e > 2$ & --- & --- 
		\\ \cline{2-4}
		& \multirow{8}{*}{$n_\o = n_\e = 2$} 
		& if $\Q_1^{\lor}$ and $\Q_2^\e$ and $\Q_2^\o$ & $(\Euler_2^\o,\Euler_2^\e) \in \ZZ^2|_{\rm p}$  \\ \cline{3-4}
		&& if $\Q_2^\e$ and $\Q_2^\o$, but not $\Q_1^{\lor}$ & $(\Euler_2^\o,\Euler_2^\e) \in \NN \times \ZZ|_{\rm p}$ \\ \cline{3-4}
		&& if $\Q_1^{\lor}$ and $\Q_2^\o$, but not $\Q_2^\e$ & $\Euler_2^{\o} \in \ZZ$ \\ \cline{3-4}		
		&& if $\Q_1^{\lor}$ and $\Q_2^\e$, but not $\Q_2^\o$ & $\Euler_2^{\e} \in \ZZ$ \\ \cline{3-4}		
		&& if $\Q_2^\o$, but not $\Q_1^{\lor}$ and not $\Q_2^\e$ & $\Euler_2^{\o} \in \NN$ \\ \cline{3-4}		
		&& if $\Q_2^\e$, but not $\Q_1^{\lor}$ and not $\Q_2^\o$ & $\Euler_2^{\e} \in \NN$ \\ \cline{3-4}		
		&& if $\Q_2^{\land}$, but not $\Q_2^\e$ and not $\Q_2^\o$ & $\SW_2 \in \ZZ_2$ \\ \cline{3-4}		
		&& if not $\Q_2^{\land}$ & ---	
		\\ \cline{2-4}
		& \multirow{4}{*}{$n_\o = 2$, $n_\e > 2$} 
		& if $\Q_1^{\lor}$ and $\Q_2^\o$ & $\Euler_2^\o \in \ZZ$ \\ \cline{3-4}
		&& 	{\color{blue} if $\Q_2^\o$, but not $\Q_1^{\lor}$} & 	{\color{blue} $\Euler_2^\o \in \NN$ \ [F] }\\ \cline{3-4}	
		&& if $\Q_2^{\land}$, but not $\Q_2^\o$ & $\SW_2 \in \ZZ_2$ \\ \cline{3-4}	
		&& if not $\Q_2^{\land}$ &  --- 
		\\ \cline{2-4}
		& \multirow{2}{*}{$n_\o$, $n_\e > 2$} 
		& if $\Q_2^{\land}$ & $\SW_2 \in \ZZ_2$ \\ \cline{3-4} 
		&& if not $\Q_2^{\land}$ & --- 
		\\ \hline 
	\end{tabular}
	\caption{
	Homotopic classification of spinless gapped band structures with $\D_4$ symmetry in two dimensions for various possible combinations of number of bands and level 0 invariants, the latter encoded in the $\Q$'s. In particular, $\Q^{SS'}_{1,a}$, $\Q^S_{1,\rho}$, $\Q_2^{\o/\e}$, $\Q_2^{\land}$ are defined in Eqs.\ (\ref{eq:condD4_SS'}), (\ref{eq:condD4_S}), (\ref{eq:condD4_integer}), and (\ref{eq:condD4_parity}), respectively. The condition $\Q_1^{\land}$ holds if $\Q^{SS'}_{1,a} \, \forall\, SS',a$ and $\Q_1^{\lor}$ holds if $\exists S$ where $Q^S_{1,\rho}$ is true $\forall \, \rho \neq B$. The set $\ZZ^2|_{\rm p}$ consists of all ordered integer pairs $(p,q)$ with $p = q \mod 2$; the set $\NN \times \ZZ_{\rm p}$ consists of all ordered integer pairs $(p,q)$ with $p \ge 0$ and $p = q \mod 2$.  The invariants are only specified for $n_\o \leq n_\e$, since the classification is symmetric under $n_\o \leftrightarrow n_\e$. The fragile topological phases are marked by \highlight{[F]}. 
	\label{tab:clfn_D4_2d}}
\end{table*}

\subsubsection{With time-reversal symmetry}
\label{sec:D4_2d_AI}
The level-0 invariants and the corresponding compatibility conditions are identical to the time-reversal-broken case. The target spaces are now real Grassmannians:
\begin{equation}
	\Xsp_\alpha = \prod_\rho \Gr_\RR(n_{\o,\rho}^\alpha, n_{\e,\rho}^\alpha)
\end{equation}
for the $p$-cell $\alpha$ with $p = 0,1$ and $\Xsp = \Gr_\RR(n_\o, n_\e)$ for the 2-cell. In contrast to $\C_4$, the presence of a two-dimensional irrep $B$ at the $\G$ and $\M$ points does not lead to a complex Grassmannian. This is because if $H_S$ transforms under the two-dimensional irrep $B$, it must commute with the representation matrices for all elements of $\D_4$, \ie, with all matrices of the form $\id_{n_B^S} \otimes \sigma^j$ for $j = 1,2,3$. Such matrices can always be written as $H_S = h_S \otimes \sigma^0$, where $h_S$ is a real symmetric matrix.

\subsubsection*{Classification on 1-cells}
There are up to six parent level-1 invariants, since there are two parity sectors for each 1-cell. The type of the level-1 invariants depends on the number $n^{SS'}_{\e,s}$ of occupied and empty bands per parity sector $a$ at each 1-cell (in contrast to $\C_2$ and $\C_4$ symmetry, where the type of the level-1 invariants is completely determined by the total numbers of occupied/empty bands). The numbers  $n^{SS'}_{\o/\e,a}$ are determined by the level-0 invariants $n^{S}_{\o/\e,\rho}$ via the splitting rules. The level-1 invariant is a first Euler invariant $\Euler_{1,a}^{SS'} = -\Euler_{1,a}^{S'S}$ if $n^{SS'}_{\o,a} = n^{SS'}_{\e,a} = 1$, a first Stiefel-Whitney invariant $\SW_{1,a}^{SS'}$ if $n^{SS'}_{\o,a} > 1$ and $n^{SS'}_{\e,a} \ge  1$ or $n^{SS'}_{\o,a} \ge 1$ and $n^{SS'}_{\e,a} >  1$, whereas no nontrivial level-1 invariant exists if $n^{SS'}_{\o,a} = 0$ or $n^{SS'}_{\e,a} = 0$.

The classification set of boundary deformations $K_1^\partial$ consists of a copy of $\pi_1[\Xsp_S]$ for each 0-cell $S$. This means that there are no topologically nontrivial boundary deformations associated with the irrep $\rho$ at $S$ if 
\begin{align}
	\! \Q_{1,\rho}^S:\ n^S_{\o,\rho} n^S_{\e,\rho}  = 0.
	\label{eq:condD4_S}
\end{align}
Boundary deformations at $S$ with irrep $\rho$ only affect level-1 invariants on a 1-cell $SS'$ in parity sector $a = \pm$ if  $\rho$ connects to the parity $a$ under the splitting rules (see Fig.~\ref{fig:D4_bandreps}). For instance, at the $\G$-point, the boundary deformation with $\rho = A_4$ only affects two level-1 invariants, \viz, the positive parity sector on $\G\M$ and the negative parity sector on $\G\X$. Boundary deformations with $\rho = B$, however, affect both parity sectors on $\G\M$ and $\G\X$, \ie, four level-1 invariants in total. Provided the boundary deformations act independently on the parent level-1 invariants, each violation of $\Q_{1,\rho}^S$ effectively removes one level-1 invariant. (An example in which boundary deformations do not act independently is discussed below.)

The level-1 compatibility relation follows from demanding that the level-1 invariant corresponding to the full target space $\Xsp$ be trivial when computed around $\partial\cwc$. Thus, for $n_{\o} = n_{\e} = 1$, the compatibility relation reads
\begin{equation}
	\sum_{a = \pm} \Big(\Euler_{1,a}^{\G\M} + \Euler_{1,a}^{\M \X} + \Euler_{1,a}^{\X \G} \Big) = 0,
	\label{eq:condD4-c1}
\end{equation}
while for $n_{\o} > 1$ or $n_{\e} > 1$, we get 
\begin{equation}
	\sum_{a = \pm} \Big(\SW_{1,a}^{\G\M} + \SW_{1,a}^{\M \X} + \SW_{1,a}^{\X \G} \Big) = 0 \mod2. 
	\label{eq:condD4-c2}
\end{equation}
In both cases, $\Euler_{1,a}^{SS'} = \SW_{1,a}^{SS'} = 0$ if $n^{SS'}_{\o,a} n^{SS'}_{\e,a} = 0$. In Eq.\ (\ref{eq:condD4-c2}), one has $\SW_{1,a}^{SS'} = \Euler_{1,a}^{SS'} \mod 2$ if $n^{SS'}_{\o,a} = n^{SS'}_{\e,a} = 1$. The integer constraint of Eq.~(\ref{eq:condD4-c1}) reduces the number of independent integer level-1 invariants by one. On the other hand, the parity constraint of Eq.~ (\ref{eq:condD4-c1}) reduces the number of independent $\ZZ_2$ level-1 invariants by one, but if an integer invariant exists, then the compatibility relation only restricts its parity.

\subsubsection*{Classification on the 2-cell}
At level-2, the parent invariants are one or two $\ZZ$-valued second Euler invariants or a $\ZZ_2$-valued second Stiefel-Whitney invariant, depending on $n_{\o,\e}$ (see Sec.\ \ref{sec:gen_Gr}). The parent classification set for boundary deformations $K_2^\partial$ contains a copy of $\pi_2[\Xsp_{SS'}]$ for each 1-cell and of $\pi_1[\Xsp_{S}]$ for each 0-cell. We now discuss their effect on $\Euler_2^{\o/\e}$ and $\SW_2$ separately. 

\para{Second Euler invariant}
The existence of the parent invariant $\Euler_2^{\o/\e} \in\ZZ$ requires $n_{\o/\e} = 2$, whereby $\Euler_2^{\o/\e} \in 2\ZZ$ if $n_{\e/\o} = 1$. Boundary deformations at an 1-cell $SS'$ and parity sector $a$ can add an arbitrary integer to $\Euler_2^{\o/\e}$ if $n_{\o/\e,a}^{SS'} = 2$ and $n_{\e/\o,a}^{SS'} \ge 2$, whereas they can add an arbitrary even integer if $n_{\o/\e,a}^{SS'} = 2$ and $n_{\e/\o,a}^{SS'} = 1$. Hence, $\Euler_2^{\o/\e}$ is robust to deformations at 1-cells if the constraint
\begin{equation}
  \Q_2^{\o/\e}:\, \Big[ n_{\o/\e,a}^{SS'} \le 1\, \lor\, n_{\e/\o,a}^{SS'} = 0 \Big]\,
  \forall\, SS',\, a,
  \label{eq:condD4_integer}
\end{equation}
holds, whereas its parity is robust to deformations at 1-cells if the condition
\begin{equation}
  \Q_2^{\land}:\, \Big[ n_{\o/\e,a}^{SS'} \le 1\, \lor\, n_{\e/\o,a}^{SS'} \le 1 \Big]\,
  \forall\, SS',\, a,
  \label{eq:condD4_parity}
\end{equation}
is satisfied. 
Nontrivial boundary deformations at a 0-cell $S$ with irrep $\rho \neq B$ may flip the orientation of the occupied and empty bases, thereby changing the signs of $\Euler_2^{\o/\e}$. The irrep $B$ at the $\G$ or $\M$ point must be excluded, since it cannot change the orientation of a subspace owing to the twofold degeneracy. Such boundary deformations do not exist if $\Q_{1,\rho}^S$ (Eq.~(\ref{eq:condD4_S})) holds $\forall\rho \neq B$ at any 0-cell. 

\ \\ \para{Second Stiefel-Whitney invariant} 
The $\ZZ_2$ parent invariant $\SW_2$ is trivialized by boundary deformations if and only if there exists a 1-cell $SS'$ and parity $a$ such that $n_{\o/\e,a}^{SS'} \ge 2$ and $n_{\e/\o,a}^{SS'} \ge 2$.  Such boundary deformations do not exist if $\Q_2^{\land}$ of Eq.\ (\ref{eq:condD4_parity}) is satisfied.

\ \\ \noindent These classification results are summarized in Tab.\ \ref{tab:clfn_D4_2d}.

\subsubsection*{Summary}
In the absence of time-reversal symmetry, there are no stable or fragile topological band structures, since $\Q_{1,a}^{SS'}$ (Eq.~(\ref{eq:condD4_SS'})) is incompatible with the rules of fragile as well as stable topology. These conditions further preclude the existence of stable representation-protected topology. 

in the presence of time-reversal symmetry, there are no stable topological band structures, since rules of stable topology violate $\Q_{1,\rho}^S$,  $\Q_2^{\o/\e}$, and $\Q_2^{\land}$. There is a fragile topological phase with a level-2 invariant $\Euler_2^{\o}\in\ZZ$ if $n_{\o} = 2$, if the occupied orbitals transform under the irrep $B$ at the $\G$ and $\M$ points.

\subsubsection{Representation-protected stable topology}	
\label{sec:D4_2d_rep}
A restriction to the irrep $B$ at $\G$ and $\M$ implies that $n^S_{\o/\e,B} = n^{SS'}_{\o/\e, \pm} = \frac12 n_{\o/\e}$ for $S = \G,\M$ and for all 1-cells $SS'$. Thus, in the stable limit, each 1-cell has two $\ZZ_2$ invariants $\SW_{1,\pm}^{SS'}$, yielding a total of six invariants. The effect of boundary deformations depends on the Wyckoff positions occupied, as we now discuss. 

If only the special Wyckoff position ``a'' is allowed, then $n^\X_{\o/\e, ++} = n^\X_{\o/\e, +-} = \frac12 n_{\o/\e}$, whereas $n^\X_{\o/\e, -+} = n^\X_{\o/\e, --} = 0$ (see Table~\ref{tab:D4_ebr}). Thus, only two independent boundary deformations exist at $\X$. These simultaneously add integers to $\SW_{1,a}^{\G\X}$ and $\SW_{1,a}^{\M\X}$, $a = \pm$ and thus do not affect the differences $\SW_{1,a}^{\G\X\M} \equiv \SW_{1,a}^{\G\X} - \SW_{1,a}^{\X\M}$.  Boundary deformations at $\G$ and $\M$ have identical effect on the remaining four $\ZZ_2$ invariants $\SW_{1,\pm}^{\G\M}$, $\SW_{1,\pm}^{\G\X\M}$, so that together they can remove one invariant. Imposing the compatibility condition, we get two independent $\ZZ_2$-valued level-1 invariants, which can be written as 
\begin{align}
	\SW^{\G\M}_1 &\equiv \SW_{1+}^{\G\M} - \SW_{1-}^{\G\M}, \nonumber \\ 
	\SW^{+}_1 &\equiv \SW_{1+}^{\G\M} + \SW_{1+}^{\M\X} + \SW_{1+}^{\X\G}. 
	\label{eq:SW_rpti1}
\end{align}
On the other hand, if only Wyckoff position ``b'' is allowed, then $n^\X_{\o/\e, -+} = n^\X_{\o/\e, --} = \frac12 n_{\o/\e}$, whereas $n^\X_{\o/\e, ++} = n^\X_{\o/\e, +-} = 0$ (see Table~\ref{tab:D4_ebr}), so that the parity labels are switched between $\G\X$ and $\X\M$. This is analogous to the case discussed above, with the second invariant redefined as 
\begin{equation}	
	\SW^{+}_1 \equiv \SW_{1+}^{\G\M} + \SW_{1+}^{\M\X} + \SW_{1-}^{\X\G}. 
\end{equation}
If both Wyckoff positions ``a'' and ``b'' are allowed, then there are three independent boundary deformation at $\X$ \footnote{
	Although there are four possible boundary deformations at $\X$ only three are independent, since the effect of simultaneous deformations at irreps $++$ and $--$ is identical to that of deformations at $+-$ and $-+$.
}, 
which remove the invariant $\SW_1^+$, leaving us with a single $\ZZ_2$-valued invariant $\SW^{\G\M}_1$. In Appendix.~\ref{app:D4-model}, we construct a minimal lattice model with nontrivial invariants $\SW^{\G\M}_1$ and $\SW^{+}_1$. 

There are no stable representation-protected level-2 invariants, since $\Q_2^{\o/\e}$ and $\Q_2^{\land}$ are incompatible with the rules of stable representation-protected topology.

\subsection{Homotopic classification in three dimensions}
\label{sec:D4_3d}

\subsubsection{Without time-reversal symmetry} 

Without time-reversal symmetry, the 3d fundamental domain is a triangular prism, as shown in Fig.~\ref{fig:fundamentalD4_3d} (left). There are no level-1 invariants, while at level-2, we get a Chern number $\Chern_z$ for the horizontal face as well as two Chern numbers $\Chern_{\pm}^{SS'}$ for each vertical face. The former is robust only when the Hamiltonian is completely fixed at $\partial\cwc_2$, \ie, if $\Q_{1,\pm}^{SS'}$ (Eq.~(\ref{eq:condD4_SS'})) is satisfied for all 1-cells. For the latter, topologically nontrivial deformations along the vertical 1-cell passing through $S$ can add arbitrary integers to $\Chern^{SS'}_\pm$ on the adjacent 2-cells following the rules identical to those for the first Euler invariant for a 2d $\D_4$-symmetric band structure with time-reversal symmetry. The Chern numbers on vertical faces further satisfy the level-2 compatibility condition 
\begin{equation}
	\sum_{a=\pm} \Big( \Chern_a^{\G\M} + \Chern_a^{\M\X} + \Chern_a^{\X\G} \Big) = 0.
\end{equation}
At level-3 and $n_\o = n_\e = 1$, we get a $\ZZ$-valued invariant if the Chern number $\Chern_z$ for a fixed-$k_z$ cross section of $\cwc$ vanishes, and a $\ZZ_{2 \Chern_z}$-valued invariant otherwise.

\subsubsection{With time-reversal symmetry}

\begin{figure}
	\includegraphics[width=0.9\columnwidth]{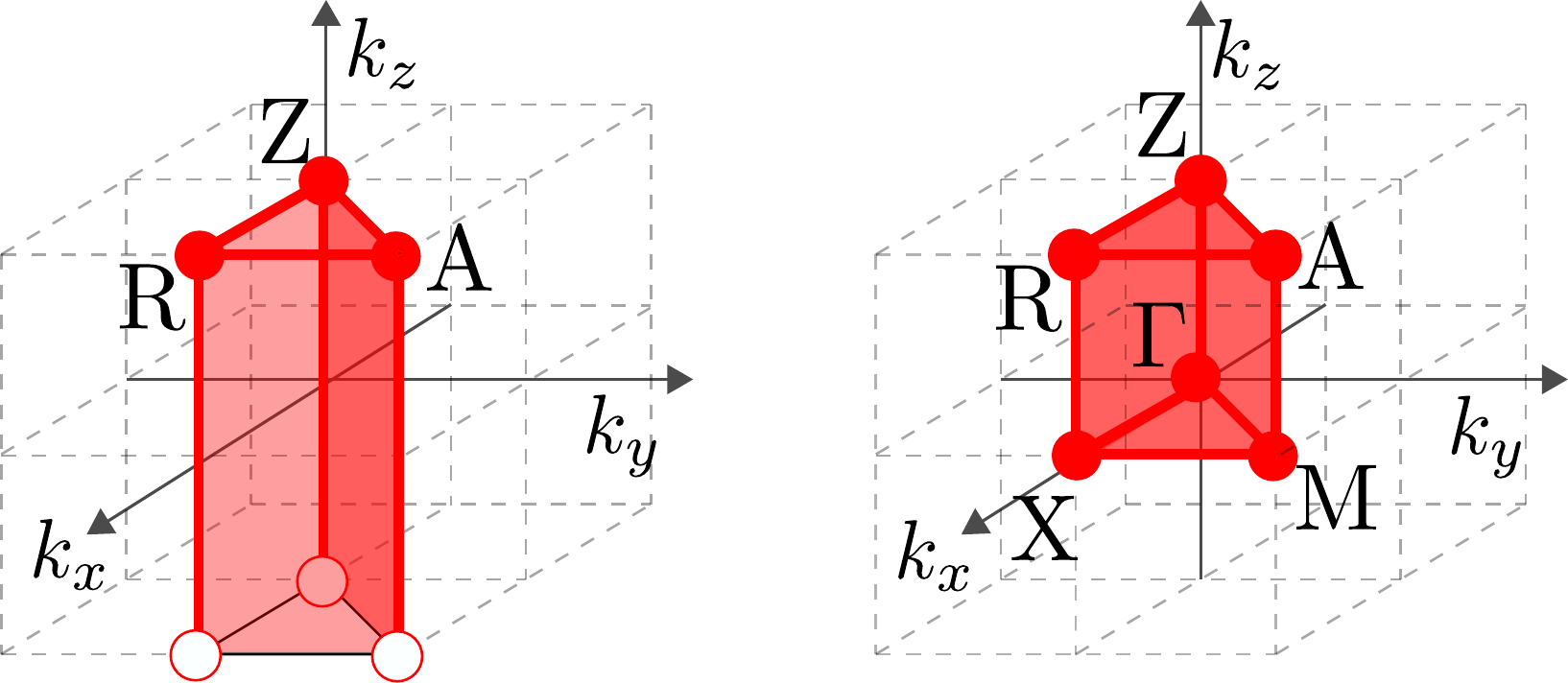}
	\caption{\label{fig:fundamentalD4_3d} Fundamental domains $\cwc$ for a three-dimensional band structure with $\D_4$ symmetry without (left) and with (right) time-reversal symmetry.}
\end{figure}

In the presence of time-reversal symmetry, the 3d fundamental domain is a triangular prism, as shown in Fig.~\ref{fig:fundamentalD4_3d} (right). At levels 0, 1, and 2, we get two copies of the 2d classification problem discussed in Sec.~\ref{sec:D4_2d}. At level-2, we additionally get Chern numbers $C_\pm^{SS'}$ for the vertical 2-cells. The conditions for their robustness and the level-2 compatibility condition are identical to the time-reversal broken case. At level-3, we get the $\ZZ$-valued Hopf invariant if $n_\o = n_\e = 1$. 

\subsubsection{Representation-protected stable topology}
A restriction to the two dimensional irrep $B$ leads to representation-protected stable topology with nontrivial level-2 invariants. In the absence of time-reversal symmetry, the parent invariants are the six Chern numbers $\Chern_{\pm}^{SS'} \in \ZZ$ computed on the vertical faces. In the presence of time-reversal symmetry, $\Chern_{\pm}^{SS'} \in \ZZ$ can be combined with the difference between the $\ZZ_2$-valued invariants $\SW_{1,\pm}^{SS'}$ computed at $k_z = 0,\pi$ to yield a $\ZZ$-valued Chern number computed for $k_z \in (-\pi,\pi]$. (The situation is analogous to the case of $\C_2 \T$-symmetry, see Sec.~\ref{sec:C2T3d}.) Time reversal symmetry thus plays no role in this case. In the following, we use $\C^{SS'}_{\pm}$ to refer to the Chern number computed for $k_z \in (-\pi,\pi]$ both without and with time-reversal symmetry.
 
The effect of boundary deformations depends on whether one or both Wyckoff positions are allowed, as in the case of the 2d representation protected phase (see Sec.~\ref{sec:D4_2d_rep}). Thus, if a single Wyckoff position is allowed, then we are left with two independent $\ZZ$-valued invariants.  For the Wyckoff position ``a'', these can be written as (\textit{cf.} Eq.~(\ref{eq:SW_rpti1}))
\begin{align}
	\Chern^{\G\M} &\equiv \Chern_{+}^{\G\M} - \Chern_{-}^{\G\M}, \nonumber \\ 
	\Chern^{+} &\equiv \Chern_{+}^{\G\M} + \Chern_{+}^{\M\X} + \Chern_{+}^{\X\G}. 
	\label{eq:SW_chern1}
\end{align}
For only the Wyckoff position ``b'', the latter is instead defined as 
\begin{equation}	
	\Chern^{+} \equiv \Chern_{+}^{\G\M} + \Chern_{+}^{\M\X} + \Chern_{+}^{\X\G}.
\end{equation}
If both Wyckoff positions are allowed, then only $\Chern^{\G\M}$ remains well-defined. 

\subsection{Anomalous boundary states} 

For the band structures with representation-protected topology, there are anomalous boundary modes at the $\D_4$-symmetric surfaces at constant $z$. For the case that only orbitals at Wyckoff position ``a'' are allowed, we find it convenient to replace the integer invariant $\Chern^+$ with	
\begin{align}
	\Chern^{\M\X\G} =&\, \Chern^{\M\X}_{+} + \Chern^{\X\G}_{+} -
	\Chern^{\M\X}_{-} - \Chern^{\X\G}_{-}
	\nonumber \\ =&\, 2 \Chern^+ - \Chern^{\G\M},
\end{align}
which, by definition, has the same parity as $\Chern^{\G\M}$. The case that only orbitals at "b" are allowed proceeds analogously, whereas $\Chern^{\M\X\G}$ is not defined if both Wyckoff positions are allowed. To relate these invariants to boundary states, we note that, since $n_{\o/\e,+}^{SS'} = n_{\o/\e,-}^{SS'}$ for each 1-cell, we can glue the two parity sectors to define a Chern insulator on $SS'$. The Chern number $\Chern^{\G\M}$ then equals the difference of the number of even-parity and odd-parity bands crossing the Fermi surface along $\G\M$, while $\Chern^{\M\X\G}$ (if defined) counts this difference along $\M\X$ and $\X\G$. An odd parity of $\Chern^{\G\M}$ (and also of $\Chern^{\M\X\G}$, if defined) implies protected band touchings at $\G$ and $\M$, similar to the case of a $\C_4$-symmetric band structure \cite{fu2011,song2020,kobayashi2021}. Even contributions to $\Chern^{\G\M}$ and $\Chern^{\M\X\G}$ can be traced back to equal numbers of even-parity and odd-parity bands crossing the Fermi level in opposite directions. Since such crossings can be gapped out away from the high-symmetry lines, they imply Dirac cones protected by mirror symmetry along $\G\M$ and, if applicable, along $\M\X\G$. If $\Chern^{\G\M}$ and $\Chern^{\M\X\G}$ are both defined, the numbers of such Dirac crossings along along $\G\M$ and $\M\X\G$, counted according parity, are independent signatures of the representation-protected topology.

From the boundary perspective, the presence of only $B$ irreps at $\sG$ and $\sM$ implies that the number of states in the occupied/empty subspaces is even, and furthermore, on each high-symmetry lines, $n_{\o,+}^{\overline{SS'}} = n_{\o,-}^{\overline{SS'}}$. In this setup, we can get anomalous gapless modes in two ways: 
\begin{itemize}
	\item A single surface band crosses the Fermi level between $\sG$ and $\sM$, so that so that $n^{\overline \G}_{\o/\e} \neq n^{\overline \M}_{\o/\e} \mod2$. This is incompatible with having only the irrep $B$ at both the $\sG$ and $\sM$ points. Since $n_{\o/\e}$ is defined for any path connecting $\sG$ to $\sM$, all such paths must have a level crossing the Fermi level, so that we get a loop of zero crossings. 
	
	\item A pair of surface bands with opposite parities cross the Fermi level on $\sG\sM$ in opposite directions. This is compatible with having only irrep $B$ at $\sG$ and $\sM$; however, the level-0 compatibility conditions at $\sG$ and $\sM$ cannot be satisfied simultaneously by only $B$ irreps. Since $n_{\o/\e}$ stay unchanged under this crossing, it can be gapped away from $\sG\sM$, leading to a Dirac cone on $\sG\sM$ protected by a mirror symmetry. 
\end{itemize}
These are precisely the two kinds of anomalous boundary states obtained from the bulk perspective. 

\section{Discussion}
\label{sec:concl}
In this article, we presented a homotopic classification of band-structures with crystalline symmetries. The homotopic classification describes the delicate topology of gapped Bloch Hamiltonians with a fixed number of occupied/empty bands $n_\o/n_\e$ and fixed numbers of orbitals for each orbital type at each Wyckoff position. A fragile classification, in which only the number of occupied bands is fixed and unoccupied bands may be added {\em ad libitum}, and a stable classification, in which both occupied and unoccupied bands may be added, can be obtained from the homotopic classification by taking the limits $n_\e\to\infty$ and $n_{\o/\e}\to\infty$, respectively.

The central object in our scheme is the interpretation as a \emph{CW-complex} of the fundamental domain of the point group, which is the minimal region in reciprocal space required to define the full Bloch Hamiltonian. A CW complex, fundamental in homotopy theory, is a topological space constructed from a collection of ``cells'' of dimension $p$ with $p=0$, \ldots, $d$ using a set of ``gluing conditions.'' We classify the Bloch Hamiltonians restricted to the $p$-cells, starting with $p=0$ and building up to higher $p$, imposing the constraints demanded from the CW-complex at each step. The structure of the classification over $p$-cells then depends on the concrete topological invariants over $q$-cells with $q<p$, thereby endowing the classification with a hierarchical structure. For a homotopic classification, the classifying space does not have a group structure. However, in the stable limit, the invariants at different levels factorize, leading to the well-known result that the full classifying spaces are products of groups, with the group operation deriving from the direct sum of bands. 

Two basic elements of our approach --- bootstrapping the full classification from that of restrictions of the Hamiltonian to individual cells, and, for each cell, the reduction of a larger (but easier to compute) ``parent'' classification group to the actual classification set by removing ambiguity from boundary deformations --- have previously been applied to the classification of gapped band structures. The former element is shared with the partial classifications based on topological quantum chemistry \cite{fu2007,fang2012point,kruthoff2017,bradlyn2017,po2017,song2018quantitative,tang2019,tang2019comprehensive,zhang2019catalogue,tang2019efficient}, which consider topological invariants of 0-cells only. The latter was used previously for symmetry class AII by Moore and Balents \cite{moore2007} as well as for the classification of inversion-symmetric insulators in the stable limit \cite{turner2012} (see also Ref.\ \cite{trifunovic2017}). 

We have illustrated the homotopic classification scheme on point groups of increasing complexity, for both time-reversal-invariant and time-reversal-broken cases. Applications to band structure without crystalline symmetry in symmetry classes A, AI, AII make a connection to existing homotopic classification approaches for gapped band structures without crystalline symmetries \cite{moore2007,kennedy2016}. The other examples we consider elucidate various aspects of the homotopic classification and illustrate that it can be performed systematically. However, this process is involved enough that we have not been able to automate the generation of delicate classification tables, which would allow us to compute the topological invariants at all levels for all space groups, as it is done for the level-0 invariants in the framework of topological quantum chemistry and symmetry indicators \cite{bradlyn2017,po2017,song2018quantitative,tang2019,tang2019comprehensive,zhang2019catalogue,tang2019efficient}. A quick look at the complexity of arguments required to produce the classification tables for a single crystalline symmetry in Secs.\ \ref{sec:C2}, \ref{sec:C4}, and \ref{sec:D4} explains why such an automatized procedure may be too much to expect.

We find that a crucial aspect of a delicate classification is that restricting only the number of occupied/empty bands does not determine the classification uniquely. Instead, it is essential to also fix the (total) number of orbitals of each type at each Wyckoff position. Typically, the Wyckoff positions determine the level-0 invariants, and all subsequent restrictions on higher-level invariants can be expressed in terms of these, so that the Wyckoff positions and the representations of orbitals do no appear explicitly in the classification tables. For the examples we consider, this is the case for the point groups $\C_2$, $\C_4$, and $\D_4$. However, the Wyckoff positions may also enter the homotopic classification directly, as is the case for $\C_2\T$-symmetric band structures. 

The existence of boundary states is a hallmark of stable topology --- a topologically nontrivial insulator protected by a lattice symmetry exhibits anomalous boundary states at a boundary that respect the requisite symmetries. In the examples that we consider here, the lattice symmetries include a rotation, so that there are no invariant boundaries in two dimensions. However, in three dimensions, there is an invariant boundary normal to the rotation axis. We establish that such a boundary indeed hosts anomalous boundary modes if the stable topology is nontrivial. Anomalous boundary modes also exist for stable representation-protected topology, whereby one must not only constrain the types (but not the number) of orbitals in the bulk, but also at the at the invariant boundary. 

In real systems, the anomalous boundary states protected by representation-protected stable topology may disappear if there is a surface reconstruction that violates the constraint on orbital types. Nevertheless, if this violation is sufficiently weak, non-anomalous boundary states may exist that can be traced back to the anomalous boundary states of representation-protected topology. A similar reasoning relating non-anomalous boundary states to anomalous boundary states protected by additional symmetries is used in the case of, \eg, simplified models with a sublattice symmetry, which have zero-energy flat bands on edges or on surfaces (such as the ``drumhead states'' of nodal-line semimetals \cite{burkov2011b} and the flat bands at zig-zag edges of graphene \cite{klein1994,fujita1996}), which do not remain flat and at zero energy once sublattice-symmetry-breaking perturbations are taken into account. 

Even in absence of \emph{individual} invariant boundaries, stable topology can lead to boundary signatures if the overall crystal is invariant under the symmetry. Such boundary signatures may be of higher-order type \cite{schindler2018hoti,langbehn2017,song2017,benalcazar2017,benalcazar2017b,fang2019} --- such as anomalous gapless states at crystal hinges --- and they are guaranteed to exist if the band structure is not of atomic limit type \cite{trifunovic2019,trifunovic2020}. For band structures of atomic-limit type, more subtle boundary signatures, such as fractional corner charges or ``twisted boundary signatures'' may exist \cite{lau2016,benalcazar2017,benalcazar2017b,vanmiert2018,schindler2019,ono2019d,song2020,peri2020,trifunovic2020b,watanabe2020,ren2021,watanabe2021,kooi2021,fang2021,takahashi2021,naito2022}, but this is not guaranteed. We expect no such higher-order boundary signatures for representation-protected topology, because it is not possible to organize orbitals according to their representation under the crystalline symmetry operations on a surface that is not itself invariant.
	
In conclusion, we have presented analytical machinery that can be used to classify band structures with crystalline symmetry at the highest possible level of granularity. Whether the application of this methodology to other symmetry classes and more complicated point groups might yield further examples of representation-protected topology, remains a topic of future study.

\acknowledgments
We would like to thank Christophe Mora, Aleksandra Nelson, Aris Alexandradinata, and Adam Yanis Chaou for stimulating discussions. We gratefully acknowledge support by the Deutsche Forschungsgemeinschaft (DFG) project grant 277101999 within the CRC network TR 183 (subprojects A02 and A03).


\appendix

\section{Symmetries of the Bloch Hamiltonian}
\label{app:symm}
We briefly recapitulate the symmetry analysis of the Bloch Hamiltonians for general crystalline symmetries. (For a detailed analysis, see, e.g., the Supplementary Material of Ref.~\cite{alexandradinata2014}.)

\subsection{Crystalline symmetries}
\label{app:symm_gen}
We consider a symmetry of the lattice, derived from an isometry of the underlying space $g\colon\RR^d\to\RR^d$. The physical systems of interest consist of a set of $\nuc$ orbitals $\alpha$ located at positions $\vr_\alpha \in (-\frac12, \frac12]^d$ within the unit cell, whose origin is labeled by $\vR\in\ZZ^d$. Under the action of $g$, an arbitrary lattice point transforms as
\begin{equation}
   g(\vR + \vr_\alpha) = g\vR + \Rshift_{\alpha} + \vr_{\alpha'} \equiv \vR' + \vr_{\alpha'},
   \label{eq:Rtransform}
\end{equation}
where, since $g\vr_\alpha$ may not lie within the unit cell, we subtract off a lattice translation $\Rshift_\alpha$ to move it back within the unit cell at the new position $\vr_{\alpha'}$. The symmetry generically maps a basis orbital at $\vr_\alpha$ to a linear combination of basis orbitals at $\vr_{\alpha'}$, so that its action on the position space basis state $\ket{\vR,\alpha}$ can be written as
\begin{equation}
  \symm(g) \ket{\vR,\alpha} = \sum_{\alpha'=1}^\nuc \symm_{\alpha\alpha'}^\ast(g) \ket{\vR',\alpha'}.
\end{equation}
The unitary matrices $\symm(g)$ for various $g\in G$ then constitute a representation of the point group $G$.

In Sec.~\ref{sec:gen}, we define the Bloch states as 
\begin{equation}
  \ket{\vk,\alpha} 
    = \sum_{\vR} \e^{\i \vk \cdot (\vR+\vr_\alpha)} \ket{\vR,\alpha}.
    \label{eq:defBloch}
\end{equation}
Under the lattice symmetry $g$, these transform as
\begin{align}
  \symm(g) \ket{\vk,\alpha}
  &= \sum_\vR \e^{\i \vk\cdot(\vR+\vr_\alpha)} \symm(g) \ket{\vR,\alpha} \nonumber \\
  &= \sum_{\alpha'=1}^\nuc \sum_\vR \e^{\i \vk\cdot(\vR+\vr_\alpha)} \symm_{\alpha\alpha'}^\ast(g) \ket{\vR',\alpha'} \nonumber \\
  &= \sum_{\alpha'=1}^\nuc \symm_{\alpha\alpha'}^\ast(g) \left( \sum_{\vR'} \e^{\i (g \vk) \cdot (\vR'+\vr_{\alpha'}) } \ket{\vR',\alpha'} \right)  \nonumber \\
  &= \sum_{\alpha'=1}^\nuc \symm_{\alpha\alpha'}^\ast(g) \ket{g\vk,\alpha'},
  \label{eq:Hbloch_symm}
\end{align}
where we have used Eq.~(\ref{eq:Rtransform}) and $\vk\cdot\vR = (g\vk) \cdot (g\vR)$. The symmetry operators are independent of $\vR'$, so that we do not need to account for the change in unit cell that a site might undergo under the symmetry operation. On the other hand, under translation by a reciprocal lattice vector $\vG$, these Bloch states pick up a phase $\e^{\i \vk\cdot\vr_\alpha}$, which can be written in form identical to Eq.~(\ref{eq:Hbloch_symm}) as 
\begin{equation}
  \symm(g_\vG) \ket{\vk,\alpha} \equiv
  \ket{\vk+\vG, \alpha} =
  \sum_{\alpha'} \trans_{\alpha\alpha'}^\ast(\vG) \ket{\vk,\alpha'},
  \label{eq:trans_def}
\end{equation}
where $\trans_{\alpha\beta} (\vk) = \e^{-\i \vk\cdot\vr_\alpha} \delta_{\alpha\beta}$. We thus consider the point group symmetries and reciprocal lattice translations together as elements of the reciprocal space group $G_\Lambda$, whose action on the Bloch states is given by Eq.~(\ref{eq:Hbloch_symm}). The Bloch Hamiltonian then satisfies the constraint
\begin{equation}
  H(g\vk) = \symm^\dg(g) H(\vk) \symm(g)
\end{equation}
for all $g\in G_\Lambda$.

We contrast this with the conventional definition of Bloch states (which does not contain the position of orbitals within the unit cell)
\begin{equation}
  \cket{\vk,\alpha} = \sum_{\vR} \e^{\i \vk \cdot \vR} \cket{\vR,\alpha}.
\end{equation}
These states are periodic under reciprocal lattice transformations, while under a lattice symmetry, they transform as
\begin{align}
  \symm(g) \cket{\vk,\alpha}
  &= \sum_\vR \e^{\i \vk\cdot\vR} \symm(g) \cket{\vR,\alpha} \nonumber \\
  &= \sum_{\alpha'=1}^\nuc \sum_\vR \e^{\i \vk\cdot\vR} \symm_{\alpha\alpha'}^\ast(g) \cket{\vR',\alpha'} \nonumber \\
  &= \sum_{\alpha'=1}^\nuc \symm_{\alpha\alpha'}^\ast(g) \e^{-\i (g\vk) \cdot \Rshift_\alpha} \cket{g\vk,\alpha'}
  \nonumber \\
  &\equiv \sum_{\alpha'=1}^\nuc \wt{\symm}_{\alpha\alpha'}^\ast(g, \vk) \cket{g\vk,\alpha'},
\end{align}
so that the unitary matrix $\wt{\symm}(g,\vk)$ now depends on $\vk$. The matrices $\wt{\symm}(g,\vk)$ can be related to the $\vk$-independent representation matrices $\symm(g)$ as $\wt{\symm}(g,\vk) = \trans(g\vk) \symm(g) \trans(\vk)^\dg$.
The two definitions of Bloch states are related by
\begin{equation}
  \ket{\vk,\alpha} = \e^{\i \vk\cdot\vr_\alpha} \cket{\vk,\alpha} = \sum_{\alpha'} \trans_{\alpha\alpha'}^\ast(\vk) \cket{\vk,\alpha'},
\end{equation}
so that the corrresponding Bloch Hamiltonians are related by
\begin{equation}
  H(\vk) = \trans(\vk) \wt{H}(\vk) \trans^\dg(\vk).
  \label{eq:HDtransform}
\end{equation}
In the following sections, we work exclusively in the Bloch basis of Eq.~(\ref{eq:defBloch}).

\subsection{Time reversal symmetry}
\label{app:symm_tr}
With the exception of Sec.\ \ref{sec:C0_AII}, we consider a time-reversal symmetry with $\T^2 = +1$, and choose a basis of orbitals at lattice sites such that the Bloch Hamiltonian satisfies
\begin{equation}
  H(\vk) = H^*(-\vk).
\end{equation}
This is equivalent to $\wt{H}^*(-\vk) = \wt{H}(\vk)$ in the conventional Bloch basis, since the unitary matrix relating the two bases (Eq.~\ref{eq:HDtransform}) satisfies 
\begin{equation}
  \trans_{\alpha\beta}^*(\vk) = \e^{\i\vk\cdot\vr_\alpha} \delta_{\alpha\beta} = \trans_{\alpha\beta}(-\vk),
\end{equation}
so that 
\begin{align}
  H^*(-\vk) 
  &= \trans^\ast(-\vk) \wt{H}^*(-\vk) \trans^*(-\vk) \nonumber \\
  &= \trans^\dagger(\vk) \wt{H}(\vk) \trans(\vk)
  = H(\vk). 
\end{align}

Let the symmetries of the lattice include twofold rotation (in 2d) or inversion (in 3d) that squares to $+1$, so that the Hamiltonian satisfies
\begin{equation}
  H(\vk) = \symm^\dagger H(-\vk) \symm
  \label{eq:H_c2symm}
\end{equation}
for some unitary matrix $C$ satisfying
\begin{equation}
  \symm^2 = \id = \symm\symm^\dagger \implies \symm = \symm^\dagger.
  \label{eq:C2_sq}
\end{equation}
Thus, the time-reversed version of Eq.~(\ref{eq:H_c2symm}) becomes
\begin{equation}
  H^* (\vk) = \symm H(\vk) \symm,
  \label{eq:c2t_symm}
\end{equation}
Complex conjugating both sides, we get
\begin{equation}
  H (\vk) 
  = \symm^* H^*(\vk) \symm^*
  = (\symm^\ast \symm) H(\vk) (\symm^* \symm)^\dagger,
\end{equation}
so that $H(\vk)$ must commute with $\symm^* \symm$.
This is trivially satisfied if $\symm^* \symm = \pm\id$, i.e., if $\symm$ is purely real or purely imaginary.

The local-in-$\vk$ condition of Eq.~(\ref{eq:c2t_symm}) can be reduced to a reality condition on the Bloch Hamiltonian by defining a unitarily equivalent Hamiltonian 
\begin{equation}
  H'(\vk) \equiv \sqrtC H(\vk) \sqrtC^\dagger,
\end{equation}
where $\sqrtC$ is unitary.
Eq.~(\ref{eq:c2t_symm}) then becomes 
\begin{align}
  H'^*(\vk) 
  &= \sqrtC^\ast\left( \symm H(\vk) \symm \right) \sqrtC^T \nonumber \\
  &= \left( \sqrtC^* \symm \sqrtC^\dagger \right) H'(\vk)  \left( \sqrtC^* \symm \sqrtC^\dagger \right)^\dagger.
\end{align}
Thus, $H'(\vk)$ is real if 
\begin{equation}
  \sqrtC^* \symm \sqrtC^\dagger = \id
  \implies \symm = \sqrtC^T \sqrtC,
  \label{eq:C_decomp}
\end{equation}
\ie, if $\sqrtC$ is a ``square root'' of $\symm$. We can readily define such a matrix $\sqrtC$ by working in the eigenbasis of $\symm$, where $\sqrtC$ is simply a diagonal matrix whose entries are square roots of the diagonal entries of $\symm$.

\section{Explicit construction of the representation matrices}
\label{app:symm_matrices}
In this section, we compute the unitary matrices $\symm(g)$ and $\trans_a$ representing the reciprocal space groups $G_\Lambda$ for the lattice symmetries discussed in this article. Specifically, for all Wyckoff position and orbitals, we derive the explicit form of the rotation matrices $\rotn_n$ and, in the case of $\D_4$, the reflection matrix $M_x$. For the translation component of $G_\Lambda$, we derive $\Lambda(\vk)$ defined in Eq.~(\ref{eq:trans_def}), from which the translation matrices can be obtained as $\trans_a = \Lambda(2\pi\ve_a)$ for $a = x, y$. 

\subsection{\texorpdfstring{$\C_2$}{C2} symmetry}
\label{app:symm_C2}
For twofold rotation symmetry, there are four special Wyckoff positions as well as the generic Wyckoff position, as listed in Table~\ref{tab:C2_wyc}. For a system with $n_\alpha$ electrons at the Wyckoff position $\alpha$, the total number of electrons in the unit cell is 
\begin{align*}
  n =&\, \sum_\alpha \text{mult}(\alpha) n_\alpha \nonumber \\
    =&\, n_\a + n_\b + n_\c + n_\d + 2n_\g. 
\end{align*}
The reciprocal space translation matrix is block diagonal in Wyckoff positions, so that   
\begin{equation}
  \trans(\vk) = \diag \left\{ \trans_\a(\vk),\trans_\b(\vk),\trans_\c(\vk),\trans_\d(\vk),\trans_\g(\vk) \right\},
\end{equation}
The translation matrices for individual Wyckoff positions can be computed using their definition in Eq.~(\ref{eq:trans_def}) and the coordinates of various Wyckoff positions listed in Table~\ref{tab:C2_wyc} as 
\begin{align*}
  \trans_\a(\vk) =&\, \openone_{n_\a},  \\
  \trans_\b(\vk) =&\, \openone_{n_\b} \e^{\i(k_x+k_y)/2}, \nonumber \\
  \trans_\c(\vk) =&\, \openone_{n_\c} \e^{\i k_x/2}, \nonumber \\
  \trans_\d(\vk) =&\, \openone_{n_\d} \e^{\i k_y/2}, \nonumber \\
  \trans_\g(\vk) =&\, \openone_{n_\g} \diag \left\{ \e^{\i(k_x a_x + k_y a_y)}, \e^{-\i(k_x a_x + k_y a_y)} \right\}.
\end{align*}
The twofold rotation operator is $\vk$-independent and block diagonal in the Wyckoff positions, i.e., 
\begin{equation}
  \symm_2 = \diag \left\{ \symm_\a,\symm_\b,\symm_\c,\symm_\d,\symm_\g \right\}.
\end{equation}
We now compute the matrices for each sector.

\begin{table}
	\centering 
	\begin{tabular}{|c|c|l|}
		\hline
		$\alpha$ & $G_\alpha$  & Location(s) within the unit cell \\ 
		\hline 
		a & \multirow{4}{*}{$\C_2$} & $(0,0)$ \\ 
		b & & $(1/2,1/2)$ \\ 
		c & & $(1/2,0)$ \\ 
		d & & $(0,1/2)$ \\ 
		\hline 
		g & --- & $(a_x, a_y)$, $(-a_x, -a_y)$ \\ 
		\hline
	\end{tabular}
	\caption{Wyckoff positions $\alpha$ at coordinate $\vr_\alpha$(s) within the unit cell and the corresponding little groups $G_\alpha$ for a 2d $\C_2$-symmetric lattice. For the generic Wyckoff position, we take $a_{x,y}\in(0,1/2)$.}
	\label{tab:C2_wyc} 
\end{table}

\begin{table}
	\centering 
	\setlength\tabcolsep{15pt}
	\begin{tabular}{|c|cccc|}
		\hline 
		Label        & $\G$   & $\M$   & $\X$   & $\Y$  \\ 
		\hline 
		$+_\a$       & $+$    & $+$    & $+$    & $+$   \\
		$-_\a$       & $-$    & $-$    & $-$    & $-$   \\
		$+_\b$       & $+$    & $+$    & $-$    & $-$   \\
		$-_\b$       & $-$    & $-$    & $+$    & $+$   \\
		$+_\c$       & $+$    & $-$    & $-$    & $+$   \\
		$-_\c$       & $-$    & $+$    & $+$    & $-$   \\
		$+_\d$       & $+$    & $-$    & $+$    & $-$   \\
		$-_\d$       & $-$    & $+$    & $-$    & $+$   \\
		\hline 
		$2_\g$       & $+-$   & $+-$   & $+-$   & $+-$  \\
		\hline
	\end{tabular}
	\caption{Elementary band representations for twofold rotation symmetry.}
	\label{tab:C2_ebr}
\end{table}

The special Wyckoff positions all have multiplicity one, and thus map back to themselves under twofold rotation. Furthermore, each of them have a little group $\C_2$, so that the orbitals at these Wyckoff positions are labeled by the corresponding irreps $\pm1$. Thus, 
\begin{equation}
  \symm_\alpha = \diag \left\{ \openone_{n_{\alpha+}},-\openone_{n_{\alpha-}} \right\}; \quad
  \alpha\in\{\text{a,b,c,d}\}. 
\end{equation}
The generic Wyckoff position has multiplicity two and thus constitutes of two points ``g1'' and ``g2'', which transform into each other under $\C_2$, so that 
\begin{equation}
  \symm_\g = \openone_{n_\g} \otimes \sigma_1.
\end{equation}
Using these definitions, we can compute the unitary matrices $\rotn_2$, $\trans_x$ and $\trans_y$ for any given combination of Wyckoff positions and orbitals. 

For symmetry class AI, we need an additional basis transformation by $\sqrtC$ to ensure that $\C_2\T$ acts as a complex conjugation, as discussed in App.~\ref{app:symm_tr}. To this end, we first rotate the basis in the subspace of generic Wyckoff positions (if present) to achieve $C_g = \id_{n_\g} \otimes \sigma_3$. The resulting representation matrix $\rotn_2$ is a diagonal matrix with entries $\pm1$. A requisite basis transformation $\sqrtC$ can then be computed using Eq.~(\ref{eq:C_decomp}) by defining $B$ as a diagonal matrix with an $\i$ wherever there is $-1$ in $\rotn_2$. 

A band representation for $\C_2$ symmetry consists of the set of rotation eigenvalues ($\pm$) at the four high-symmetry points. The EBRs, corresponding to single orbital at a single Wyckoff position, are listed in Table~\ref{tab:C2_ebr}. 

\begin{table}
	\centering 
	\setlength\tabcolsep{8pt}
	\begin{tabular}{|c|c|l|}
		\hline
		$\alpha$ & $G_\alpha$  & Location(s) within the unit cell ($\vr_\alpha$)\\ 
		\hline 
		a & \multirow{2}{*}{$\C_4$} & $(0,0)$ \\ 
		b & & $(1/2,1/2)$ \\ 
		\hline 
		c & $\C_2$ & $(1/2,0)$, $(0,1/2)$ \\ 
		\hline
		g & --- &$(a_x,a_y)$, $(-a_y,a_x)$, $(-a_x,-a_y)$, $(a_y,-a_x)$  \\ 
		\hline
	\end{tabular}
	\caption{Wyckoff positions $\alpha$ at coordinate $\vr_\alpha$(s) within the unit cell and the corresponding little groups $G_\alpha$ for a 2d $\C_4$-symmetric lattice. We take $a_{x,y}\in(0,1/2)$. }
	\label{tab:C4_wyc} 
\end{table}

\subsection{\texorpdfstring{$\C_4$}{C4} symmetry}
\label{app:symm_C4}
For fourfold rotation symmetry, there are three special Wyckoff positions as well as the generic Wyckoff position, as listed in Table~\ref{tab:C4_wyc}. For a system with $n_\alpha$ electrons at the Wyckoff position $\alpha$, the total number of electrons in the unit cell is 
\[ 
  n = n_\a + n_\b + 2n_\c + 4n_\g. 
\]
The reciprocal space translation matrix is 
\begin{equation}
  \trans(\vk) = \diag \left\{ \trans_\a(\vk),\trans_\b(\vk),\trans_\c(\vk),\trans_\g(\vk) \right\} ,
\end{equation}
where
\begin{align}
  \trans_\a(\vk) =&\, \openone_{n_\a}, \nonumber \\
  \trans_\b(\vk) =&\, \openone_{n_\b} \e^{\i(k_x+k_y)/2}, \nonumber \\
  \trans_\c(\vk) =&\,
  \openone_{n_\c} \diag \left\{
  \e^{\i k_x/2}, \e^{ik_y/2} \right\}   \\
  \trans_\g(\vk) =&\,
  \openone_{n_\g} \diag \left\{ \e^{\i (a_x k_x + a_y k_y)},
  \e^{\i(-a_y k_x+a_x k_y)}, \right. \nonumber \\ 
  &\qquad\qquad\left. \e^{\i(-a_x k_x - a_y k_y)},
  \e^{\i(a_y k_x - a_x k_y)} \right\}. \nonumber 
\end{align}
The fourfold rotation operator is block diagonal in the Wyckoff positions, i.e., 
\begin{equation}
  \symm_4 = \diag \left\{ \symm_\a,\symm_\b,\symm_\c,\symm_\g \right\}.
\end{equation}
We now compute the matrices for each sector.

The Wyckoff positions ``a'' and ``b'' have multiplicity one, so that they map back onto themselves under $\C_4$. The orbitals at these Wyckoff positions are of $s$, $d$ or $p_\pm$ type, corresponding to a $\C_4$ eigenvalue 1, $-1$ and $\pm\i$. Thus, in absence of time-reversal symmetry, 
\begin{equation}
   \symm_\alpha = \diag \left\{\openone_{n_{s\alpha}},-\openone_{n_{d\alpha}},i \openone_{n_{p_+\alpha}},-i \openone_{n_{p_-\alpha}} \right\},
\end{equation}
where $\alpha\in \{\text{a, b}\}$ and $n_\alpha = n_{s\alpha} + n_{d\alpha} + n_{p_+\alpha} + n_{p_-\alpha}$. In the presence of time-reversal symmetry, the $p_\pm$ orbitals come together, so that we can set $n_{p_+,\alpha} = n_{p_-,\alpha} \equiv n_{p,\alpha}$. We write them in the real basis as $p_x, p_y$, so that 
\begin{equation}
   \symm_\alpha = \diag \left\{\openone_{n_{s\alpha}},-\openone_{n_{d\alpha}},\i \sigma_y \otimes \openone_{n_{p\alpha}}  \right\},
\end{equation}
with $n_\alpha = n_{s\alpha} + n_{d\alpha} + 2n_{p\alpha}$. 
The two points constituting Wyckoff position ``c'' are labeled by their inversion parity and transform into each other under $\C_4$, so that 
\begin{equation}
  \symm_\c^2 = \id_2 \; c_\c; \quad  c_\c = \diag\left\{ \openone_{n_{+\c}},-\openone_{n_{-\c}} \right\},
  \label{eq:C4_C2op}
\end{equation}
where $\id_2$ corresponds to the two sites of Wyckoff position ``c.'' The $\C_4$ operator, which is a square root of this matrix, depends on the relation between the basis states at ``c1''  and  ``c2''. We choose the basis at ``c1'' as the eigenbasis of $\C_2$, and define the basis states at ``c2'' as the images under $\C_4$ of those at ``c1''. We can thus write 
\begin{equation}
  \C_4 
  \begin{pmatrix} 
    \ket{\vR, \text{c1}} \\ 
    \ket{\vR, \text{c2}} 
  \end{pmatrix}
  = 
  \begin{pmatrix} 
    \ket{\vR, \text{c2}} \\ 
    c_\c \ket{\vR-\ve_x, \text{c1}} 
  \end{pmatrix},  
\end{equation}
where  $\ket{\vR, \text{c1}}$ to denote a column vector consisting of $n_\c$ orbitals at ``c1''. The matrix multiplying the RHS can be identified as $\symm_4^T$, so that
\begin{equation}
  \symm_\c(\vk) =
  \begin{pmatrix} 
    0 & c_\c \\ 
    \openone_{n_\c} & 0 
  \end{pmatrix}, 
\end{equation}
squaring which yields Eq.~\ref{eq:C4_C2op}. 

Finally, the four points that constitute the generic Wyckoff position ``g'' transform under $\C_4$ as  
\begin{align}
  &\C_4 \vr_{g1} = \vr_{g2},  
  &\C_4 \vr_{g2} = \vr_{g3},  \nonumber \\ 
  &\C_4 \vr_{g3} = \vr_{g4},  
  &\C_4 \vr_{g4} = \vr_{g1},
\end{align}
so that 
\begin{equation}
  \symm_\g(\vk) =
  \begin{pmatrix} 
    0 & 0 & 0 & \openone_{n_\c} \\ 
    \openone_{n_\c} & 0 & 0 & 0 \\ 
    0 & \openone_{n_\c} & 0 & 0 \\ 
    0 & 0 & \openone_{n_\c} & 0 
  \end{pmatrix}.
\end{equation}
In the presence of time-reversal symmetry, we can again diagonalize the $\rotn_2$ operator and perform a basis transformation, as described in App.~\ref{app:symm_C2}, to ensure that $\C_2\T$ acts as a complex conjugation. 

A band representation for $\C_4$ symmetry consists of the $\rotn_4$  eigenvalue, (i.e., the orbital type $s$, $p_\pm$, $d$) at the $\G$ and $\M$ points and the $\rotn_2$ eigenvalue ($\pm$) at the $\X$ point. The EBRs for $\C_4$ symmetry are listed in Table~\ref{tab:C4_ebr}. 

\begin{table}
  \centering
  \begin{tabular}{|c|ccc|}
    \hline 
    Label        & $\G$    & $\M$      & $\X$  \\ 
    \hline 
    $s_\a$       & $s$     & $s$       & $+$   \\
    $d_\a$       & $d$     & $d$       & $+$   \\
    $p_{\pm \a}$ & $p_\pm$ & $p_\pm$   & $-$   \\
    $2p_\a$      & $p^2$   & $p^2$     & $--$  \\
    \hline 
    $s_\b$       & $s$     & $d$       & $-$   \\
    $d_\b$       & $d$     & $s$       & $-$   \\
    $p_{\pm \b}$ & $p_\pm$ & $p_\mp$   & $+$   \\
    $2p_\b$      & $p^2$   & $p^2$     & $++$  \\
    \hline 
    $+_\c$       & $sd$    & $p^2$     & $+-$  \\
    $-_\c$       & $p^2$   & $sd$      & $+-$  \\
    \hline 
    $4_\g$       & $sdp^2$ & $sdp^2$   & $++--$\\
    \hline 
  \end{tabular}
  \caption{
    Elementary band representations for fourfold rotation symmetry. in the presence of time-reversal symmetry, the two $p$ orbitals transform as a doublet, denoted by $2p_\a$. 
  }
  \label{tab:C4_ebr}
\end{table}

\subsection{\texorpdfstring{$\D_4$}{D4} symmetry}
\label{app:symm_D4}
For the fourfold dihedral symmetry,  there are five special Wyckoff positions as well as the generic Wyckoff position, as listed in Table~\ref{tab:D4_wyc}. For a system with $n_\alpha$ electrons at the Wyckoff position $\alpha$, the total number of electrons in the unit cell is 
\[ 
  n = n_\a + n_\b + 2n_\c + 4n_\d + 4n_\e + 8n_\g. 
\]
The reciprocal space translation matrix is 
\begin{equation}
  \trans(\vk) = \diag \left\{ \trans_\a (\vk),\trans_\b (\vk),\trans_\c (\vk),\trans_\d (\vk),\trans_\e (\vk),\trans_\g (\vk) \right\},
\end{equation}
where 
\begin{align}
  \trans_\a(\vk) =&\, \id_{n_\a}, \nonumber \\
  \trans_\b(\vk) =&\, \id_{n_\b} \e^{\i(k_x+k_y)/2}, \nonumber \\
  \trans_\c(\vk) =&\, \id_{n_\c} \diag \left\{ \e^{\i k_x/2}, \e^{ik_y/2} \right\}   \nonumber \\
  \trans_\d(\vk) =&\, \id_{n_\d} \diag \left\{ \e^{\i a k_x}, \e^{\i a k_y}, \e^{-\i a k_x}, \e^{-\i a k_y} \right\} \nonumber \\ 
  \trans_\e(\vk) =&\,
  \id_{n_\e} \diag \left\{ \e^{\i a (k_x + k_y)}, \e^{\i a (-k_x + k_y)}, \e^{\i a (-k_x - k_y)},  \right. \nonumber \\ 
  &\qquad\qquad \left. \e^{\i a (k_x - k_y)}  \right\}, 
  \nonumber  \\
   \trans_\g(\vk) =&\, 
   \id_{n_\g}\diag \Big\{
   \e^{\i (a_x k_x + a_y k_y)}, \e^{\i (-a_x k_x + a_y k_y)},   \nonumber \\
   &\qquad\qquad  \e^{\i (a_x k_x - a_y k_y)}, \e^{\i (-a_x k_x - a_y k_y)}, \nonumber \\ 
   &\qquad\qquad  \e^{\i (a_y k_x + a_x k_y)}, \e^{\i (-a_y k_x + a_x k_y)}, \nonumber \\
   &\qquad\qquad  \e^{\i (a_y k_x - a_x k_y)}, \e^{\i (-a_y k_x - a_x k_y)}  \Big\}. 
\end{align}
Since the group $\D_4$ is generated by two operators, we need to compute a fourfold rotation operator $\rotn_4$ and the mirror operator $\mirror_x$, both of which are block diagonal in the Wyckoff positions, i.e.,
\begin{align}
  \rotn &= \diag \left\{ \rotn_\a,\rotn_\b,\rotn_\c,\rotn_\d,\rotn_\e,\rotn_\g \right\}, \nonumber \\ 
  \mirror &= \diag \left\{ \mirror_\a,\mirror_\b,\mirror_\c,\mirror_\d,\mirror_\e,\mirror_\g \right\}.
\end{align}
We now compute the matrices for each sector.

\begin{table}
  \centering 
  \setlength\tabcolsep{6pt}
  \begin{tabular}{|c|c|l|}
    \hline
    $\alpha$ & $G_\alpha$  & Location(s) within the unit cell ($\vr_\alpha$)\\ 
    \hline 
    a & \multirow{2}{*}{$\D_4$} & $(0,0)$ \\ 
    b & & $(1/2,1/2)$ \\ 
    \hline 
    c & $\D_2$ & $(1/2,0)$, $(0,1/2)$ \\ 
    \hline 
    d & \multirow{2}{*}{$\ZZ_2$} & $(a,0)$,  $(0,a)$, $(-a,0)$, $(0,-a)$  \\
    e & & $(a,a)$, $(-a,a)$, $(-a,-a)$, $(a,-a)$ \\ 
    \hline
    \multirow{2}{*}{g} & \multirow{2}{*}{---} & $(a_x,a_y)$, $(-a_y,a_x)$, $(-a_x,-a_y)$, $(a_y,-a_x)$ , \\ 
      & & $(a_x,-a_y)$, $(a_y,a_x)$, $(-a_x,a_y)$, $(-a_y,-a_x)$   \\ 
    \hline 
  \end{tabular}
  \caption{Wyckoff positions $\alpha$ at coordinate $\vr_\alpha$(s) within the unit cell and the corresponding little groups $G_\alpha$ for a 2d $\D_4$-symmetric lattice. Here, $\D_n$ denotes the group formed by $n$-fold rotation along with $n$ reflections. We take $a,a_{x,y}\in(0,1/2)$.}
  \label{tab:D4_wyc} 
\end{table}

The Wyckoff positions ``a'' and ``b'' have multiplicity one and thus map to themselves under both $\C_4$ and $\M_x$. 
The orbitals at  Wyckoff positions ``a'' and ``b'' are labeled by the irreps of $\D_4$ ( see Table~\ref{tab:D4_chr}). 
Thus, for $\alpha\in \{\text{a, b}\}$, 
\begin{align}
   \rotn_\alpha &= \diag \left\{     
       \id_{n_{A_1\alpha}},
       \id_{n_{A_2\alpha}},
      -\id_{n_{A_3\alpha}},
      -\id_{n_{A_4\alpha}},
       \i\sigma_2 \id_{n_{B\alpha}}
  \right\},  \nonumber \\ 
   \mirror_\alpha &=  \diag \left\{
       \id_{n_{A_1\alpha}},
      -\id_{n_{A_2\alpha}},
       \id_{n_{A_3\alpha}},
      -\id_{n_{A_4\alpha}},
       \sigma_3 \id_{n_{B\alpha}}
  \right\},  
\end{align}
with $n_\alpha = n_{A_1\alpha} + n_{A_2\alpha} + n_{A_3\alpha} + n_{A_4\alpha} + 2n_{B\alpha}$, and we have chosen the explicit matrices for irrep $B$ as in Eq.~(\ref{eq:irrepB}). 

The two points constituting the Wyckoff position ``c'' transform into each other under $\rotn_4$ but into themselves under $\mirror_x$. The orbitals at these points are labeled by $(\pm\pm)$, their parity under $\rotn_2$ and $\mirror_x$, respectively. Thus, in the simultaneous eigenbasis of $\rotn_2$ and $\mirror_x$, we get 
\begin{align}
   c_\c &= \diag \left\{     
       \id_{n_{++\c}},
       \id_{n_{+-\c}},
      -\id_{n_{-+\c}},
      -\id_{n_{--\c}}
  \right\},  \nonumber \\ 
   m_\c &=  \diag \left\{          
       \id_{n_{++\c}},
      -\id_{n_{+-\c}},
       \id_{n_{-+\c}},
      -\id_{n_{--\c}}
  \right\},  
\end{align}
with $n_\c = n_{++\c} + n_{+-\c} + n_{-+\c} + n_{--\c}$. As for $\C_4$, we choose the basis at ``c2'' as the images of those at ``c1'' under a rotation, so that the rotation and mirror operators are explicitly given by
\begin{equation}
  \rotn_\c = 
  \begin{pmatrix} 
    0 & c_\c \\ 
    \id_{n_\c} & 0 
  \end{pmatrix}, \quad  
  \mirror_\c =
  \begin{pmatrix} 
    m_\c & 0  \\ 
    0 & m_\c c_\c
  \end{pmatrix}.
\end{equation}
where in the latter case, $m_c$ denotes the mirror eigenvalues of $M_y = M_x \C_2$. 

\begin{table*}
	\centering
	\begin{tabular}{|c|ccc|ccc|}
		\hline  
		Label     & $\G$   & $\M$   & $\X$    & $\G\M$  & $\M\X$   & $\G\X$  \\ 
		\hline 
		$(A_1)_\a$ & $A_1$  & $A_1$  & $(++)$    & $+$     & $+$      & $+$     \\ 
		$(A_2)_\a$ & $A_2$  & $A_2$  & $(+-)$    & $-$     & $-$      & $-$     \\ 
		$(A_3)_\a$ & $A_3$  & $A_3$  & $(++)$    & $-$     & $+$      & $+$     \\ 
		$(A_4)_\a$ & $A_4$  & $A_4$  & $(+-)$    & $+$     & $-$      & $-$     \\ 
		$(B)_\a$ & $B$  & $B$  & $(++)(+-)$& $+-$    & $+-$     & $+-$    \\ 
		\hline 
		$(A_1)_\b$ & $A_1$  & $A_4$  & $(-+)$    & $+$     & $+$      & $+$     \\ 
		$(A_2)_\b$ & $A_2$  & $A_3$  & $(--)$    & $-$     & $-$      & $-$     \\ 
		$(A_3)_\b$ & $A_3$  & $A_2$  & $(-+)$    & $-$     & $+$      & $+$     \\ 
		$(A_4)_\b$ & $A_4$  & $A_1$  & $(--)$    & $+$     & $-$      & $-$     \\  
		$(B)_\b$ & $B$  & $B$  & $(-+)(--)$& $+-$    & $+-$     & $+-$    \\
		\hline 
		$(++)_\c$  & $A_1A_3$   & $B$       & $(++)(-+)$    & $+-$     & $++$      & $++$     \\ 
		$(+-)_\c$  & $A_2A_4$   & $B$       & $(+-)(--)$    & $+-$     & $--$      & $--$     \\ 
		$(-+)_\c$  & $B$        & $A_1A_3$  & $(++)(--)$    & $+-$     & $+-$      & $+-$     \\ 
		$(--)_\c$  & $B$        & $A_2A_4$  & $(+-)(-+)$    & $+-$     & $+-$      & $+-$     \\ 
		\hline 
		$(+)_\d$  & $A_1A_3B$   & $A_1A_3B$ & $(++)(++)(-+)(--)$  & $++--$  & $+++-$ & $+++-$   \\ 
		$(+)_\d$  & $A_2A_4B$   & $A_2A_4B$ & $(+-)(+-)(-+)(--)$  & $++--$  & $+---$ & $+---$   \\ 
		\hline 
		$(+)_\e$  & $A_1A_4B$   & $A_1A_4B$ & $(++)(+-)(-+)(--)$  & $+++-$  & $++--$ & $++--$   \\ 
		$(+)_\e$  & $A_2A_3B$   & $A_2A_3B$ & $(++)(+-)(-+)(--)$  & $+---$  & $++--$ & $++--$   \\ 
		\hline    
		\multirow{2}{*}{ $8_\g$ }   & \multirow{2}{*}{$A_1A_2A_3A_4B^2$}  & \multirow{2}{*}{$A_1A_2A_3A_4B^2$}  & $(++)(++)(-+)(-+)$    & $++++$     & $++++$      & $++++$     \\ 
		&              &              & $(+-)(+-)(--)(--)$    & $----$     & $----$      & $----$     \\ 
		\hline 
	\end{tabular}
	\caption{Elementary band representations of the square lattice with $\D_4$ symmetry. The representations of the little groups $G_\G$ and $G_\M$ are labeled by the irrep labels of $\D_4$ (see Tab.~\ref{tab:D4_chr}), while representations of $G_\X$ are labeled by the eigenvalue of $\rotn_2$ and $\mirror_x$.}  
	\label{tab:D4_ebr}
\end{table*}

The four points constituting the Wyckoff position ``d'' transform as 
\begin{align}
  C_4 \vr_{\d1} &= \vr_{\d2}, 
  &M_x \vr_{\d1} &= \vr_{\d1}, \nonumber \\ 
  C_4 \vr_{\d2} &= \vr_{\d3}, 
  &M_x \vr_{\d2} &= \vr_{\d4}, \nonumber \\ 
  C_4 \vr_{\d3} &= \vr_{\d4}, 
  &M_x \vr_{\d3} &= \vr_{\d3}, \nonumber \\ 
  C_4 \vr_{\d4} &= \vr_{\d1}, 
  &M_x \vr_{\d4} &= \vr_{\d2}.
\end{align} 
The orbitals at  Wyckoff position ``d'' are labeled by $\pm$, corresponding to the eigenvalue under reflection by the $x$ (resp. $y$) axis. Thus, 
\begin{align}
   m_\d &= \diag \left\{     
       \id_{n_{+\d}},
      -\id_{n_{-\d}}
  \right\},  
\end{align}
with $n_\d = n_{+\d} + n_{-\d}$. 
We choose the rotated version of the orbitals at ``d1'' as the orbitals at the remaining three points constituting the Wyckoff position ``d''. Thus, 
\begin{equation}
  \rotn_\d = 
  \begin{pmatrix} 
    0 & 0 & 0 & \id_{n_\d} \\ 
    \id_{n_\d} & 0 & 0 & 0 \\ 
    0 & \id_{n_\d} & 0 & 0 \\ 
    0 & 0 & \id_{n_\d} & 0 
  \end{pmatrix}
\end{equation}
and
\begin{equation}
  \mirror_\d =
  \begin{pmatrix}
    m_\d & 0 & 0 & 0  \\  
    0 & 0 & 0 & m_\d  \\
    0 & 0 & m_\d & 0  \\
    0 & m_\d & 0 & 0 
  \end{pmatrix}.
\end{equation}

The four points constituting the Wyckoff position ``e'' transform as 
\begin{align}
  C_4 \vr_{\e1} &= \vr_{\e2}, 
  &M_x \vr_{\e1} &= \vr_{\e4}, \nonumber \\ 
  C_4 \vr_{\e2} &= \vr_{\e3}, 
  &M_x \vr_{\e2} &= \vr_{\e3}, \nonumber \\ 
  C_4 \vr_{\e3} &= \vr_{\e4}, 
  &M_x \vr_{\e3} &= \vr_{\e2}, \nonumber \\ 
  C_4 \vr_{\e4} &= \vr_{\e1}, 
  &M_x \vr_{\e4} &= \vr_{\e1}.
\end{align} 
with $n_\e = n_{+\e} + n_{-\e}$. 
We again choose the rotated version of the orbitals at ``e1'' as the orbitals at the remaining three points constituting the Wyckoff position ``e''. Thus, 
\begin{equation}
  \rotn_\e = 
  \begin{pmatrix} 
    0 & 0 & 0 & \id_{n_\e} \\ 
    \id_{n_\e} & 0 & 0 & 0 \\ 
    0 & \id_{n_\e} & 0 & 0 \\ 
    0 & 0 & \id_{n_\e} & 0 
  \end{pmatrix}
\end{equation}
and 
\begin{equation}
  \mirror_\e =
  \begin{pmatrix}
    0 & 0 & 0 & m_\e \\
    0 & 0 & m_\e & 0 \\ 
    0 & m_\e & 0 & 0 \\ 
    m_\e & 0 & 0 & 0  
  \end{pmatrix}.
\end{equation}

Finally, the eight points constituting the generic Wyckoff position transform as 
\begin{align}
  C_4 \vr_{\g1} &= \vr_{\g2}, 
  &M_x \vr_{\g1} &= \vr_{\g5}, \nonumber \\ 
  C_4 \vr_{\g2} &= \vr_{\g3}, 
  &M_x \vr_{\g2} &= \vr_{\g8}, \nonumber \\ 
  C_4 \vr_{\g3} &= \vr_{\g4}, 
  &M_x \vr_{\g3} &= \vr_{\g7}, \nonumber \\ 
  C_4 \vr_{\g4} &= \vr_{\g1}, 
  &M_x \vr_{\g4} &= \vr_{\g6}, \nonumber \\ 
  C_4 \vr_{\g5} &= \vr_{\g6}, 
  &M_x \vr_{\g5} &= \vr_{\g1}, \nonumber \\ 
  C_4 \vr_{\g6} &= \vr_{\g7}, 
  &M_x \vr_{\g6} &= \vr_{\g4}, \nonumber \\ 
  C_4 \vr_{\g7} &= \vr_{\g8}, 
  &M_x \vr_{\g7} &= \vr_{\g3}, \nonumber \\ 
  C_4 \vr_{\g8} &= \vr_{\g5}, 
  &M_x \vr_{\g8} &= \vr_{\g2},
\end{align} 
so that 
\begin{equation*}
  \rotn_\g = 
  \begin{pmatrix} 
    0 & 0 & 0 & \id_{n_\g} & 0 & 0 & 0 & 0  \\ 
    \id_{n_\g} & 0 & 0 & 0 & 0 & 0 & 0 & 0  \\ 
    0 & \id_{n_\g} & 0 & 0 & 0 & 0 & 0 & 0  \\ 
    0 & 0 & \id_{n_\g} & 0 & 0 & 0 & 0 & 0  \\ 
    0 & 0 & 0 & 0 & 0 & 0 & 0 & \id_{n_\g} \\ 
    0 & 0 & 0 & 0 & \id_{n_\g} & 0 & 0 & 0 \\ 
    0 & 0 & 0 & 0 & 0 & \id_{n_\g} & 0 & 0 \\ 
    0 & 0 & 0 & 0 & 0 & 0 & \id_{n_\g} & 0 
  \end{pmatrix}
\end{equation*}
and 
\begin{equation*}
  \mirror_\g =
  \begin{pmatrix} 
    0 & 0 & 0 & 0 & \id_{n_\g} & 0 & 0 & 0  \\ 
    0 & 0 & 0 & 0 & 0 & 0 & 0 & \id_{n_\g}  \\ 
    0 & 0 & 0 & 0 & 0 & 0 & \id_{n_\g} & 0  \\ 
    0 & 0 & 0 & 0 & 0 & \id_{n_\g} & 0 & 0  \\ 
    \id_{n_\g} & 0 & 0 & 0 & 0 & 0 & 0 & 0 \\ 
    0 & 0 & 0 & \id_{n_\g} & 0 & 0 & 0 & 0 \\ 
    0 & 0 & \id_{n_\g} & 0 & 0 & 0 & 0 & 0 \\ 
    0 & \id_{n_\g} & 0 & 0 & 0 & 0 & 0 & 0 
  \end{pmatrix}.
\end{equation*}

A band representation for a $\D_4$ symmetric lattice thus consists of the irrep label at the $\G$ and $\M$ points, the twofold rotation and mirror eigenvalues ($\pm$) at the $\X$ point, and the eigenvalues $\pm$ of $\mirror_\alpha$ for the 1-cells $\alpha = \G\X, \M\X, \G\M$. The EBRs for $\D_4$-symmetric band structures are listed in Table~\ref{tab:D4_ebr}.

\section{Homotopic classification of maps defined on CW complexes}
\label{app:CW}
In this Section, we describe the classification procedure outlined in Sec.~\ref{sec:gen_clfn} more rigorously. 

\ \\ \para{CW-complex}
The fundamental domain of the $d$-dimensional Brillouin zone has the structure of a CW-complex $\cwc$. Formally, $\cwc$ is a collection of ``$p$-cells'' with $0 \leq p \leq d$, each of which is  homotopic to $(0,1)^p$, or, equivalently, to a $p$-ball. The CW-complex structure further includes a gluing map, which, given a $p$-cell $\alpha$, assigns to its boundary $\partial\alpha$ --- homotopically equivalent to the $(p-1)$-sphere $S^{p-1}$ --- a collection of $q$-cells with $q<p$. Note that a given $q$-cell $\beta$ may be glued multiple times to the boundary $\partial\alpha$. We define the boundary $\partial \cwc$ of the CW complex as the union of boundaries of individual $d$-cells. Note that for a $d$-dimensional CW-complex $\cwc$ with multiple $d$-cells, lower-dimensional cells at the boundary between two $d$-cells are also considered to be part of $\partial\cwc$.

\ \\ \para{Bloch Hamiltonians and loop spaces}
A Bloch Hamiltonian is defined as a map $h\colon \cwc \to \Xsp$, which constitutes of a set of continuous maps $\alpha \to \Xsp_\alpha$ for each $p$-cell $\alpha$, so that $h$ is compatible with the CW-complex structure of $\cwc$. We denote the space of such maps by $\Omega^0(\cwc,\Xsp)$.

For the classification procedure, we also need $m$-parameter families or ``$m$-loops'' of Bloch Hamiltonians. We recursively define such an $m$-loop $h^m$ as a map $h^m: \cwc \times [0,1]^m \to \Xsp$ that is compatible with the CW-complex structure of $\cwc$ for each fixed $(t_1,\ldots,t_m) \in [0,1]^{m}$ and which for $m \ge 1$ satisfies the condition 
\begin{equation}
	h^{m}(\vk;t_1,\ldots,t_{m-1},t_{m}) = h^{m-1}_{\rm ref}(\vk;t_1,\ldots,t_{m-1}),
\end{equation}
with a $h^{m-1}_{\rm ref}$ a reference $(m-1)$-loop if $m > 1$ and a reference Bloch Hamiltonian $h^0_{\rm ref} \equiv h_{\rm ref}$ if $m = 1$. With this notation, a Bloch Hamiltonian can be considered a ``$0$-loop''. The space of such $m$-loops is denoted $\Omega^{m}(\cwc,\Xsp)$. We can similarly define the loop spaces for $\partial \cwc$ and any other subcomplex of $\cwc$. Our end goal is the computation of the homotopy set $\pi_0[\Omega^0(\cwc,\Xsp)]$ enumerating homotopy classes of Bloch Hamiltonians. However, we will need homotopy sets of $m$-loops as an intermediate step. 

\ \\ \para{Classification on $\partial\cwc$} 
The computation of $\pi_0[\Omega^m(\cwc,\Xsp)]$ is simplified by the observation that two maps $g^m, h^m \in \Omega^m(\cwc,\Xsp)$ cannot be topologically equivalent if the restrictions $g^m \rvert_{\partial\cwc}, h^m\rvert_{\partial\cwc} \in \Omega^m(\partial\cwc,\Xsp)$ are topologically inequivalent. Furthermore, not all $h^m_\partial\in\Omega^m(\partial\cwc,\Xsp)$ can occur as the boundary restriction of some $h^m\in\Omega^m(\cwc,\Xsp)$. To see this, consider a $d$-cell $\alpha$. Since $\alpha$ is homotopically isomorphic to the $d$-ball, $\partial \alpha$ is homtopically equivalent to $S^{d-1}$, so that the restriction $h^m\rvert_{\partial\alpha}$ defines a map $S^{d-1} \to \Omega^m(\Xsp_\alpha)$. Since $\alpha$ is contractible, $h^m\rvert_{\partial\alpha}$ must be trivial in  $\pi_{d-1}[\Omega^m(\Xsp_\alpha)]$ if there exists a $h^m\in\Omega^m(\cwc,\Xsp)$ such that $h^m_\partial$ is the boundary restriction $h^m$. We refer to this restriction as a \emph{compatibility condition}. We denote the subspace of $\Omega^m(\partial\cwc,\Xsp)$ that satisfies the compatibility condition for all $d$-cells by $\Omega^m(\partial\cwc,\Xsp)\rvert_0$. 

Being in $\Omega^m(\partial\cwc,\Xsp)\rvert_0$ is not only \emph{necessary}, but also a \emph{sufficient} condition for a $h^m_{\partial} \in \Omega^m(\partial\cwc,\Xsp)$ to be the boundary restriction of some $h^m\in\Omega^m(\cwc,\Xsp)$. This is because satisfying the compatibility condition implies that for each $d$-cell $\alpha \in \cwc$ there exists a homotopy $H_\partial^m:\partial\alpha \times [0,1] \to \Xsp$ that interpolates between $h_\partial^m$ and the trivial (\ie, constant) map. Using the homotopic equivalence of $\alpha$ and the $p$-ball, we can interpret $H_\partial^m$ as an extension of $h_\partial^m$ to the full $d$-cell $\alpha$. Continuing this for each $\alpha$, we can thus define an extension of $h_\partial^m$ to $\Omega^m(\cwc,\Xsp)$.

We next compute the equivalence classes of $\Omega^m(\cwc,\Xsp)$ for a fixed homotopic class on $\partial \cwc$, \ie, for a fixed element of $\pi_0\!\left[\,\Omega^m(\partial\cwc,\Xsp)\rvert_0 \,\right]$. We denote the corresponding subspaces of $\Omega^m(\cwc,\Xsp)$ and $\Omega^m(\partial\cwc,\Xsp)\rvert_0$ by $\overline \Omega^m(\cwc,\Xsp)$ and $\overline \Omega^m(\partial\cwc,\Xsp)$, respectively. Hence, by construction $\overline \Omega^m(\partial \cwc,\Xsp)$ is a connected space. The remaining classification problem amounts to the computation of $\pi_0 [\overline \Omega^m(\cwc,\Xsp)]$, which enumerates the connected components of $\overline \Omega^m(\cwc,\Xsp)$.

\ \\ \para{Classification on $\cwc$} 
We choose a reference $m$-loop $h^m_{\partial,{\rm ref}} \in \overline \Omega^m(\partial\cwc,\Xsp)$ and define $\overline \Omega_{\rm ref}^m(\cwc,\Xsp)$ as the set of maps $h^m$ satisfying $h^m \rvert_{\partial\cwc} = h^m_{\partial,{\rm ref}} \forall (t_1,\ldots,t_m) \in [0,1]^{m}$. If we further fix a reference $m$-loop $h^m_\text{ref} \in \overline \Omega_{\rm ref}^m(\cwc,\Xsp)$, then for any $h^m \in \overline \Omega_{\rm ref}^m(\cwc,\Xsp)$ and any $d$-cell $\alpha$, the pair $(h^m, h^m_\text{ref})$ defines a map from $(\alpha\times\alpha) \times [0,1]^m \to \Xsp_\alpha$. Since $h^m$ and $h^m_\text{ref}$ agree on the boundary of $\alpha$, we can identify them on $\partial \alpha$, thereby forming an $m$-loop defined on the $d$-sphere $S^d$. Such $m$-loops can be classified by 
\begin{equation}
	\pi_d[\Omega^m(\Xsp_\alpha)] \cong \pi_{m+d}[\Xsp_\alpha],
\end{equation}
where we have used the Freudenthal suspension theorem \cite[Corollary 4.24]{hatcher2002}. Combining classification spaces from all $d$-cells in $\cwc$, we get 
\begin{equation}
 	\pi_0[\overline\Omega_{\rm ref}^m(\cwc,\Xsp)] = \prod_{d-{\rm cells}\, \alpha\in\cwc} \pi_{m+d} [\Xsp_\alpha]. 
	\label{eq:pi_0_def_ref}
\end{equation}

To compute $\pi_0[\overline\Omega^m(\cwc,\Xsp)]$, we use that the natural inclusion $i\colon \overline\Omega^m_{\rm ref}(\cwc,\Xsp) \hookrightarrow \overline\Omega^m(\cwc,\Xsp)$ induces a map
\begin{equation}
	\pi_0\! \left[ \overline\Omega^m_{\rm ref}(\cwc,\Xsp) \right] \xrightarrow{\;\;\; i_* \;} \pi_0\! \left[ \overline\Omega^m(\cwc,\Xsp) \right]. \label{eq:istar}
\end{equation}
This map is surjective, because, by construction, the restriction $h^m|_{\partial \cwc}$ of any $m$-loop $h^m \in \overline \Omega^m(\cwc,\Xsp)$ can always be continuously deformed to $h^m_{\partial,{\rm ref}}$ and ths deformation can be extended to a homotopy between the full $m$-loop $h^m$ and an $m$-loop in $\overline \Omega_{\rm ref}^m(\cwc,\Xsp)$ using the construction of Fig.\ \ref{fig:boundary_transform}. Since $i_*$ is surjective, it follows that
\begin{equation}
	\pi_0\! \left[ \overline\Omega^m(\cwc,\Xsp) \right] \cong \pi_0\! \left[ \overline\Omega^m_{\rm ref}(\cwc,\Xsp) \right] / \sim,  
	\label{eq:pi_0_def}
\end{equation}
where $\sim$ denotes the equivalence classes of elements of  $\pi_0\! [ \overline\Omega^m_{\rm ref}(\cwc,\Xsp) ]$ whose images under $i_\ast$ are identical.

\begin{figure}
	\includegraphics[width=0.99\columnwidth]{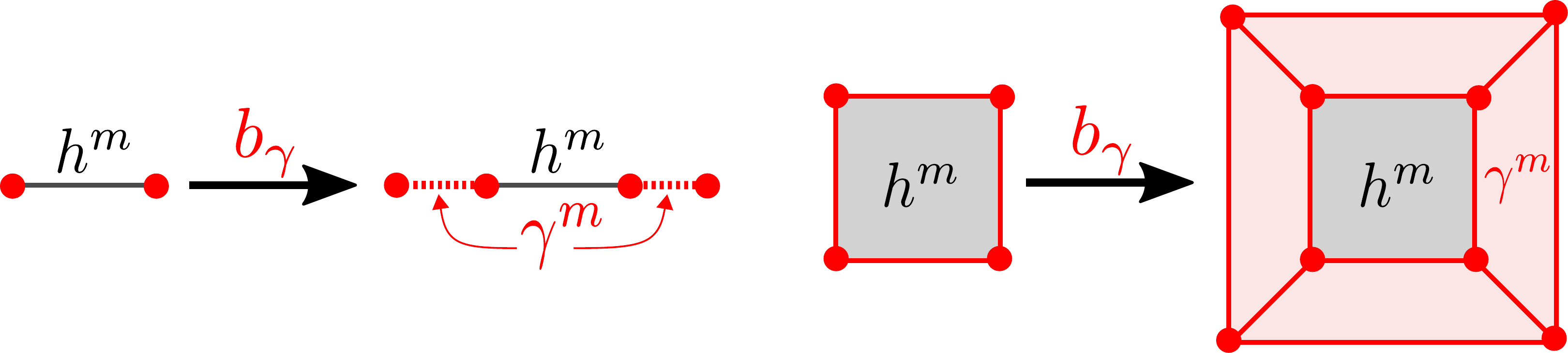}
	\caption{\label{fig:bijection} Graphical representation of the bijection $b_{\gamma*}$ on $\pi_0 \left[\Omega_{\rm ref}^m(\cwc,\Xsp) \right]$ induced by an element a loop $\gamma$ on $\Omega^m(\partial\cwc,\Xsp)$ for a 1-cell (left) and a 2-cell(right).
	}
\end{figure}

\ \\ \para{Quotienting out the boundary deformations}
To characterize the equivalence classes defined in Eq.~(\ref{eq:pi_0_def}), we first consider a pointed loop $\gamma^m$ in $\overline \Omega^{m+1}(\partial\cwc, \Xsp)$ with the reference $h^{m}_{\partial,{\rm ref}}$, \ie, a family of $m$-loops $\gamma^m(t) \in \Omega^{m}(\partial\cwc, \Xsp)$ parameterized by $t \in [0,1]$, such that $\gamma^m(0) = \gamma^m(1) = h^{m}_{\partial,{\rm ref}}$. Given $h^m \in \overline\Omega^m_{\rm ref}(\cwc,\Xsp)$, we can ``attach'' $\gamma^m$ to $h^m$ as shown graphically in Fig.~\ref{fig:bijection} for one- and two-dimensional CW-complexes. We term this process a {\em boundary deformation} of $h^m$.

More formally, boundary deformation corresponding to the pointed loop $\gamma^m \in \overline \Omega^{m+1}(\partial\cwc, \Xsp)$ induces the map 
\begin{equation}
	\overline\Omega^m_{\rm ref}(\cwc,\Xsp) \xrightarrow{\;\;\; b_\gamma \;} \overline\Omega^m_{\rm ref}(\cwc,\Xsp).
	\label{eq:b_def}
\end{equation}
This further induces a map between the homotopy sets  
\begin{equation}
	\pi_0\! \left[ \overline\Omega^m_\text{ref}(\cwc,\Xsp) \right] \xrightarrow{\;\;\; b_{\gamma\ast} \;}  \pi_0\! \left[ \overline\Omega^m_\text{ref}(\cwc,\Xsp) \right]. \label{eq:b_hom}
\end{equation} 
This map is well defined, since if $g^m, h^m \in \overline\Omega_{\rm ref}^m(\cwc,\Xsp)$ are homotopically equivalent, they remain so after identical boundary deformations. 
Since any loop $\gamma^m$ has an inverse obtained by traversing it in the opposite direction, $b_{\gamma\ast}$ is a bijection.

We now show that for $g^m$, $h^m \in \overline\Omega_{\rm ref}^m(\cwc,\Xsp)$, the topological equivalence classes $[g^m] \sim [h^m]$ under the equivalence relation of Eq.~(\ref{eq:pi_0_def}) \emph{if and only if} there exists a pointed loop $\gamma^m \in \overline\Omega^{m+1}(\partial\cwc, \Xsp)$ such that $b_{\gamma\ast}([g^m]) = [h^m]$. To this end, we first note that attaching the loop $\gamma^m$ to $h^m \in \overline\Omega^m_{\rm ref}(\cwc,\Xsp)$ naturally provides a homotopy between $h^m$ and $b_{\gamma}(h^m)$ in $\overline\Omega^m(\cwc,\Xsp)$ (but not necessarily in $\overline\Omega^m_{\rm ref}(\cwc,\Xsp)$), so that the images in $\pi_0\! [ \overline\Omega^m_\text{ref}(\cwc,\Xsp)]$ of the topological equivalence classes $[h^m]$ and $b_{\gamma\ast}[h^m]$ under $i_\ast$ are identical. To show that the converse is also true, we consider  $g^m$, $h^m \in \overline\Omega^m_\text{ref}(\cwc,\Xsp)$ such that $i_\ast[g^m] = i_\ast[h^m]$, \ie, $g^m$ and $h^m$ are homotopically equivalent in $\overline\Omega^m(\cwc,\Xsp)$. Thus, there exists a homotopy $I^m(t)$ such that 
\[
	I^m(\,\cdot\,, 0) = g^m, \qquad 
	I^m(\,\cdot\,, 1) = h^m. 
\]
Since the boundary restrictions of $g^m$ and $h^m$ both equal $h^m_{\partial, \text{ref}}$, the restriction $I^m \rvert_{\partial\cwc}$ defines a boundary deformation in $\overline\Omega^{m+1}(\partial\cwc, \Xsp)$. Setting $\gamma^m \equiv I^m \rvert_{\partial\cwc}$, the existence of $I$ implies that $h^m$ can be homotopically deformed into $b_\gamma(g^m)$, \ie, $b_{\gamma\ast} \colon [g^m] \mapsto [h^m]$. 

From the definition of $b_{\gamma\ast}$ we note that $b_{\gamma\ast}$ depends only on the homotopic equivalence class of $\gamma^m$. These homotopic equivalence classes are classified by 
\begin{equation}
	\pi_1 \! \left[ \overline\Omega^m(\partial\cwc, \Xsp) \right] \cong	\pi_0 \! \left[ \overline\Omega^{m+1}(\partial\cwc, \Xsp) \right].
\end{equation}
Hence, to compute the equivalence class under $\sim$, it is sufficient to consider the induced map $b_{\gamma\ast}$ for one boundary deformation $\gamma^m$ of each equivalence class in $\pi_0 \! [ \overline\Omega^{m+1}(\partial\cwc, \Xsp) ]$. 

\ \\ \para{The full classification}
We can complete the classification by computing $\pi_0[\overline\Omega^0(\cwc,\Xsp)]$ using Eqns.~(\ref{eq:pi_0_def_ref}) and (\ref{eq:pi_0_def}) for each element of $\pi_0\![\,\Omega^0(\partial\cwc,\Xsp)\rvert_0 \,]$, \ie, for each possible lower-dimensional invariant. This requires knowledge of $\pi_0\!\left[\,\Omega^0(\partial\cwc,\Xsp)\rvert_0 \,\right]$, \ie., of the homotopic classification on boundary $\partial \cwc$. Further, as discussed above, to evaluate the quotient in Eq.~(\ref{eq:pi_0_def}), we need to know $\pi_0 \!\left[ \Omega^{1}(\partial\cwc,\Xsp) \right]$. Since $\partial \cwc$ is itself a $(d-1)$-dimensional CW complex, both $\pi_0\![\,\Omega^0(\partial\cwc,\Xsp)\rvert_0 \,]$ and $\pi_0 \![ \Omega^{1}(\partial\cwc,\Xsp) ]$ can be calculated using the procedure described above. Proceeding recursively, we get classification sets with increasing loop order $m$, while lowering the dimension of the CW-complexes. We eventually end up with a 0-dimensional CW-complex, \ie, a disjoint set of points, the classification on which can be performed by elementary means.  

This recursive procedure is greatly simplified by noting that it is not necessary to precisely determine $\pi_0 \!\left[ \Omega^{m+1}(\partial\cwc,\Xsp) \right]$ in order to obtain $\pi_0 \!\left[ \Omega^m(\cwc,\Xsp) \right]$ from Eq.~(\ref{eq:pi_0_def}). Instead, it suffices to use the easier-to-compute classification sets $\pi_0 \![ \overline\Omega_\text{ref}^{m+1}(\partial\cwc,\Xsp) ]$, thereby terminating the recursive procedure after one step. This simplification is possible because $i_*$ is surjective and it is sufficient to consider a collection of boundary deformations $\gamma^m \in \overline\Omega^{m+1}(\partial\cwc, \Xsp)$ that is large enough to include {\em at least one} boundary deformation in each topological equivalence class.

\section{Homotopy groups of Grassmannians}
\label{app:Gr}
The Grassmannian $\Gr_\FF(m,n)$ is a manifold that parameterize the $n$-dimenional linear subspaces of a $(m+n)$-dimensional vector space over the field $\FF = \RR,\CC,\mathbb{H}$. They can be explicitly written as a quotient of groups as 
\begin{equation}
  \Gr_F(m,n) = \frac{G(m+n)}{G(m)\times G(n)},  
\end{equation}
where $G$ denotes the orthogonal, unitary or symplectic group for $F = \RR,\CC,\mathbb{H}$, respectively. The homotopy spaces of the Grassmannians can thus be computed by starting with the short exact sequence of groups 
\begin{align*}
	1 \to G(m)\times G(n)
	\to G(m+n)
	\to \Gr_\FF(m,n) 
	\to 1. 
\end{align*}
This defines the long exact sequence of homotopy spaces \cite[{Sec 4.2}]{hatcher2002}
\begin{align}
	\dots &\to \pi_p[G(m)] \times \pi_p[G(n)]
	\to \pi_p[G(m+n)] \nonumber \\ 
	&\to \pi_p[\Gr_F(m,n)] \to \pi_{p-1}[G(m)] \times \pi_{p-1}[G(n)] \to \dots \nonumber
\end{align}
The homotopy groups of $G(n)$ for $G = \mathrm{O,U,} \text { and } \Sp$ are well-known. They can all be computed in terms of the known homotopy groups of spheres using the short exact sequence \cite[{Sec 4.2, Example 4.55}]{hatcher2002}
\begin{equation}
  1 \to G(n) \to G(n+1) \to S^{r(n)} \to 1,
\end{equation}
where $r(n) = n, 2n+1,$ and $4n+3$ for $G = \mathrm{O,U,} \text { and } \Sp$, respectively. This leads to a long-exact sequence of homotopies 
\begin{align}
	\dots \to \pi_p[G(n)] \to \pi_p[G(n+1)] \to \pi_p[S^{r(n)}] \to\dots \nonumber 
\end{align}
But since $\pi_p[S^k] = 0$ for $k>p$, this long exact sequence splits into 
\begin{equation*}
	0 \to \pi_p[G(n)] \to \pi_p[G(n+1)] \to 0. 
\end{equation*}
for $p > r(n)$. Thus, $\pi_p[G(n)] \cong \pi_p[G(n+1)]$ for $p > r(n)$, which is the stabilization of the homotopy groups of $G(n)$ for $n\to\infty$.

\section{The Hopf invariant}
\label{app:Hopf}
The continuous maps from the 3-sphere $S^3$ to the complex Grassmannian $\Gr_{\CC}(1,1) \cong \CC\mathbb{P}^1 \cong S^2$ are classified by the third homotopy group 
\begin{equation}
	\pi_3[\Gr_{\CC}(1,1)] = \pi_3[S^2] = \ZZ.
	\label{eq:pi3Hopf}
\end{equation}
The corresponding topological invariant is termed the \emph{Hopf invariant} $\Hopf$. For a given map $h\colon S^3 \to S^2$, it can be calculated as the linking number of the preimages under $h$ of two distinct points in $S^2$. 

More explicitly, given a point $x\in S^2$, the preimage $\gamma = h^{-1}(x)$ defines a (single or multiple) closed loop(s). We take two distinct points $x'$ and $x''$ infinitesimally close to $x$ and denote their preimages under $h$ by $\gamma'$ and $\gamma''$, respectively. We then assign a directionality to $\gamma$ for each point on it by computing the cross product of the vectors pointing towards $\gamma'$ and $\gamma''$. The same direction may be assigned to $\gamma'$ and $\gamma''$, as these curves are (infinitesimally) close to $\gamma$. The Hopf number $\Hopf$ is then given by the linking number of the directed curves $\gamma$ and $\gamma'$. Note that $\gamma''$ only serves to define the direction along $\gamma$ and $\gamma'$, but is otherwise not involved in the the calculation of the linking number.

An example of a map with $\Hopf=1$ can be constructed by taking $S^3$ as the 3-ball $B^3 = \{ \vr|\, |\vr| \le 1\}$ with its surface identified to a single point. On $B^3$, we then consider the ``band structure'' 
\begin{equation}
	h(\vr) = \e^{-\i \pi \vr \cdot \vtau} h_0 \e^{\i \pi \vr \cdot \vtau},
	\label{eq:HB3}
\end{equation}
where the $\tau_{\alpha}$, $\alpha = 1, 2, 3$ are the Pauli matrices and $h_0 = \vh_0\cdot\vtau$ is a fixed gapped Hamiltonian. More explicitly,
\begin{equation}
	h(\vr) = \cos(2\pi r) h_0 - \sin(2\pi r) \vh_0 \times \vr 
\end{equation}
Setting $h_0 = \tau_3$ in the following, the preimages $\gamma$, $\gamma'$, and $\gamma''$ of $h = -\tau_3$, $h' = -\tau_3 + \pi^2 \eta \tau_2$, and $h'' = -\tau_3 + \pi^2 \eta \tau_1$, with $\eta \downarrow 0$ a positive infinitesimal are the curves parameterized by
\begin{align}
	\vr(\phi) =&\, (1/2)(\cos \phi,\sin \phi,0), \nonumber \\
	\vr'(\phi) =&\, (1 + \eta \cos \phi) \vr(\phi) + \eta \sin \phi (0,0,1), \nonumber \\
	\vr''(\phi) =&\, (1 - \eta \sin \phi)\, \vr(\phi) + \eta \cos \phi (0,0,1),
\end{align}
with $0 \le \phi < 2 \pi$. 
The curves are directed in the positive $\phi$ direction. The Hopf invariant $\Hopf = 1$ is the linking number of the curves $\gamma$ and $\gamma'$.

\section{Boundary deformations} 
\label{app:deform}
This Appendix contains details on boundary deformations that may change the parent topological invariants at level-$p$. We thus consider families of Hamiltonians $H_\partial(\vk_\partial,t)$, where $t\in[0,1]$ and $\vk_\partial$ the coordinate along the boundary of a $p$-cell $\alpha$, such that for any fixed $t$, $H_\partial$ satisfies the symmetry constraints at the $q$-cells constituting $\partial\alpha$.

\begin{figure}
	\includegraphics[width=0.9\columnwidth]{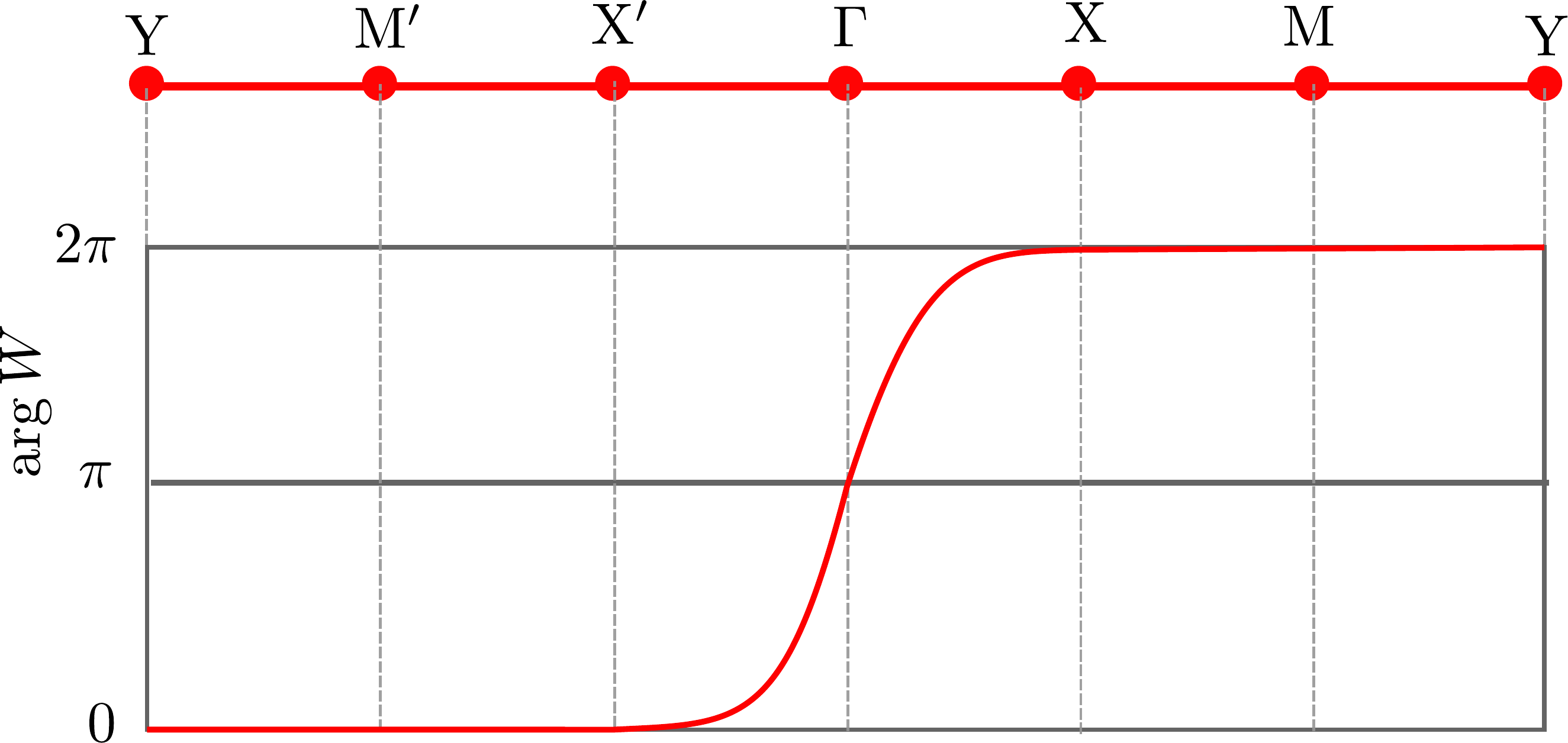}
	\caption{\label{fig:wilson_winding} 
		The winding of the Wilson loop for a boundary deformation with nontrivial topology at $\G$. Since $\arg\wilson$ is pinned to $\pi$ at $\G$ and time-reversal requires antisymmetry about it, the total winding must be odd. (
		see Fig.\ \ref{fig:fundamental_general_d2} for the definition of the high-symmetry points)}
\end{figure}

\subsection{The Chern number}
\label{app:deform_Chern}
The $\ZZ$-valued Chern number is the level-2 parent invariant when the Hamiltonian is complex-valued. In the following, we consider the Chern number over the 2-cell for a two-dimensional system with spinless time-reversal symmetry (symmetry class AI). We first present general arguments that a boundary deformation may change the parent Chern number $\Chern$ by an arbitrary integer and then construct an explicit deformation that changes $\Chern$ by one. 

Consider a general boundary deformation $H_\partial(\vk_\partial,t)$, where $\vk_\partial$ denotes the momentum along the boundary and $t$ is the homotopy parameter. Time-reversal symmetry implies that $H_\partial(\vk_\partial,t) = \uop_\T^\pdg H^\ast_\partial(-\vk_\partial,t) \uop_\T^\dagger$ for some unitary $\uop_\T$, where $-\vk_{\partial}$ is the time-reversed momentum of $\vk_{\partial}$, up to a translation by a reciprocal lattice vector. Thus, the Wilson loops  $\wilson_{\partial}(\vk_\partial)$ describing parallel transport along $t$ at fixed $\vk_{\partial} \in \partial \cwc$ are related to those at $-\vk_\partial$ as $\wilson_{\partial}(\vk_\partial) = \wilson_{\partial}^*(-\vk_\partial)$, so that  
\begin{equation}
	\arg \det \wilson_{\partial}(\vk_\partial) = -\arg\det \wilson_{\partial}(-\vk_\partial) \mod 2\pi.
\end{equation}
Thus, the windings from boundary segments related by time-reversal {\em add up}. This observation immediately allows one to construct boundary deformations with even Chern numbers. 

To construct a boundary deformation with odd Chern numbers, we  choose a basis where $\uop_\T = \id$, so that the Hamiltonian is real at the high-symmetry points $S$. The corresponding Wilson loops are orthogonal, so that $\det\wilson_{\partial}(\vk_S) \in \{\pm1\}$.  If $\det \wilson(\vk_S) = -1$ at an odd number of 0-cells, then $\wilson(\vk_{\perp})$ must have an odd winding number. This is illustrated in Fig.~\ref{fig:wilson_winding} for a minimal boundary transformation, which has winding number $1$. 

The existence of such a deformation can be traced back to nontrivial topology at one of the 0-cells, since $\pi_1[\Xsp_S] = \pi_1[\Gr_{\RR}(n_{\o},n_{\e})] \cong \ZZ_2$ in the stable limit. Thus, we can construct a boundary deformation carrying a nonzero Chern number that is constant everywhere except in a neighborhood of a single high-symmetry point $S$. We denote the 1d coordinate along the boundary near $S$ by $\theta$ such that $S$ lies at $\theta=0$, and define a deformation $H_\partial(\theta, t)$ as $\tau_3$ (independent of $t$) for $\abs\theta \geq \pi$ and  
\begin{align}
	H_\partial(\theta, t) =&\, \e^{\i \pi t \Gamma\left(\theta/2\right)} \tau_3 \e^{-\i \pi t \Gamma\left(\theta/2\right)},
\end{align}
for $\theta \in (-\pi,\pi)$, where $\Gamma(\theta) = \tau_2\cos\theta - \tau_3\sin\theta$. We can verify that 
\[
H_\partial(\pm\pi, t) = H_\partial(\theta,0) = H_\partial(\theta,1) = \tau_3
\]
and $H^\ast_\partial(\theta,t) = H_\partial(-\theta, t)$, so that the boundary constraint and the time-reversal symmetry are satisfied. The wavefunction for the occupied band is given by 
\begin{align*}
	\psi_\o(\theta,t) =&\, \e^{\i \pi t \Gamma\left(\theta/2\right)} 
	\begin{pmatrix} 0 \\ 1 \end{pmatrix}
	\nonumber \\
	=&\, \e^{\i \pi t \sin\left(\theta/2\right)} \e^{\i \pi t \cos\left(\theta/2\right) \tau_2} 
	\begin{pmatrix} 0 \\ 1 \end{pmatrix}. 
\end{align*}
The corresponding Wilson loop is 
\begin{equation}
	\wilson^\o (\theta) = - \e^{\i \pi \sin\left(\theta/2\right)}.
\end{equation}
Thus, at the high symmetry point $\theta=0$, we have $\wilson^\o(0) = -1$. Upon taking $\theta$ over the interval $[-\pi,\pi]$, the phase of $\wilson^o(\theta)$ winds once, corresponding to a Chern number $+1$.

\subsection{The second Euler/Stiefel-Whitney number}
The $\ZZ$-valued second Euler invariants $\Euler_2^{\o/\e}$ are the level-2 parent invariants when the Hamiltonian is real-valued and the occupied/empty subspace consists of exactly two bands. In the following, we construct a boundary deformation that is nontrivial only in the neighborhood a high-symmetry point and carries a nonzero second Euler invariant. Such a deformation thus adds integers of the same parity to $\Euler_2^{\o/\e}$. This boundary deformation is relevant for $\C_2$ and $\C_4$ symmetric band structure, whereby the rotation symmetry relates the two 1-cells adjacent to the high-symmetry point.

Explicitly, we assume that the rotation operator at the high symmetry point $\rotn_S$, restricted to the subspace of interest $\rotn_S^{\o/\e}$, contains one positive and one negative eigenvalue. We work in the eigenbasis of the rotation operator, so that $\rotn_S = \rho_3\otimes\tau_0$, where $\rho_\mu$ and $\tau_\mu$ are Pauli matrices. We again choose the 1d coordinate along the 1-cells near $S$ by $\theta$, with $S$ located at $\theta = 0$. We then consider a boundary deformation
\begin{align}
	H^{\pm}_{t}(\theta) =&\,
	\e^{-\i (\pi/4) \sin \left(\theta/2\right) \Sigma_{12}} \, 
	h^{\pm}_{t}(\theta) \, \e^{\i (\pi/4) \sin \left(\theta/2\right) \Sigma_{12}}, 
	\label{eq:hplusminus}
\end{align}
where $\Sigma_{12} = \rho_1 \otimes \tau_2 - \rho_2 \otimes \tau_1$ and 
\begin{equation}
	h'^{\pm}_{t}(\theta) = \e^{\i \pi t \Gamma_\pm(\theta)}(\rho_0 \otimes \tau_3) \e^{-\i \pi t \Gamma_\pm(\theta)}
\end{equation}
with 
\begin{align}
	\Gamma_-(\theta) =&\, \rho_2 \otimes \tau_0 \sin(\theta/2) + \rho_3 \otimes \tau_2 \cos(\theta/2), \nonumber \\
	\Gamma_+(\theta) =&\, \rho_2 \otimes \tau_3 \sin(\theta/2) + \rho_0 \otimes \tau_2 \cos(\theta/2).
\end{align}
One verifies that $H^{\pm}_{t}(\theta) = \rho_3 H_{t}(-\theta) \rho_3$ and that $H^{\pm}_0(\theta) = H^{\pm}_1(\theta)$. At the high-symmetry point $\theta = 0$, $H^{\pm}_{t}(\theta)$ satisfies the symmetry constraints and has one occupied and one unoccupied band of each symmetry type. At $\theta=\pm \pi$, one has $H^{\pm}_{t}(\pi) = \rho_3 \otimes \tau_0$, independent of $t$. As a function of $t$, the two blocks of $H^{\pm}$ at $\theta = 0$ have winding numbers 1 and $\pm 1$. To find the Wilson loops at fixed $\theta$, we note that they are the same for $H^{\pm}$ and $h^{\pm}$, since $H^{\pm}$ and $h^{\pm}$ differ only by a $t$-independent basis transformation. Hence, 
\begin{align}
	\wilson_{\pm}^\o (\theta) &= - \e^{\mp i \pi \rho_2 \sin(\theta/2)}, \nonumber \\  
	\wilson_{\pm}^\e (\theta) &= - \e^{\i \pi \rho_2 \sin(\theta/2)}.	
\end{align}
At $\theta = 0$, all Wilson loops are equal to $-\rho_0$, which is topologically nontrivial and consistent with the winding numbers of $h^{\pm}$ being odd. Looking at the full family of Wilson loops $\wilson^{\o/\e} (\theta)$for $-\pi \le \theta \le \pi$, we see that $H^{\pm}$ generates winding number pairs $(1,1)$ and $(-1,1)$ for occupied and unoccupied orbitals, thereby yielding the desired result.

\subsection{The Hopf invariant}
The $\ZZ$-valued Hopf invariant is the parent invariant at level-3 when the Hamiltonian is complex-valued on the 3-cell and $n_\o = n_\e = 1$. In the following, we describe boundary deformations that may affect $\Hopf$. 

\ \\ \para{No additional symmetries} 
In this case, the effect of boundary deformations on $\Hopf$ depends on the level-2 invariants, \ie, the Chern numbers $\C_{x,y,z}$. To describe a general deformations, we first note that while the preimages of points for the Hopf map $S^3 \to S^2$ are closed curves in $S^3$, those for a 3-cell may start and end at the boundaries of $\cwc$. The difference between the number of times a component of a preimage $\gamma$ starts and ends at the 2-cell perpendicular to the coordinate direction $\alpha = x$, $y$, $z$ is equal to the Chern number $\Chern_{\alpha}$. Since $\Hopf$ is given by the linking number of a pair of preimages, we must construct a deformation that changes this linking number. 

\begin{figure}[t]
	\includegraphics[width=0.5\columnwidth]{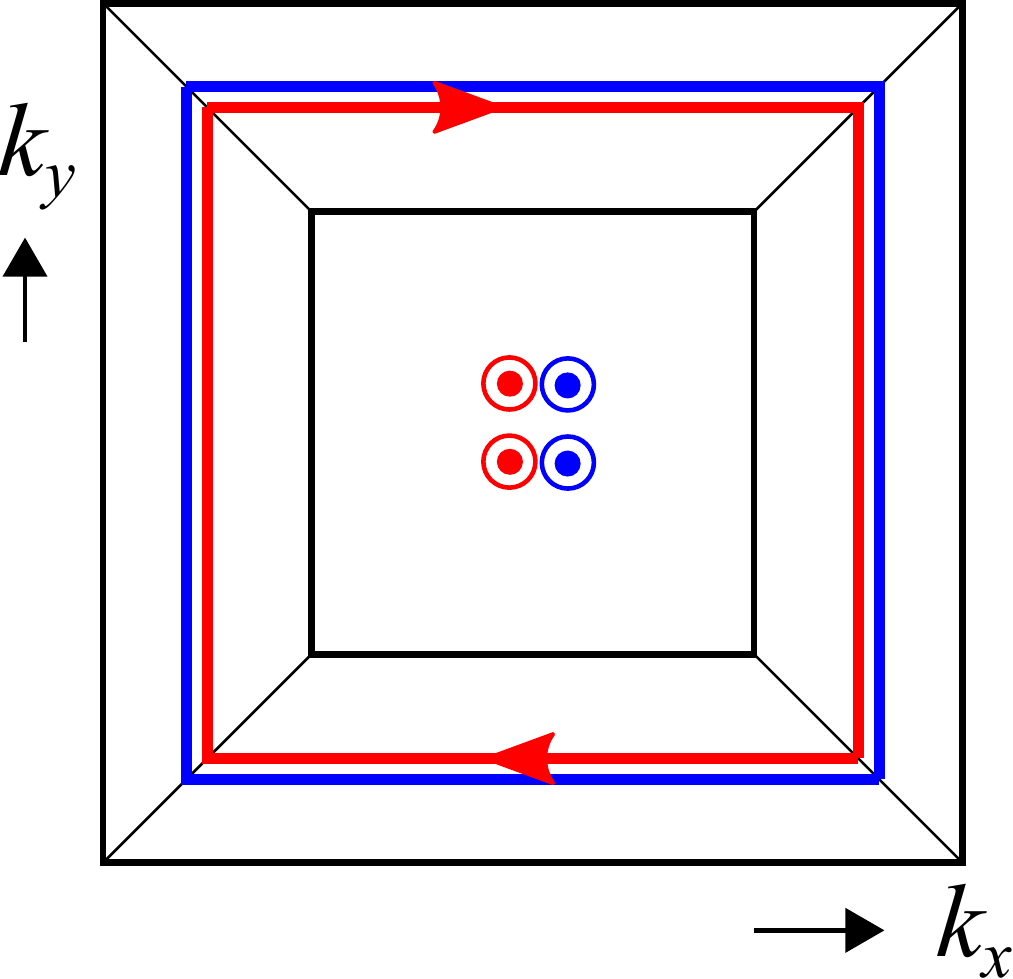}
	\caption{\label{fig:Hopf_deformation} Boundary deformation to change the Hopf invariant for a nonzero Chern number $C_z = 2$. The square represents a fixed-$k_z$ cut of the 3-cell. The Chern number implies the existence of two directed segment pairs (shown red and blue) along the $k_z$-axis. The deformation nucleates a closed curve pair at the boundary,which encircles two curve pair segments running in the $z$ direction, thereby changing the linking number  --- and hence the Hopf invariant --- by $2\Chern_z = 4$. 
	}
\end{figure}

In Fig.~\ref{fig:Hopf_deformation}, we depict such a deformation for the case of $\Chern_x = \Chern_y = 0$, $\Chern_z = 2$, which implies the existence of two pairs of directed segments along $z$. We can then define a boundary deformation of the vertical faces, so that the deformation of each 1-cell carries a Chern number $+1$. This necessitates the existence of a pair of closed preimage curves, along the boundary. These curves encircle the two curve pairs running along $z$, thereby changing the linking number by $2 \Chern_z$. A similar deformation can be defined for Chern numbers in the other directions, so that the Hopf invariant can be changed by $n = 2 \mbox{gcd}(\Chern_x, \Chern_y, \Chern_z)$. 



\ \\ \para{With time-reversal symmetry} 
We explicitly construct a boundary deformation that changes $\Hopf$ by one and is nontrivial only in the vicinity of a high symmetry point $S$ at $\vk_S = (0,0,\pi)$.  We introduce polar coordinates $(\rho,\phi)$ in the $k_x$-$k_y$ plane with the origin at $S$ and define a deformation $H_t(\rho,\phi)$ that is independent of $t$ for $\rho>1$. Explicitly, we take 
\begin{equation}
	H_t(\rho,\phi) = \e^{\i \pi \Gamma(t,\rho,\phi)} \tau_3 \e^{-\i \pi \Gamma(t,\rho,\phi)},
\end{equation}
with
\begin{equation}
	\Gamma(t,\rho,\phi) = \cos(\pi t) \tau_2 + \rho \sin(\pi t)(\tau_1 \cos \phi + \tau_3 \sin \phi),
\end{equation}
This deformation then meets all required constraints, $H_t(\rho,\phi)= \tau_3$ for $t = 0$, 1 and for $\rho = 1$ and $H_t(\rho,\phi) = H_t(\rho,\phi + \pi)^*$. The boundary deformation $H_t(\rho,\phi)$ is a continuous deformation of the band structure $h$ of Eq.\ (\ref{eq:HB3}). Hence, it has unit Hopf number. 
This can be seen by noting that $H_t(\rho,\phi)$ is a deformation of the matrix-valued function $h(\vr) = \e^{\i \pi \vr \cdot \vtau} \tau_3 \e^{-\i \pi \vr \cdot \vtau}$, which is defined on the 3-ball $|\vr| \le 1$ and which the paradigmatic example of a map with unit Hopf number.

\ \\ \para{With $\C_2 \T$ symmetry}
In this case, the Hamiltonian is real for $k_z = 0$ and $k_z = \pi$. On the one hand, this implies that the Chern number $C_z = 0$, so that boundary deformations along the $xz$ and $yz$ faces of the fundamental domain do not change the Hopf number. On the other hand, this prohibits a boundary deformation along the $xy$ and $xz$ faces or along the $xy$ and $yz$ faces to nucleate a closed curve pair that encloses the fundamental domain. Hence, neither boundary deformations can change the Hopf number, so that Hopf number is a robust topological index if $n_\o = n_\e = 1$.

\ \\ \para{With $\C_2$ or $\C_4$ symmetry} 
In this case, there can be a curve pair penetrating the fundamental domain in the $k_z$ direction only, so that boundary deformations associated with faces parallel to the $k_z$ axis cannot change the parent Hopf number. Hence, with broken time-reversal symmetry the Hopf number is defined modulo $2 \Chern$, $\Chern$ being the Chern number associated with a plane at constant $k_z$.
With time-reversal symmetry, the Hamiltonian is real for $k_z = 0$ and $k_z = \pi$, which also rules out curve pairs penetrating the fundamental domain in the $k_z$ direction. Hence, in this case the Hopf invariant is an integer.

\section{Lattice model for the $\D_4$-symmetric representation protected stable topological insulator}
\label{app:D4-model}
In this section, we construct lattice models that exhibit the $\D_4$-symmetric representation protected stable topological phase described in Sec.~\ref{sec:D4}. 

\subsection{Two dimensions} 
Consider a system with two atoms at the Wyckoff position ``a'', each with a doublet that transforms under the 2d irrep $B$ of $\D_4$. The symmetry operators are thus 
\begin{equation}
	\rotn_4 = \i\sigma^2\tau^0, \qquad 
	\mirror_x = \sigma^3 \tau^0,
\end{equation}
where the Pauli matrices $\sigma^\mu$ and $\tau^\mu$ denote the orbital and site degrees of freedom, respectively. The most general 4-band Bloch Hamiltonian can be written as 
\begin{equation}
	H(\vk) = h_{\mu\nu}(\vk) \sigma^\mu \tau^\nu, \quad \mu,\nu \in \{0,1,2,3\}, 
	\label{eq:lattice_hlt_gen}
\end{equation}
where the functions $h_{\mu\nu}(\vk)$ are real-valued and periodic in $\vk$, since we only consider the Wyckoff position ``a'' (see Appendix~\ref{app:symm_D4}). Next, as $\rotn_2 \propto \id_4$, $\C_2\T$ that $H(\vk)$ be real-valued, restricting the allowed matrices in Eq.~(\ref{eq:lattice_hlt_gen}) to $\sigma^a \tau^b, \, a,b\in\{0,1,3\}$ and $\sigma^2\tau^2$. Finally, under  $\rotn_4$ and $\mirror_x$, $\sigma^\mu \tau^\nu$ transform up to a sign, so that the functions $h_{\mu\nu}(\vk)$ must have the same behavior. A convenient set of such functions is given by: 
\begin{align}
	f_{p,q}^\pm(\vk) &\equiv \cos(p k_x) \cos(q k_y) \pm \cos(q k_x) \cos(p k_y), \nonumber \\ 
	g_{p,q}^\pm(\vk) &\equiv \sin(p k_x) \sin(q k_y) \pm \sin(q k_x) \sin(p k_y), 
	\label{eq:fg_def}
\end{align}
where $0 \leq p \leq q \in\ZZ$.  Under $\mirror_x$, $f_{p,q}^\pm$ remain invariant while $g_{p,q}^\pm$ change sign, and under $\rotn_4$, $f_{p,q}^+$ and $g_{p,q}^-$ remain invariant while $f_{p,q}^-$ and $g_{p,q}^+$ change sign. Taking these into account, we can write $h_{\mu\nu}(\vk)$ as a linear combination of only one kind of functions defined in Eq.~(\ref{eq:fg_def}). 

Accounting for the constraints described above, we set 
\begin{align}
	H(\vk) =&\, \left[ m - f^+_{0,1}(\vk) \right] \sigma^0\tau^1 + t g^+_{1,2}(\vk) \sigma^1 \tau^3 
  \nonumber \\ &\, \mbox{}
+ t' f^-_{0,2}(\vk) \sigma^3 \tau^3 \nonumber \\ 
	=&\, \left( \cos k_x + \cos k_y \right) \left[ -\sigma^0 \tau^1 + 2 t \sin k_x \sin k_y \sigma^1 \tau^3 \right. \nonumber \\ 
	&\, \quad\left. - 2 t' (\cos k_x - \cos k_y)\sigma^3 \tau^3 \right] + m \sigma^0 \tau^1.   \label{eq:H2d_def} 
\end{align}
We now show that $H(\vk)$ has two nontrivial level-1 invariants defined by Eq.~(\ref{eq:SW_rpti1}). These can be computed from the two parity sectors of $H(\vk)$ at the high-symmetry lines, which correspond to eigenvalues $\pm1$ of the mirror operators (see Eq.~(\ref{eq:D4_M_SS'})) given by 
\[ 
	\mirror_{\G\M} = \sigma^1\tau^0, \quad 
	\mirror_{\G\X} = \sigma^3\tau^0, \quad 
	\mirror_{\X\M} = -\sigma^3\tau^0.
\]
The two parity sectors can be explicitly written as 
\begin{align}
	h^\pm_{\G\M}(k) &= (m - 2\cos k) \tau^1 \pm 4 t \sin^2 k \cos k \,\tau^3, \nonumber  \\ 
	h^\pm_{\G\X}(k) &= (m - 1 - \cos k) \tau^1 \pm 2 t' \sin^2 k \,\tau^3, \nonumber  \\ 
	h^\pm_{\X\M}(k) &= (m + 1 - \cos k) \tau^1 \pm 2 t' \sin^2 k \,\tau^3,
\end{align}
where the argument $k \in [0,\pi]$ of $h^\pm_{SS'}$ denotes the coordinate along $SS'$. 
For $t,t' \neq 0$, all of these are gapless when $m=0, \pm2$ and trivial for $m\to\pm\infty$, so that we may get a topological phase for $0 < \abs{m} < 2$.

\begin{figure}
	\includegraphics[width=0.7\columnwidth]{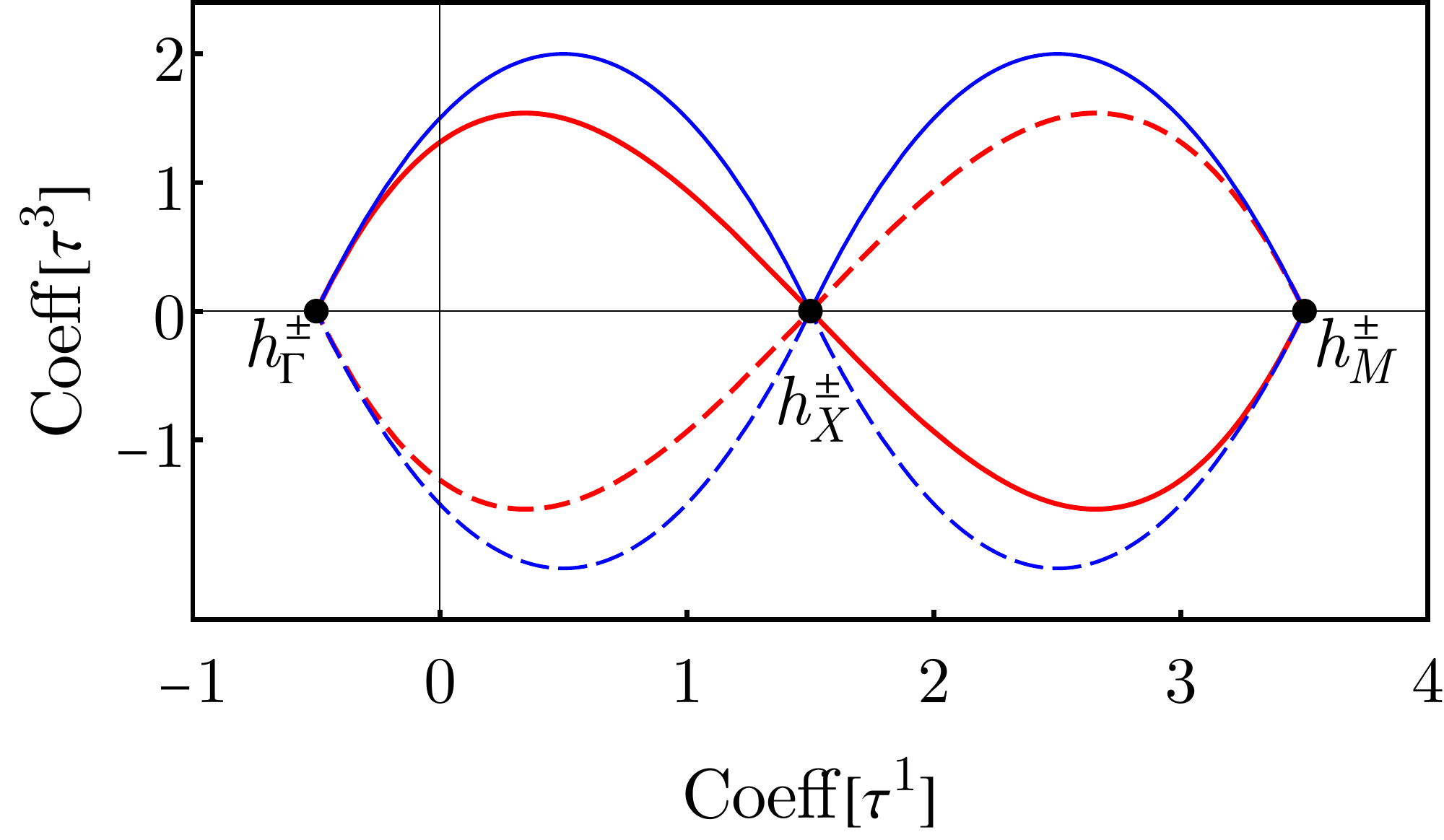} 
	\caption{\label{fig:D4_winding} 
	The windings of $h_{\G\M}^\pm(k)$ for $m = 1.5$ for the positive (solid) and negative (dashed) subspaces along the $\G\M$ (red) and $\G\X\M$ (blue). The two parities on $\G\M$ taken together have a winding number $-1$, while the positive parities taken together have a winding number 0.  
	}
\end{figure}

The level-1 invariants are defined in terms of the Wilson loops on the 1-cells, which we can compute as the parity of the winding numbers associated with the $2\times2$ Hamiltonians $h^\pm_{SS'}(k)$ in the $\tau^1$-$\tau^3$ plane. Thus, to compute the invariant $\SW_1^{\G\M} = \SW_{1,+}^{\G\M}-\SW_{1,-}^{\G\M}$, we consider the difference of the winding numbers for $h^+_{SS'}$ and $h^-_{SS'}$, which is well defined, because $h^+_{SS'}(k)=h^-_{SS'}(k)$ for $k=0,\pi$, \ie, for the $\G$ and $\M$ points. For $0 < \abs{m} < 2$, this winding number is given by $\sgn\!(mt)$. To compute $\SW_1^+$, we need the winding number of the positive parity sectors for the full loop $\G\M\X\G$, which is $\sgn\!(mt')$ if $tt'$ is negative and zero otherwise (see Fig.~\ref{fig:D4_winding}). Thus, for $0 < \abs{m} < 2$
\begin{equation}
	\SW_1^{\G\M} = 1, \qquad 
	\SW_1^+ = \Theta(-tt'),
\end{equation}
while both invariants are trivial for $\abs{m}>2$.

\begin{figure}
	\includegraphics[width=0.35\columnwidth]{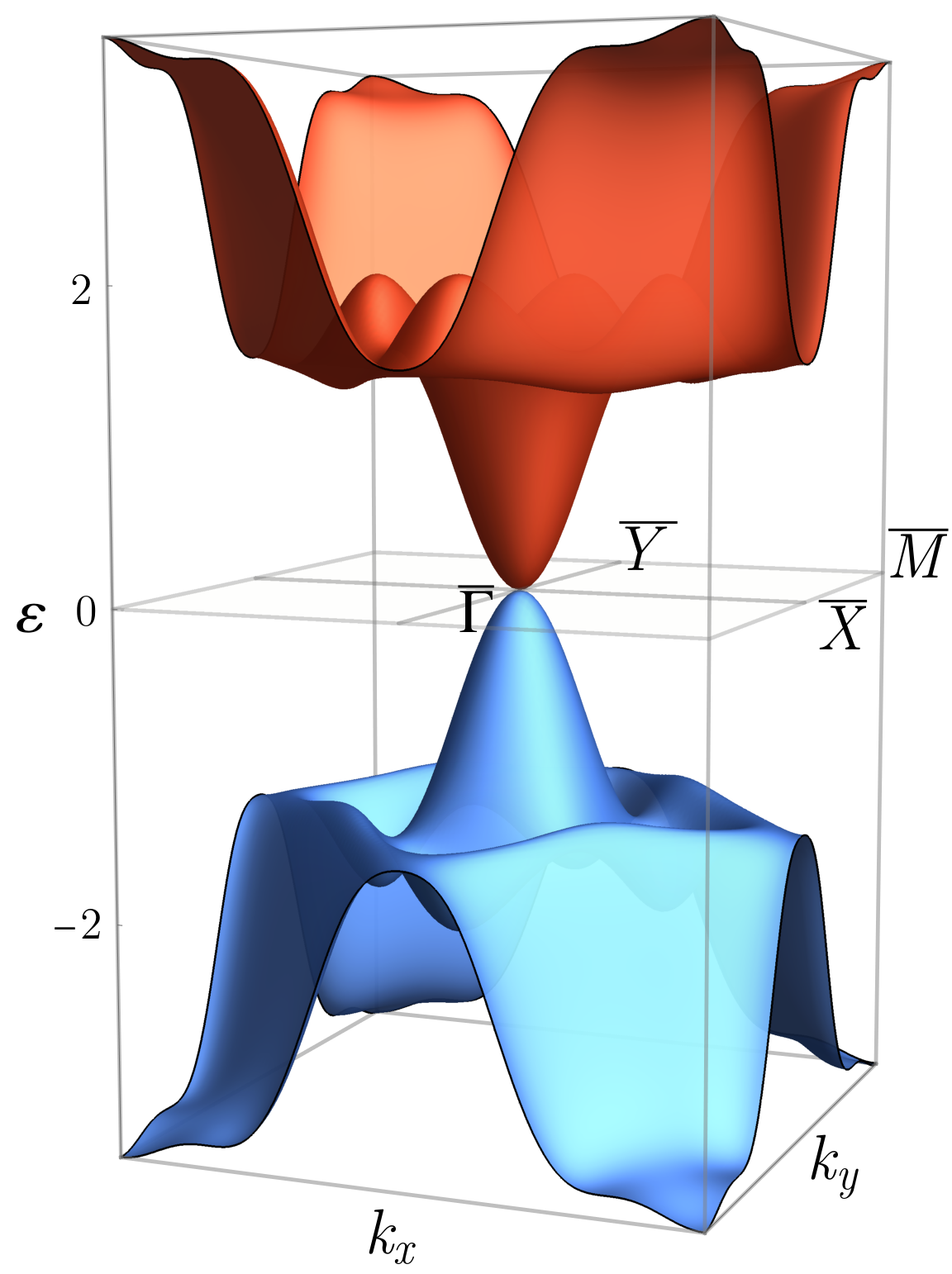}
	\includegraphics[width=0.3\columnwidth]{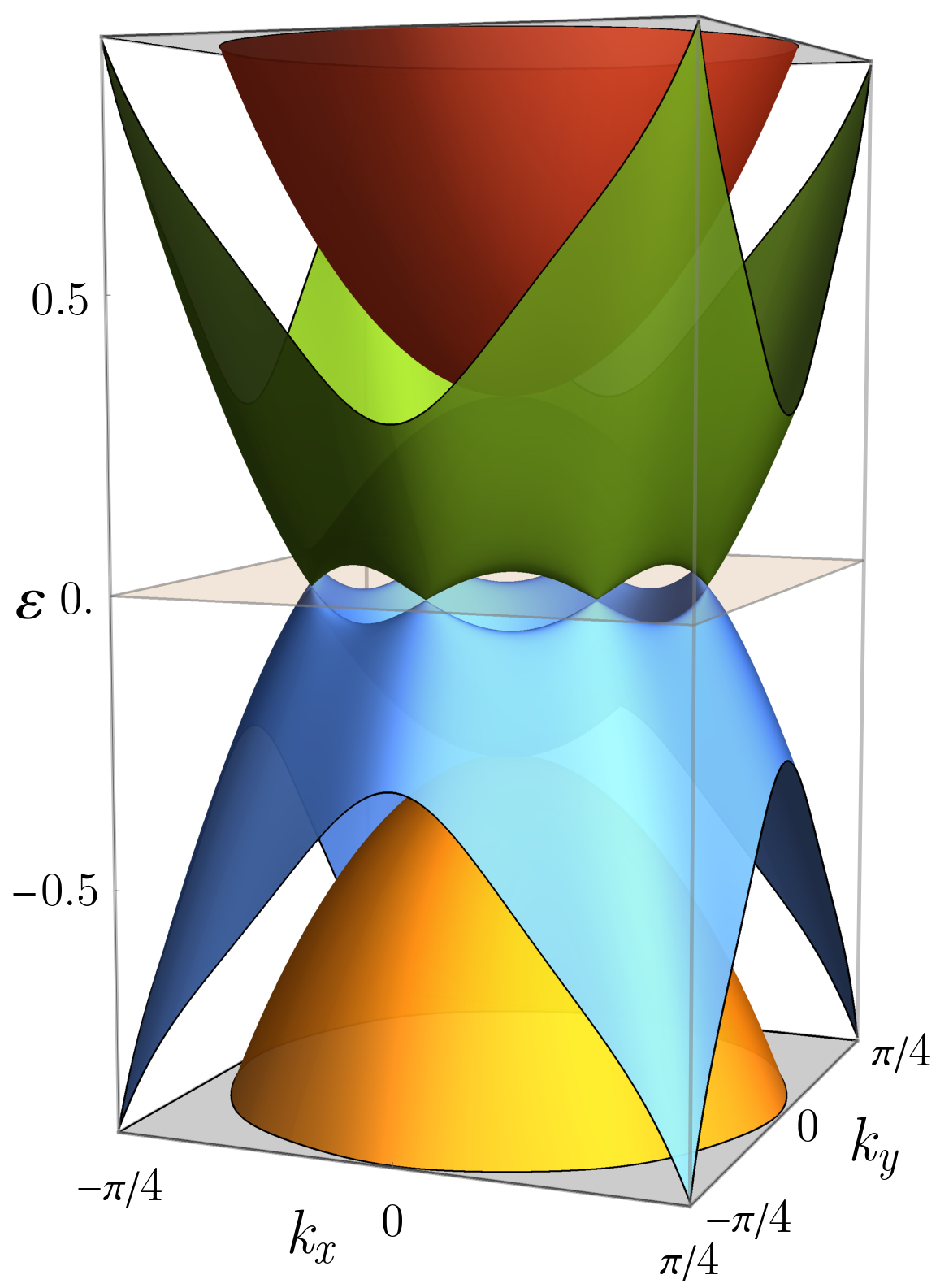}
	\includegraphics[width=0.3\columnwidth]{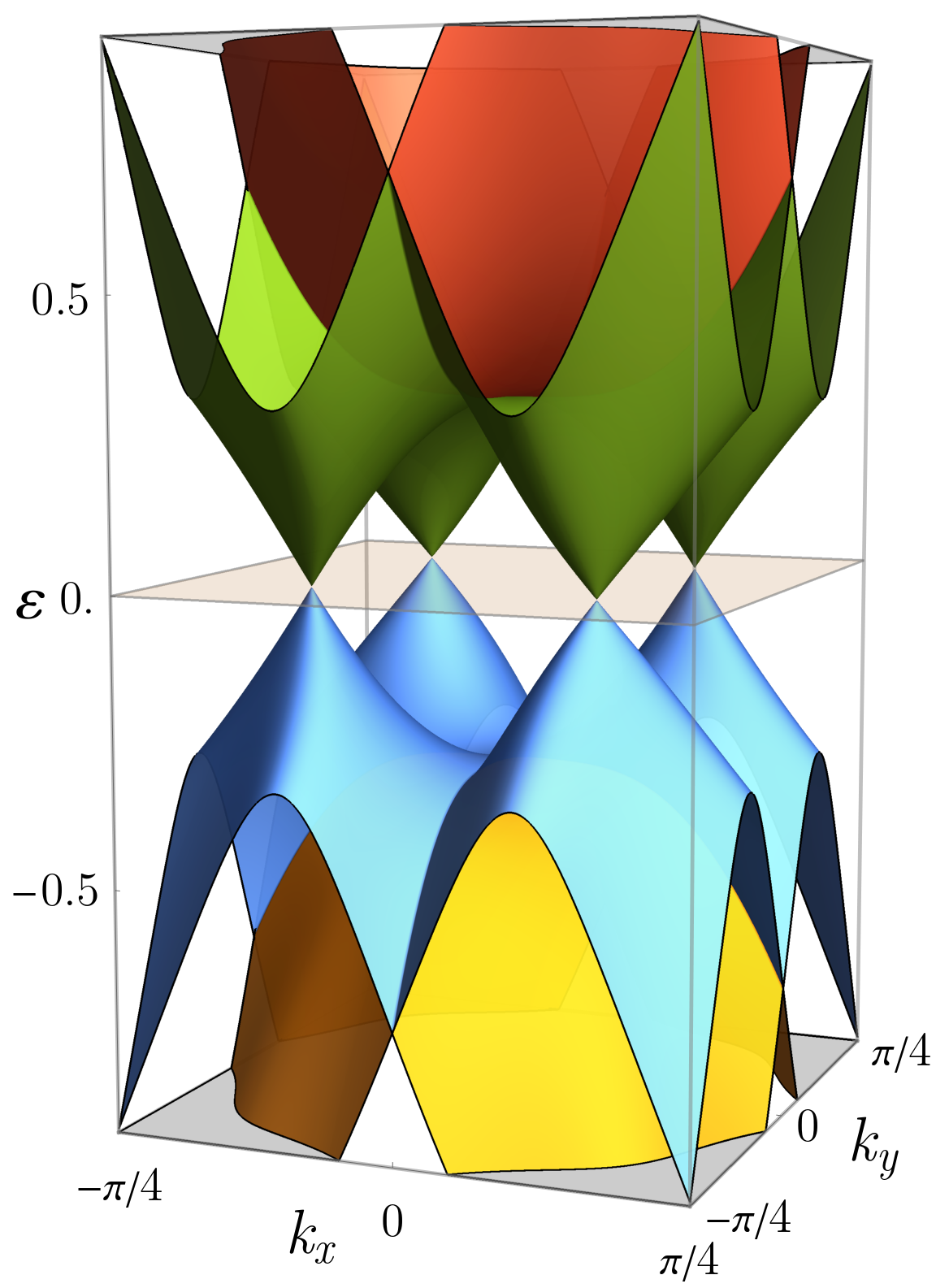}
	\caption{\label{fig:D4_spec} 
		(Left) Surface spectrum showing a quadratic band touching at the $\sG$ point for the model defined in Eq.~(\ref{eq:H3d_def}) with parameters $t = t' = 1$ and $m = 2.5$. 
		The remaining plots show the surface spectrum near the $\sG$-point for two copies of this model with $t'=1$ on both copies (center) and $t'=\pm1$ for the two copies (right). The two copies are coupled by the Hamiltonian defined in Eq.~(\ref{eq:Hmix}) with $\lambda_1 = 0.3$ and $\lambda_2 = 1.0$. The two cases show eight and four Dirac nodes in the surface Brillouin zone. 
	}
\end{figure}

\subsection{Three dimensions}
We consider the 3d Bloch Hamiltonian 
\begin{equation}
	H_{3d}(\vk) = H(\ksurf) +  \sigma^0 \otimes \left[ \cos k_z \tau^1 + \sin k_z \tau^2\right],  \label{eq:H3d_def} 
\end{equation}
where $\vk_\perp = (k_x,k_y)$. For $1<\abs{m}<3$, the restriction of $H_{3d}(\vk)$ to $k_z = 0$ and $k_z = \pi$ are in different topological phases, so that $H_{3d}(\vk)$ has nontrivial level-2 invariants
\begin{equation}
	\Chern^{\G\M} = \sgn\!(mt), \qquad 
	\Chern^+ = \sgn(mt') \Theta(-tt').
\end{equation}
For this model, we only get $\Chern^{\G\M} = \Chern^{\M\X\G} = \pm1$, so that we expect a quadratic band touching at a surface normal to $k_z$ for $t,t'\neq 0$ for any $1 < \abs{m} < 3$ (see the discussion in Sec.~\ref{sec:D4_3d}). This is indeed what we see in numerical exact diagonalization, as shown in Fig.~\ref{fig:D4_spec}(left). 

The boundary modes can be computed analytically following the generalized transfer matrix approach \cite{dwivedi2016tmat}. We follow the notation of Sec. V of  Ref.~\cite{dwivedi2016tmat} and identify the hopping and on-site matrices as \cite[see][Eq.~(124)]{dwivedi2016tmat}
\begin{equation}
J = \begin{pmatrix}
		0 & \id_2 \\ 
		0 & 0
	\end{pmatrix}, \quad 
M(\vk_\perp) = \begin{pmatrix}
		\vb\cdot\vsigma & \mu\sigma^0  \\ 
		\mu\sigma^0  & -\vb\cdot\vsigma
	\end{pmatrix}, 
\end{equation} 
where $\mu(\vk_\perp) = m - \cos k_x - \cos k_y$ and 
\[
	\vb(\vk_\perp) =  2(\cos k_x + \cos k_y) 
	\begin{pmatrix}
		t \sin k_x \sin k_y \\ 
		0 \\ 
		t' (\cos k_y - \cos k_x) 
	\end{pmatrix}. 
\]
The transfer matrix is given by \cite[Eq.~(131)]{dwivedi2016tmat}
\begin{equation}	
	T(\vk_\perp) = \frac1{\mu}
	\begin{pmatrix}	
		\left( \varepsilon^2 - b^2 - \mu^2 \right) \sigma^0 & -\varepsilon \sigma^0 + \vb\cdot\vsigma \\ 
		\varepsilon \sigma^0 + \vb\cdot\vsigma & - \sigma^0
	\end{pmatrix}.
\end{equation}
The modes localized at the surface satisfy 
\begin{equation}
	\varepsilon_\text{bdry}(\vk_\perp) = \pm\abs{\vb(\vk_\perp)}, \qquad 
	\abs{\mu(\vk_\perp)} < 1. 
\end{equation}
The first equation is obtained by imposing the Dirichlet boundary condition at the surface, while the inequality follows from demanding that the states decay into the bulk.  Near the $\sG$ and $\sM$ points, to lowest order, $\varepsilon_\text{bdry}(\vk_\perp) = 2\abs{\delta\vk_\perp}^2$, where $\delta\vk_\perp$ is measured from the gapless point. Since $\mu = m \mp2$ for the $\sG$ and $\sM$ points, we get quadratic band touching at $\sG$ for $1<m<3$ and at $\sM$ for $-3<m<-1$.  

The nature of the anomalous boundary modes is independent of $\Chern^+$ so far. However, we can see its effect by taking two copies of this model. From the transfer matrix, the surface Hamiltonian near the $\sG$ point to the quadratic order in $\vk_\perp$ is given by 
\begin{equation}
	H_\text{surf}(\vk_\perp) \approx 2 t k_x k_y \sigma^1 - 2 t' (k_x^2 - k_y^2) \sigma^3. 
\end{equation}
Assuming $m,t>0$, we get $\Chern^{\G\M} = 1$ and $\Chern^{+} = \Theta(-t')$. We now consider the doubled Hamiltonian with the symmetry allowed mixing term 
\begin{equation}
	\delta H(\vk_\perp) = \lambda_0 \sigma^0 \rho^1 + \lambda_2 k_x k_y (k_x^2 - k_y^2) \sigma^2 \rho^2,
	\label{eq:Hmix}
\end{equation}
where $\rho^\mu$ are Pauli matrices in the space of the two copies. For two identical copies of $H_\text{surf}(\vk_\perp)$ with $t'=1$, 
\[ 
	\Chern^{\G\M} = 2, \;
	\Chern^{+} = 0 \implies 
	\Chern^{\G\X\M} = 2\Chern^{+} - \Chern^{\G\M} = -2. 
\]
Thus, we get one Dirac cone on both $\G\M$ and $\G\X\M$, leading to eight Dirac cones in the surface Brillouin zone, as shown in Fig.~\ref{fig:D4_spec}(center). On the other hand, for two distinct copies of $H_\text{surf}(\vk_\perp)$ with $t'=\pm1$, 
\[ 
	\Chern^{\G\M} = 2, \;
	\Chern^{+} = 1 \implies 
	\Chern^{\G\X\M} = 2\Chern^{+} - \Chern^{\G\M} = 0. 
\]
Thus, we get four Dirac cones in the surface Brillouin zone, as shown in Fig.~\ref{fig:D4_spec}(right).

\bibliography{refs}
  
\end{document}